# Insights into the use of polyethylene oxide in energy storage/conversion devices: A critical review

**Anil Arya and A L Sharma***

Centre for Physical Sciences, Central University of Punjab, Bathinda-151001, Punjab, India

E-Mail address:alsharmaiitkgp@gmail.com

**Abstract**

In this review, latest updates in the poly (ethylene oxide) based electrolytes are summarized. The ultimate goal of researchers globally is towards the development of free standing solid polymeric separator for energy storage devices. This single free standing solid polymeric separator may replace the liquid and separator (organic/Inorganic) used in existing efficient/smart energy technology. As an example polyethylene oxide (PEO) consist of an electron donor rich group which provides coordinating sites to the cation for migration. Owing to this exclusive structure PEO exhibits some remarkable properties such as; low glass transition temperature, excellent flexibility and ability to make complexation with various metal salts which are unattainable by another polymer host. Hence, the PEO is the most emerging candidate that have been examined or is currently under audition for application in energy storage devices. So, this review article first provides the detailed study of the PEO properties, characteristic of constituents of polymer electrolyte and suitable approaches for the modification of polymer electrolytes. Then, the synthesization and characterizations techniques are outlined. The structures, characteristics, and performance during charge-discharge of four types of electrolyte/separators which are Liquid, Plasticized, and dispersed/intercalated electrolyte are highlighted. The suitable ion transport mechanism proposed by researchers in different renowned group have been discussed for better understanding of the ion dynamics in such systems.

## 1. Introduction

Energy is the lifeblood of modern society. Global warming, finite fossil-fuel supplies and city pollution conspire to make the use of renewable energy, together with electric transportation, a global imperative [1]. Due to increased extreme dependency of modern human being in recent years on fossil fuels of finite supply and uneven global distribution, this leads to two problematic consequences: (1) vulnerability of nation states to fossil-fuel imports and (2) $CO_2$ emissions that are acidifying our oceans and creating global warming [2].

So, controlled environment/climate has drawn the attention of scientific community towards development and replacement of fossil fuels by an alternative/efficient energy source such as solar, tidal and wind. In our modern society, an extensive global attention has been focused toward fulfilling the demand of alternative and miniaturized renewable energy storage/conversion devices (ESCDs) with high performance, durability to balance supply with demand. Energy storage is essential in order to restore it as electricity and the perfect approach is to convert chemical energy into electrical energy. The most convenient energy storage devices are batteries having portability of stored chemical energy with the ability to deliver this energy as electrical energy with high conversion efficiency without gaseous exhaust as with fossil fuels [1, 3].

Lithium ion batteries (LIB) are expected to witness remarkable growth as an energy source for modern 24/7 lifestyle because of their application in the consumer market: cell phones, digital cameras, laptops, smartphones,UPS etc., also due to their 2-3 times energy storage per weight and volume as compared to earlier batteries. The first breakthrough in LIB done by John Goodenough and commercialization by the Sony corp. in 1991 has drawn much attention of researchers during the past two decades due to their much higher energy density than alkali metals. Goodenough said in a statement. "*Cost, safety, energy density, rates of charge and discharge and cycle life are critical for battery-driven cars to be more widely adopted*". A battery converts chemical energy into electrical energy and structure is composed of positive electrode as a cathode ($LiCoO_2$), a negative electrode as an anode (Graphite) and electrolyte, as shown in figure 1. Simultaneous movement of ions and electrons occurs in the battery system; former one flow through the electrolyte while later one is generated at the anode and flow towards the cathode via external circuit respectively. The amount of charge of a battery depends on the storage capacity of ions between the electrodes while the capacity is examined by rate of lithium ion migration (to and fro) through the electrolyte. So, the electrolyte is a critical component of the LIB, and it guarantees to the transport of ions as well as provides a physical barrier to prevent the internal short circuiting of the battery. Table 1 summarizes the requirements for selection of electrolyte cum separator for efficient lithium-ion battery system [4].

At, present, LIB is the ideal choice for the automotive sector by Companies including Tesla Inc. and Volkswagen AG due to ease of transportation, environmentally friendly, less cost [5]. Allied Market Research published a report entitled, "*World Lithium-Ion Battery Market: Opportunities and Forecasts, 2015-2022*" and the main point was that the global LIB market is expected to generate revenue of $46.21 billion by 2022, with a CAGR of 10.8% during the forecast period (2016-2022). At a 37 percent revenue share, Li-ion is the battery of choice for portable devices and the electric vehicles. There are no other systems till now that threaten the market of LIB today [figure 2].

Recently in 2017, John Goodenough, inventor of Li-ion battery has developed the first all-solid-state battery cells using sodium- or lithium-coated glass electrolyte and claimed that new battery cells have three times more energy density (1,200 charge-discharge cycles) in comparison to existing rechargeable batteries. He concluded that they store and transmit energy at temperatures (-20 $^{o}$C to 60 $^{o}$C) lower than traditional lithium-ion packs and can be made using globally abundant supplies of sodium. One advantage of this battery is prevention from dendrite growth due to the glass-based electrolyte which enhances the safety of the battery. Also, the Companies including Tesla Inc. and Volkswagen AG have set their sights on lithium-ion to escort in a new generation of plug-in vehicles [7]. The ultimate goal of both academics and industry researchers is to develop an advance polymer electrolyte (PE) system with high ionic conductivity, longer shelf life, high energy density, leak proof, wide operational window, easy process ability, flexible geometry, safety and light weight. Polymer electrolytes act as a bridge between conventional solvent based liquid systems and solvent-free electrolytes [8]. They are an outstanding candidate due to their key application in ESCDs such as solid-state batteries, satellites, electric vehicles, super capacitors, consumer portable electronic and telecommunications markets and will power the implantable biomedical devices, hybrid electric vehicles and military/national security communication and surveillance equipment of tomorrow [9-11]. Figure 3 describes the solutions to increase safety and cell design of LIB using shut-down and redox shuttles additives and Table 2 provides a glimpse of the merits and demerits of the lithium-ion battery.

So, in this review article the recent progress in polyethylene oxide based polymer electrolytes for application in energy storage devices are elaborated fully. The discussion will be focused towards the material selection criteria along with characteristics of constituents of polymer electrolytes, synthesis methods and ionic liquid/plasticizer incorporated polymer electrolyte (GPE) and dispersed/intercalated type solid polymer electrolytes (SPE). Some, latest developments in the field of Gel/Liquid/Dispersed/Intercalated polymer electrolytes with improved and significant results are elaborated. At last, the role of nanofiller, clay and ionic liquid in the enhancement of electrochemical properties with suitable proposed transport mechanisms are included.

**Definitions**

The following terms are used for studying the performance of an energy storage device.

A *battery* is an electrochemical device that converts chemical energy into the electrical energy. It comprises of two electrodes (anode/cathode) with a separator sandwiched between both.

The *cathode* is a positive electrode and reduction of chemical reaction occurs in it, along with acceptance of electrons from an external circuit.

A *separator* provides a physical barrier to prevent the internal short circuiting of the battery and can be a solid/liquid/gel electrolyte. It allows transport of ions by providing a medium for flow.

The *anode* is the negative electrode where oxidative chemical reactions occur, along with the release of electrons in an external circuit.

*Open circuit voltage* is the thermodynamic voltage in the device when no external current is flowing within the system.

*Closed circuit voltage* is the voltage of device when external current flows in the system.

*The charge* is a mechanism of loading of energy when system reverses back.

*Discharge* is the mechanism of unloading of stored energy to the system via an external load.

*Specific energy* is the maximum energy of a cell that can be used per mass of the active material.

*Energy density* is the maximum energy of a cell that can be stored per unit volume.

*Coulombic efficiency* is the ratio of discharging time and charging time.

*C-rate* is the measure of charging and discharging current of an electrochemical cell.

*Glass transition temperature* is the temperature at which an amorphous polymer changes from glassy to rubbery phase and leads to enhancement in polymer flexibility.

## 2. Performance and Classification of Polymer Electrolytes

Polymer electrolytes have received remarkable research investment over last decades due to extreme demand of practical material application in devices and are an alternative approach to replace organic liquid based electrolytes. Lithium ion batteries are suitable as ESCDs due to growing demand in daily life. But some safety issues are concerned with them like poor thermal stability, flammable reaction products and leakage of electrolyte and internal short circuits [12].

Although the polymers are insulators, Wright and co -workers report in 1973 on poly ethylene oxide (PEO) and alkali metal salts (NaI) system confirmed non negligble conductvity in the polymers and that boosted the focus toward the developmet of the polymer electrolytes. Their technological importance was realized by Armand in early 1980's due to flexibility and deformability and introduced as solid polymer electrolytes [13]. To date, the high molecular weight Poly (ethylene oxide) (PEO)-based polymeric electrolytes are one of the most promising candidates for the polymer electrolyte preparation due to their semi crystalline nature and presence of ether group which supports faster ionic transport due to beneficial polymer flexibility. Therefore, PEO-based SPE is the most preferred polymer host in research system due to its flexible backbone and ability to solvate lithium ions, with the coordination number dependent upon the salt concentration and identity of the anion.

Also, the main advantage of PEO is high solvation power. Hence it can form complex easily with many alkalis salts and provides a direct path for cation migration due to the presence of $(–CH_2–CH_2–\ddot{O}–)_n$ in the polymer backbone. But the low conductivity value ($10^{-7}$-$10^{-6}$ S $cm^{-1}$) and poor mechanical properties of PEO at the ambient temperature limit their use in devices. In order to overcome the drawback, the most common approach is altering of the morphology using the nanofiller/plasticizer/ionic liquid and salt with large anion in the polymer matrix. Polymer blending is one of the best adopted techniques to improve the properties of host polymer and is a physical mixing of two individual polymers together. This gives in hand superior properties than individual polymer and can be easily controlled by varying the composition of polymer used [14]. A further boost in electrical and mechanical properties can be done by addition of nanofiller and plasticizer. Another approach is the use of organic solvents, but resulting electrolytes film suffer from the high volatility and thus flammability of the solvent as well as from its reaction with lithium metal electrodes and restricts its use in efficient energy storage device. The ability to control and enhance the mechanical strength independently, ionic conductivity, voltage stability and thermal properties of an electrolyte are highly desirable in the development of electrochemical devices. Generally, in a polymer blend presence of two phases occurs, former one provides a conduction path to the ion and later one provides mechanical stability to the electrolyte. In this review, first of all the polymer electrolyte are divided into different parts followed by the principal requirements for the polymer host, salt, solvent, nanofiller, ionic liquid and nanoclay are discussed. Then the preparation technique and characterization techniques for polymer electrolytes are described in a systematic manner. And the reported results till date based on PEO based system with different modification approaches for improving the properties are introduced. In the last section proposed transport mechanism in the case of different systems of electrolytes are outlined.

### ***2.1.*** *Intrinsic properties of Polyethylene oxide (PEO)*

So far, PEO is the best polymer for the preparation of polymer electrolytes for fast ion transport due to its structure. Another man advantage of PEO is high solvation power, so it can form ccomplex easily with many alkalis salts and provides a direct path for cation migration due to the presence of $(–CH_2–CH_2–\ddot{O}–)_n$ in the polymer backbone.

### ***2.1.1.*** *Basic Structure*

PEO is obtained by the ring-opening polymerization of ethylene oxide ($-CH_2-CH_2-\ddot{O}-$; Mol. Wt. 44 g/mol) (figure 4 a) [15]. Pristine PEO shows a helical structure and a thread of 1.93 nm per unit quadratic cell (figure 4 b). PEO is a linear polymer containing polar ether group (-C= Ö-) with a lone pair of electrons and has strong tendency to coordinate with alkali metal salts to form polymer electrolytes. The ether group of PEO having oxygen atom at center makes it suitable for polymer electrolytes due to effective interaction with the conductive species and provides good solvation properties as suggested by Armand *et al* [16].

### ***2.1.2.*** *Chemical properties*

The high dielectric constant of PEO (~4-5) evidences its suitability to dissolve various salts. Another superior property of PEO is high chain flexibility; it supports the faster transport of cation which is the core requirement for a fast ionic conductor.

### ***2.1.3.*** *Morphological properties*

It is a semi crystalline material with about 70-85 % crystallinity and amorphous elastomeric phase at room temperature. PEO is a semi-crystalline polymer having crystalline phase as well as amorphous phase. As high ionic conductivity occurs in the amorphous regions above $T_g$ (glass transition temperature) and segmental motion of polymer chains promotes the ionic conductivity. The crystalline phase of PEO prevents the ion migration as it is supported by amorphous phase. Since, an amorphous arrangement of molecules lacks the long-range order and it reduces the chain reorganization tendency which supports fast ion transport. So, the main strategy at present is to suppress the crystallinity of PEO and enhancement of amorphous phase so that it can be employed as separator/ionic conductor in energy storage devices [19].

### ***2.1.4.*** *Physical properties*

Poly (ethylene oxide) is a interesting polymer host due to its low glass transition temperature (-60 $^oC$), high melting point (65 $^oC$), excellent flexibility and low toxicity. Low glass transition temperature supports faster ion transport as above glass transition temperature polymer phase changes from rigid to a viscous phase which in turn enhances the polymer flexibility. The thermal stability of PEO can be improved by dispersing nanofiller or plasticizer which will decrease the $T_g$ and increase the degradation temperature of the polymer. This decrease in $T_g$ leads to fast ionic transport or the high ionic conductivity and degradation temperature increase enhances the thermal stability window.

*Approaches for modification of PEO*

Although, PEO has several advantageous such as electrochemical, chemical, and physical properties, poor ionic conductivity makes them inadequate for applications. So, to overcome the above drawbacks various approaches are proposed based on the addition of materials with different properties (figure 5). Among many PNCs, two obvious attempts dominate for improving the properties of PE. The first approach is the incorporation of nanofiller in polymer salt matrix which prevents the polymer chain reorganization, and free volume is produced. This increases the ionic conductivity of polymer electrolytes. Also, the mutual interaction between nanofiller supports more dissociation of salt and conductivity increases due to increase in free charge carriers. The second approach is the incorporation of a low viscosity and high molecular plasticizer Improvement in conductivity can be achieved by following methods:

### ***2.2.*** *Extrinsic properties of PEO electrolyte system*

A simple battery system basically consists of three components, two electrodes (cathode and anode) and electrolyte/separator which play a dual role. So, during operation of a battery polymer electrolyte plays an important role as it provides a path for swimming of ion between both electrodes. For a successful battery system it becomes important to select suitable materials used for the preparation of polymer electrolytes such as polymer host, salt, solvent, nanofiller, nanoclay and an ionic liquid. Different materials have different properties which affect the overall function of electrolyte to do best in battery system and are discussed below:

#### ***2.2.1.*** *Properties of Polymer Host*

In LIB system electrolyte is sandwiched between electrodes and allows migration of Li ions during battery operation. So, the electrolyte is a major factor that affects the electrochemical properties. Polymer host must have some properties so that it can fulfill our criteria of suitable polymer electrolyte:

#### ***2.2.2.*** *Properties of Solvent*

As a good solvent is energetic for the better dissolution of salt and polymer via dipole-dipole interactions (Polymer----Solvent, Salt----Solvent). So, the solvent must possess basic characteristics such as high dielectric constant for batter dissolution of salt, low viscosity for faster ion motion, high electrochemical stability window, inert toward cell components and principle one is it must provide ionic conductive solid electrolyte interface layer on electrodes. The high dielectric constant of solvent is an important parameter as it prevents the formation of the ion pairs since; ion pairs do not play any role in ionic conductivity. The solvent must fully swell and stretch the polymer chains within this so that more favorable interaction can occur between the cation and electron donor group of host polymer [21].

#### ***2.2.3.*** *Properties of Salt*

For fulfilling the criterion of faster ion transport and high ionic conductivity salt plays an important role for the same. Besides solvent and polymer dissociation rate of salt affects directly the ionic conductivity. So, the salt must possess suitable properties such as; low lattice energy for more dissociation of salt, lower cation size for faster movement in the polymer matrix etc. One important requirement is the stability of anion towards electrodes and non-toxic nature. Also the high ionic conductivity and voltage stability window of the salt directly affects the overall conductivity of the electrolyte. As the mobility of ion plays an important role in determining conductivity so salt with smaller cation size and larger anion size are beneficial for obtaining optimal performance of an electrolyte. For better visibility all the properties have been coined in the pictorial diagram as shown below.

### ***2.2.4.*** *Properties of Ionic Liquid*

The incorporation of ILs is recent promising approach adopted to enhance the mechanical and electrochemical properties. Basically, the ILs are molten salts with large anion at room temperature and bulky anion also supports the dissociation of salt in a polymer matrix which affects the electrochemical properties. On, the basis of their composition there are three types of ILs aprotic, protic and zwitter, and used for lithium batteries/supercaapcitor, fuel cells and ionic-liquid-based membranes, respectively [22-23]. Another interesting advantage of the ionic liquid is that the change in the structure of both cation and anion in a very extensive and fashionable way allows us tailoring of the ILs properties accordingly to the foreseen applications [24]. The addition of ionic liquids in the polymer salt system also enhances the interfacial contact of electrodes with electrolyte and enhances the properties of the system.

### ***2.2.5.*** *Properties of Plasticizer*

Another effective approach to enhance the ionic conductivity of a polymer electrolyte based system is the incorporation of a plasticizer such as EC, DEC, PEG. Plasticizers are basically the low molecular weight substances and when added in polymer matrix increases its flexibility. The addition of plasticizer increases the free volume required for transport by penetrating between polymer chains. Since, plasticizers are of a small size so they easily penetrate in between the polymer chain. This modifies the polymer chain arrangement by disrupting the cohesive force between polymer chains. This increases the segmental motion of polymer chains and more free volume is created. As for fast ion transport polymer matric must possess high amorphous content. So to achieve the above the incorporation of plasticizer is the best approach which makes a significant enhancement of amorphous phase as well as better dissociation of salt into free cations and anions. The overall increase in the flexibility is an indication of decrease $T_g$ glass transition temperature and it evidences the increase in conductivity. So, the plasticizer must possess some properties for better performance of electrolyte in the application [25].

### ***2.2.6.*** *Properties of nanofiller*

The addition of nanofiller in polymer electrolyte (PEs) not only enhances the conductivity but also improves the mechanical properties such as physical strength of PEs. The filler acts as a solid plasticizer that enhances the transport properties and reduces the crystallinity, glass transition temperature due to increase in the dielectric constant. Fillers with Lewis acid surface group interacts with both polymer and ion and reduces the ion coupling. The polymer provides the path for conduction of ions and filler affects the polymer stiffness to support the ionic transport.

#### ***2.2.7.*** *Properties of nanoclay*

Among the large numbers of layered solids, clay minerals especially the smectite group are most widely used for the reinforcement of polymer matrices and thus for the fabrication of polymer nanocomposites because of their unique structure and reactivity together with their high strength and stiffness and high aspect ratio of each platelet. The clay layered silicate composites prevent the anion entry in polymer chain while support cation intercalation and increases the conductivity. The low ionic conductivity is generally due to concentration polarization because of simultaneous mobility of ions (cations and anions). So, the above problem can be resolved by nanoclay having high cation exchange capacity which permits the only cation in the polymer matrix. This suppresses the concentration polarization and formation of ion pairs which provides the suitable path for cation migration through intercalated channels of the clay [26].

Table 2 summarizes the characteristics of all constituents of the polymer electrolyte together for the better guide of the eye.

### 2.3. Types of polymer electrolytes

Polymer electrolytes are the best candidate till date over the liquid electrolyte based systems in the commercialized market. A lot of research is focused toward the improvement in various properties of polymer electrolytes. There are three different types of polymer electrolytes based on their composition (i) Ionic liquid based polymer electrolyte, (ii) gel or plasticized polymer electrolyte and (iii) solid polymer electrolyte which is further classified into the dispersed and intercalated type solid polymer electrolyte on the basis of the formation of polymer matrix. All the types are discussed in detail as follows;

#### ***2.3.1.*** *Ionic liquid based polymer electrolytes*

ILs are basically room temperature molten salts which are composed of bulky asymmetric organic cations and organic/inorganic anions. ILs plays the role of plasticizers (which increase the amorphicity and flexibility of the SPEs) and also provide free charge carriers for conduction in SPEs. Unlike inorganic salts that require a solvent for dissociation into cation and anion, ionic liquids do not require any solvent for dissociation and are entirely composed of dissociated organic cations and inorganic/organic anions. Incorporation of ILs is the latest approach as a replacement for plasticizers for improving the conductivity of polymer electrolytes due to their beneficial unique properties such as; low vapor pressure, excellent chemical and thermal stability, non-flammability, and large ionic conductivities. ILs consists of weak Lewis acidic cations and weak Lewis basic anions, causing the interaction between the ions and the polymers to be weak relative to ion-coordinating conventional electrolytes [27].

#### ***2.3.2.*** *Gel polymer electrolytes*

Since the first report by Feuillade and Perche in 1975 gel polymer electrolyte become an attractive system for LIB applications [28]. Basically, GPE is an intermediate state between solid and liquid with a semisolid structure where the liquid electrolyte is incorporated into a solid polymer matrix and unlike the SPE the host polymer itself does not contribute to the conductivity or dissolving the ions. Composite gel/solid polymer electrolyte has the same features, but in which the nano-sized ceramic fillers are dispersed in the polymer matrix. GPE has the ability to suppress the formation of dendrite on the surface of lithium metal and to improve safety. Gel electrolytes have high ionic conductivities and enhanced interfacial contacts between electrode and electrolyte due to encapsulation of liquid electrolyte in the polymer matrix. GPE provides us improved shape flexibility and safety. The addition of plasticizers in the polymer matrix results in an overall enhancement in the ionic conductivity due to penetration between polymer chains which increase polymer flexibility. GPE provides us with the simultaneous presence of cohesive properties of the polymer matrix and diffusive properties of liquid electrolytes [29]. The GPE provides in hand the improved thermal and electrochemical properties.

### ***2.3.3.*** *Solid polymer electrolyte (SPE)*

SPE act as the separator, solvent as well as the electrolyte. SPE is basically a solvent free system with a polymer matrix with the cation of smaller radii as mobile species and bulky anion attached to a high molecular weight polymer chain. The optimized system of SPE is formed by dissolving a low lattice energy salt, host polymer with electron donating group in a solvent with a high dielectric constant for better dissolution of salt. Since, the first report by Wright in 1975 on poly (ethylene oxide) + alkali metal salt complexes (PEO) boosted the research towards the solid state ionic conductors. Further the application of polymer electrolyte in LIB by Armand in 1978 inspired the worldwide researchers. The polymer host provides the coordinating sites for cation migration in the electrolyte system. The interaction of salt with polymer dissociates it in cation/anion and when an electric filed is applied cation moves via segmental motion of polymer chains. The amorphous phase is a basic requirement for ion transport while crystallinity hinders the migration path of action. The ion conduction in SPE is mainly via the two mechanisms, (i) hopping, (ii) via polymer chain segmental motion.

#### *Dispersed type SPE*

These types of electrolytes are also called dispersed solid electrolytes which are high conducting multiphase solid systems. They have a two phase mixture in which the first phase contains an ion conducting solid such as AgI, CuI etc. and the second phase contains inorganic insulating nanofiller such as $Al_2O_3$, $SiO_2$, $TiO_2$, $BaTiO_3$ and zeolite powders. Without changing the structural and chemical nature of compounds, a significant improvement in mechanical strength, hardness and stiffness of the polymer matrix has been achieved with the addition of nanofiller [18]. The ceramic fillers due to its large surface area, prevent local chain reorganization with the result of locking in at ambient temperature, a high degree of disorder characteristic of the amorphous phase, which favors faster transport. Another important parameter is particle size of nanofiller, as it directly affects the ionic conductivity.

*Intercalated/Exfoliated type SPE*

The pioneering work carried out by Toyota Central Research Laboratories on Nylon 6-Montmorillonite (MMT) nanocomposites in 1950. They obtained encouraging enhancements in mechanical properties along with heat distortion temperature and a decrease in permeability, with only few weight percentage content of MMT, by an in situ-polymerization method. Intercalated nanocomposites are formed when the polymer chains are inserted into the layered silicate structure with fixed inter-layer spacings. On the other hand, exfoliated nanocomposites are formed when the individual silicate layers are individually dispersed in the polymer matrix [30-31].

## 3. Polymer electrolyte material technologies

### ***3.1.*** *Polymer Electrolytes Preparation Technique*

#### ***3.1.1.*** *Solution casting technique*

Solution Cast Technique is one of the easiest traditional methods for the preparation of polymer electrolytes. In this method, the polymer is dissolved in the compatible solvent (low boiling point and high dielectric constant e.g. acetonitrile, dimethyl form amide etc.) at room temperature followed by adding the required amount of non-aqueous salt in the polymer matrix. The quantity of salt is obtained the w.r.t amount of polymer host. After complete dissolution of polymer and salt a homogenous mixture is obtained. Then the nanofiller/plasticizer is dispersed in the mixture and further stirring is done till a homogenous solution is obtained. Then the resulting solution is casted on well cleaned petri-dishes (glass/poly propylene) in a controlled atmosphere and dried under room temperature for few days. Further the film can be dried in a vacuum oven at some temperature for complete removal of the solvent. Finally, a free standing mechanical stable polymer film with flexibility is obtained and kept in a vacuum desiccator to avoid from moisture absorption for further characterization [32].

#### ***3.1.2.*** *Hot-press technique*

Hot-press technique for preparation of polymer electrolyte is one of most emerging technique due to the advantage of no need of solvent. Generally, in this technique appropriate amount of polymer and salt are mixed physically and a homogenous powder obtained is heated to a temperature up to the melting point of polymer hoist with continuous mixing. Then the soft slurry obtained is pressed by placing it between two stainless steel electrodes for obtaining good quality polymer films. The Same procedure is followed for dispersion of nanofiller in the polymer salt system. First an optimized composition of polymer-salt is chosen for preparation of composite polymer electrolyte [33-34].

### ***3.2.*** *Characterization Techniques*

Characterization is an essential part to study the requirements of polymer electrolyte before complete application in energy storage device. However, it is customary to employ various techniques for the characterization of polymer nanocomposite materials such as, high mechanical strength, shape variation and high electrochemical properties. So, it becomes important to study the surface properties, electrical properties, structural properties and interfacial properties. Characterization results are helpful in correlating the ionic conductivity, crystallinity and transport properties of polymer electrolytes in a change of morphology and structural properties. Some technique often recommended for characterizing PNCs film is discussed briefly in sub-sections below:

*Morphological/Microstructural study*

X-ray diffraction is used to determine the atomic arrangement within a material. XRD can also be used to calculate the interpleader spacing and crystalline size which helps in understanding the various physical properties of materials. In the case of polymer electrolytes, XRD is used for calculating crystallinity and amount of amorphous phase content. Also the presence of sharp peaks indicates crystalline nature of sample while broads region shows amorphous content. The exterior and interior morphological studies of polymer electrolytes are studied by field emission scanning electron microscope (FESEM), while the presence of a various type of molecular intercation in the polymer system is identified by fourier transform infra-red spcetroscopy (FTIR). FTIR enabled us to obtain moleculat structure withour destructing the sample and position of interacion of the cation with the polymer in the case of the polymer blend. FTIR (Fourier Transform Infra-red Spectroscopy) also used for studying local molecular structure and functional group analysis and interaction between polymer chains and formation of complex

FESEM provides us smothness, homogenity of the surface of the sample. FESEM (Field Emission Scanning Electron Microscope) also used for investigating surface morphology, texture, topography (Done in high vacuum $10^{-5}$-$10^{-7}$ torr with 100 eV to 40 keV electron beam). The sample for FESEM requires is of conductive narure so on insulating samples ultrathin gold coating is done prior to mesuarement. Another advantage of the coating is that surafe resolution is improved. EDX can also be used along with FESEM to investigate the elemental composition of dopant in the polymer matrix. Elemental mapping is also of great interest to know the uniform dipsersion of nanopartilces added in the polymer host.

The crystalline size, crystallinity and amorphous phase region can be obtained from the XRD. The crystallinity is obtained using the equation 1;

$$X_c(\%) = \frac{A_c}{A_c + A_a} \times 100\ \% \qquad (1)$$

Where, $A_c$ = area of crystalline peaks and $A_a$ = area of amorphous peaks

*Electrical Analysis*

The ionic conductvity of the polymer electrolyte is measured by impedance plot (Z'' vs. Z') by placing polymer electrolyte film between the two stainless steel (SS) blocking electrodes in the configuration SS|PE|SS. Prior to the meausrement the thiclness and of film and area of elecrodes need to be measured. Admittance analysis and dielectrc analysis is done by obtaining dielectric constant, dielctric loss using the real (Z') and imaginary part (Z") of impedance (figure 17). The ionic conductvity is calculated using the equation 2;

$$\sigma = \frac{1}{R_b}\frac{t}{A}\ \mathrm{S\ cm^{-1}} \qquad (2)$$

Where,'t' is thickness of sample, A is area of electrodes and $R_b$is bulk resistance obtained by the intercept of spike or semicircle on the real axis.

Further transforming the Nyquist plot data in to the dielectric data, real and imaginary parts of the complex conductivity can be obtained using the given equations 3 a & b:

$$\sigma' = \omega\varepsilon_o\varepsilon'' \qquad (3\ a)$$

$$\sigma'' = \omega\varepsilon_o\varepsilon' \qquad (3\ b)$$

Where, $\sigma'$ is a real part of ac conductivity and $\sigma''$is imaginary part of ac conductivity, ω is 2πf, $\varepsilon'$ is a real part of dielectric constant and $\varepsilon''$ is imaginary part of dielectric constant

*Transference Number analysis*

As studied earlier that ionic conductivity is an important parameter for better performance of electrolytes. Another crucial parameter is ion transference number which for ideal electrolyte is unity. In the case of polymer electrolytes transport is usually by the movement of ions and ionic conductivity dominates while electronic conductivity is negligible. So, transference number enables us to calculate the contribution of both electrons and ions. For an ideal electrolyte $t_{ion}$=1. Wagner AC impedance technique is used to obtain ion transference number by placing polymer film between stainless steel electrodes (SS‖PE‖SS). DC polarization technique is used to calculate the cation transference number by measuring the current and steady state resistance before and after polarization (Li‖PE‖Li). As low cation transference number affects the energy density and power density of battery and supercaapcitor due to less to and from the jump of the ion. Following equation4 & 5 are used for calculating the ion and cation trenaference number.

$$t_{ion} = \left(1 - \frac{I_e}{I_t}\right) \qquad (4)$$

Where, $I_t$, is total initial current due to ions and electrons contribution and $I_e$ is residual current due to electrons contribution only.

For cation transference number

$$t_{cation} = \left(\frac{I_s(V - I_oR_o)}{I_o(V - I_sR_s)}\right) \qquad (5)$$

Where, V is the applied dc voltage for sample polarization, $I_o$ and $I_s$ are the currents before and after polarization, $R_o$ and $R_s$ are the initial and steady-state resistance of the passivation layers

*Stability Analysis*

The thermal, mechanical and voltage stability of polymer electrolyte is an essential requirement duering cell fabrication for application in energy storage devices. Electrochemical stability window is obtained by linear sweep voltammetry (LSV) from the variation of current and voltage and it should be equal to electrode stability window. In this technique, a fixed potential is applied and breakdown voltage at the abrupt change of current is ESW. Since mechnaical stability desirable for a safe device. Mechanical properties of polymer electrolute are characterized by obtaing young modulus by stress-starin curev and storage modulus /loss modulus by dynamical mechanical analysis. As higher young modulus value indicates improved the mechanical strength of electrolyte. So, in raegard to as said above the Stress-strain Curve is used for calculating Young's modulus and Dynamical Mechanical Analysis (DMA) for obtaining storage modulus and loss modulus

*Thermal analysis*

Differential scanning calorimetry (DSC) is a thermal analytical technique used to quantitatively interrogate the thermal phase transitions in an inert atmosphere and may be of two types endothermic or exothermic; endothermic are melting, glass transitions, or decompositions while the exothermic transition includes crystallization of the material in the polymeric materials. DSC measures the temperatures and heat flow associated with transitions in materials as a function of time and temperature in a controlled atmosphere Ar or $N_2$. In the case of polymer electrolytes glass transition temperature, melting point and crystallinity enable us to obtain information regarding the transport of ions in the polymer matrix. Glass transition temperature ($T_g$) is when polymer phase changes by a step from glassy to rubbery or viscous phase and polymer flexibility increases. Glass transition temperature is a crucial parameter since it is linked with the polymer flexibility and directly correlated with the enhancement in the ionic conductivity. From the experimental measurements further parameters can be obtained that will be beneficial foe the study of polymer materials and correlation with the XRD and the impedance results.

The crystallinity of the synthesized polymer can be obtained using equation 6 and is an important parameter since it is linked with the XRD which provide key insights for enhancement in the electrochemical properties.

$$X_c(\%) = \frac{\Delta H_m^{sample}}{\Delta H_m^{o}} \times 100\ \% \qquad (6)$$

Where, $\Delta H_m^{sample}$ is experimental melting enthalpy and $\Delta H_m^o$ is melting enthalpy for 100 % crystalline host polymer

Another important technique for investigating the thermal stability and thermal degradation of polymer electrolytes is the Thermal Gravimetric Analysis (TGA) and it provides us the decomposition range of polymer salt and the whole polymer matrix. It can be operated in the dynamical and isothermal mode. Former one deals with the simple heating of the sample placed inside the alumina crucible up to ~500 $^oC$ at a fixed rate. While in the later one as the name suggests that isothermal means constant temperature. So, in this sample is kept for the sometime at constant temperature to study the thermal stability of the materials. One important parameter during the measurement of the both technique is scale rate value. Although it is not specified any value lower scan rate is always preferred for the proper analysis of the sample.

*Electrochemical Analysis*

The electrochemical performance of a battery and supercaapcitor enables us to measure the energy density, power density, specific capacity and cyclic stability. Galvano-static charge/discharge and cyclic voltammetry provide us above information's regarding the battery and supercaapcitor. Cyclic voltammetry (CV) is done for measuring the specific power ($P_{sp.}$), specific energy ($E_{sp.}$) at different scan rate in a user defined voltage range. The galvano-static charge/discharge (CD) is beneficial for obtaining discharge capacitance and cyclic stability of the battery. Specific capacitance in the case of supercaapcitor can also be obtained by measuring the discharge time interval ($\Delta t$) and current applied (i) as given below by equation 7-9;

$$C_d = \frac{2i \times \Delta t}{m.\Delta V} \quad (7)$$

Where, i = discharge current, $\Delta t$ = discharge time interval, m= mass of active material and ESR is inward resistance of cell

$$Energy_{spec.} = \frac{C_d V^2}{8} \quad (8)$$

$$Power_{spec.} = \frac{V^2}{8m \times ESR} \quad (9)$$

## 4. Reported results (Current Updates)

The remarkable progress for a safe and alternative energy source with high energy density/capacity for the commercial industry can be achieved by lithium ion battery. But, there are safety issues faced by LIB using liquid or gel polymer electrolyte and can be accomplished using solid polymer electrolyte. Therefore, many methods have been reported to optimize the electrochemical properties and mechanical/morphological properties of polymer electrolytes which can satisfy the need of all solid-state lithium polymer electrolyte batteries. A review of the significant research findings on polymer electrolytes is reported below: The pioneering work on polymer electrolyte was started by Wright and co-workers who were the first to report the ionic conduction in PEO-alkali metal salt complexes. Later Armand in 1978 explored the various device applications of these materials.

### *4.1. Polymer in salt system*

From last four decades a lot of salts were used to prepare the polymer salt system and The addition of salt in the polymer hots is one of simpel approach to enhance the conductvity, transference nymber and stability window. The role of salt is that it reduces the crystallinity and enhances the polymer chain segmentakl motion which supports fast ionic transport. One, an important parameter is glass transition tempearture which is directly related with the ionic conductvity. Above glass tranition tempearture a rigid syetm change into a rubbery system which provides shape flexibility to polymer electrolytes. So, over all salt plays a major role in polymer electrolyte system.

The influence of viscosity-molar masses of PEO and fraction of salt was investigated on the ionic conductivity, thermal properties, crystalline structure and intermolecular interaction [35]. DSC analysis concluded that the solvation of salt in the polymer is independent of the molar mass of PEO and almost negligible differences are observed in $T_g$ as well as $\Delta C_p$ for different molar masses of PEO, as observed. All SPEs shows amorphous phase content for lower salt content and is almost independent of the molar mass of PEO. XRD pattern of pure PEO exhibits two distinct peaks at 19° ((120) plane) and at 24° ((032) and (112) planes) and confirms the monoclinic crystalline structure. The pure polymer system and polymer salt system undergoes phase separation as evidenced by XRD (figure 6). FTIR spectrum shows no shifting in the two shoulders at 1144 and 1061 $cm^{-1}$ and is an indication of almost no influence of polymer molar mass. For SPEs prepared with morphologies close to equilibrium shows no significant difference of $\sigma_{DC}$ for PEO with different molar masses at $W_S$ =const. Further the mobility and diffusion coefficient of the charge carriers in PEO host polymer are also independent of the molar mass of PEO for the morphologies close to equilibrium while for non-equilibrium morphologies increased the value of the diffusion coefficient is observed for PEO with $M = 4 \times 10^6$ g $mol^{-1}$.

Boschin *et al* [36] synthesized solid polymer electrolyte $NaFSI(PEO)_n$ and $NaTFSI(PEO)_n$ with different ether oxygen to sodium (O:Na) molar ratios (n). The glass transition temperature ($T_g$) obtained in the first scan from DSC analysis for $NaTFSI(PEO)_9$ was at 237 K and during the second scan the $T_g$ increases to 242 K, and may be attributed to increased $Na^+$ concentration in the amorphous region. For $NaFSI(PEO)_{20}$ and $NaFSI(PEO)_6$ based SPE no change in $T_g$ was observed. DSC analysis concluded that for low salt concentration (n = 20 and n = 9) there was almost little or no impact on the dynamics as observed by $T_g$.

The highest ionic conductivity at 293 K was $4.5\times10^{-6}$ $Scm^{-1}$ for NaTFSI$(PEO)_9$, while at 343 K both NaTFSI $(PEO)_n$ and NaFSI$(PEO)_n$ with n = 20 and n = 9 show similar conductivities and follows a non-linear, Vogel-Tammann-Fulcher (VTF) behavior on cooling which evidences the presence of amorphous phase. The conductivities for NaTFSI $(PEO)_{20}$ and NaTFSI$(PEO)_6$ overlap in the range 283–312 K, and at 343 K the conductivity of NaTFSI$(PEO)_{20}$ ($3\times 10^{-4}$ $Scm^{-1}$) is slightly higher than for NaTFSI$(PEO)_6$ ($1.3\times 10^{-4}$ S $cm^{-1}$). The observed difference in conductivity may be due to two reasons; one is the difference in $T_g$ (about 20 K) and second is a number of ion pairs. Raman spectra (Figure 7) shows bands at 844 $cm^{-1}$ ($CH_2$ rocking) and 860 $cm^{-1}$ (C-O stretching), typical of neat PEO for both NaX$(PEO)_9$ and NaX$(PEO)_6$. While, for NaTFSI$(PEO)_9$ and NaTFSI$(PEO)_6$ the 740 $cm^{-1}$ band is assigned to "free" TFSI, containing two TFSI conformers, $C_1$ (cis) and $C_2$ (trans). $T_e$ band observed at 744 $cm^{-1}$ is assigned to $Na^+$-TFSI contact ion pairs, The deconvoluted spectra evidence larger amount of ion pairs (44±16) % NaTFSI$(PEO)_6$ as compared to both NaTFSI $(PEO)_9$ and NaTFSI$(PEO)_{20}$, is in correlation with the ionic conductivity data. It was concluded that the NaFSI $(PEO)_{20}$ and NaFSI$(PEO)_9$ shows almost similar relative amounts of contact ion pairs, while for higher concentration, n = 6, the FSI based SPE indicates the absence of ion aggregation.

Polu *et al* [37] prepared the solid polymer electrolyte using the high molecular weight poly(ethylene oxide) (PEO) complexed with lithium difluoro(oxalato)borate (LiDFOB) by using solution casting technique. XRD analysis suggests the decrease of crystallinity on the addition of salt in PEO and is due to destruction effect of the lithium salt on the ordered arrangement of the polymer chains. This modification in the structure of host polymer indicates the enhancement of the amorphous content. DSC analysis evidences the decrease of crystallinity and glass transition temperature which indicates the increase in polymer flexibility. Further thermal stability was upto 240 $^oC$ and is better than pure PEO. The highest ionic conductivity was $3.18\times10^{-5}$ S $cm^{-1}$ for Ö/Li=10 and may be due to increasing in a number of charge carriers (figure 8). The decrease in ionic conductivity at high salt content is due to ion association or charge multiplet formation. As it is well known that ionic conductivity dominates in the amorphous phase so impedance data is in good agreement with both XRD and DSC.

Recently, the preparation of novel flexible and freestanding solid polymer electrolyte based on polyethylene oxide (PEO) and poly (vinyl pyrrolidone) (PVP) complexed with lithium nitrate ($LiNO_3$) was reported using solution casting method by Jinisha *et al* [32]. The absence of corresponding peaks corresponding to salt in XRD pattern indicates complete dissociation of salt. The $Li^+$ and $NO_3^-$ ions of the salt also interact with the polymer back bone of PEO which improves the order/alignment of PEO structure and is evidenced by the small increase of peak intensity for characteristic peaks of PEO beyond 15 wt. % salt content. FTIR spectrum confirms the interaction of $Li_+$ with the carbonyl group in PVP and results in the disruption of crystallization of the complex. Also, the absence of vibration bonds for the salt in all SPEs confirms the complete dissolution of salt in the polymer blend. The addition of PVP in PEO changes the surface with micro-cracks of pure PEO to a porous surface. Further addition of salt smoothened the surface morphology indicating the decrease of crystallinity of PEO and is in good agreement with FTIR and XRD analysis. TGA analysis confirms the improved stability of SPE above 400 $^{o}$C as compared to PEO having 340 $^{o}$C. DSC analysis shows a decrease of crystallinity upto 15 wt. % $LiNO_3$ (34.04 %) and at higher than this increase of crystallinity is due to the ion aggregation. The interaction of salt with polymer disturbs the order of polymer chains which in turn affects the relative crystallinity and results in a flexible SPE. The highest ionic conductivity is $1.13\times10^{-3}$ S cm$^{-1}$ for 15 wt. % $LiNO_3$ and further addition of salts restricts the ionic and polymeric segmental mobility due to the formation of ion pairs and ion clusters. $Li^+$ transference number obtained by Evand and Vincent method was 0.332 and is in good agreement with theoretical value (figure 9 a & b). The electrochemical stability window was of the order of 5 V and makes them a suitable candidate for high voltage Li ion battery.

Another report by Karuppasamy *et al* [38] deals with the preparation of solid polymer electrolyte by adding bulky anion based novel lithium bisnonafluoro-1-butanesulfonimidate salt in PEO polymer matrix by solution cast method. ATR-FTIR analysis confirmed the complexation of LiBNFSI with PEO host. The addition of salt in polymer matrix shifts the peak toward lower wavenumber side which indicates the interaction between cation and ether group of the polymer. The deconvoluted spectra of C-S of stretching of LiBNFSI evidence the increase of free ion concentration with increase of salt concentration. The XRD patter confirms the complete dissolution of salt in the polymer matrix and a decrease of peak intensity indicates the decrease of crystallinity. The increased amorphous phase supports fast ion conduction. DSC analysis also supported the XRD data and may be due to the weakening of interaction between polymer chains and salt which in turn reduces the amount of energy consumed to break the bond. The decrease of melting point toward lower temperature shows an increase of amorphous content (figure 9 c). The mechanical strength was higher for Ö/Li=20 (2 MPa) and decreases with the increase of salt due to an increase of polymer flexibility. The highest conductivity was $2.2\times10^{-4}$ S cm$^{-1}$ (at 333 K) for Ö/Li=14 due to the faster migration of ions. The cation transference number obtained was 0.31 and is suitable for the electrolyte in application devices. The electrochemical stability window was 5.4 V as observed from linear sweep voltammetry.

Anilkumar *et al* [39] investigated the solid polymer blend electrolyte based on poly (ethylene oxide) (PEO) – Poly (vinyl pyrrolidone) (PVP) as polymer and $Mg(NO_3)_2$ as salt. A free standing and flexible film was obtained with decreased intensity of crystalline peaks as evidence by XRD. The salt was completely dissociated in the polymer matrix and enhanced amorphous region is observed due to PVP. FESEM analysis of PEO shows cracks in the film which gets modified on blending with PVP and results in a flexible film. Further addition of salts in polymer matrix results in microstructure with pores (1.5 µm) which are required for fast ionic transport. The microstructure topography gets changed and is in correlation with XRD data. Thermo gravimetric analysis shows film stability up to 380 ºC for all SPEs (figure 9 d). The maximum ionic conductivity was observed for Mg 30 around $5.8 \times 10^{-4}$ S $cm^{-1}$ at RT and is four orders higher than pure polymer blend [40]. Also, the activation energy was the minimum for Mg 30 around 0.31 eV. The ionic transference number was 0.997 and cation transference number 0.33. The electrochemical stability window was apprx. 4 V obtained from CV and open circuit voltage for Mg ion cell was 1.46 V.

The ionic conductivity, structure, and segmental motions of two types of high-crystallinity $PEO/NaPF_6$ electrolytes with ethylene oxide Ö/Na=8, 6 was investigated by the Luo *et al* [41]. DSC analysis concludes that high melt-point crystalline phase forms during the isotherm at 343 K while the low melt-point crystalline phase forms at 295 K on cooling and melts at 325 K upon heating. XRD analysis indicates complete complex formation as no corresponding peak for PEO and salt was observed. Further, a $^1H$ static NMR spectrum was used to calculate the crystallinity at room temperature. The ionic conductivity of $PEO_8:NaPF_6$ is $7.7 \times 10^{-7}$ S $cm^{-1}$ and is five times more than that of $PEO_6:NaPF_6$ ($1.4 \times 10^{-7}$ S $cm^{-1}$). Also, NMR analysis shows a high segmental mobility in $PEO_8:NaPF_6$ due to the low activation energy of 0.31 eV, which is 3–4 times lower than those found in the $PEO_6:NaPF_6$. The cooperative motion of several ethylene oxide monomers occurs due to strong coordination between the polymer segments and the $Na^+$ ions, while the large-angle segmental motion in PEO-salt complex along coil axis enhances the ion motion or conductivity.

Zhnag *et al* [42] prepared the solid polymer electrolyte with lithium bis(fluorosulfonyl)imide (LiFSI) as salt and high molecular weight poly(ethylene oxide) as host polymer. DSC analysis shows that the LiFSI-based polymer electrolyte shows a lower $T_g$ (−45 ◦C) than that of the corresponding LiTFSI-based one (−36 ◦C). The lower value of glass transition temperature suggests a stronger plasticizing effect of LiFSI as comared to the LiTFSI, The polymer electrolyte of LiFSI/PEO exhibits ionic conductivity as high as $1.3\times10^{-3}$ S $cm^{-1}$ at 80 ˚C. The cation transference number obtained was 0.14 for LiFSI/PEO and 0.18 for LiTFSI/PEO at 80 ºC. The electrochemical stability window obtained for LiFSI/PEO is about 5.3 V vs. $Li^+/Li$, and LiTFSI/PEO shows improved stability window of about 5.7 V vs. $Li^+/Li$ (figure 10). The electrochemical analysis of cell using LiFSI/PEO electrolyte at 80 ◦C shows a specific charge capacity of 159 mAh $g^{-1}$ and discharge capacity of 146 mAh $g^{-1}$ at the first cycle at C/5 rate with coulombic 99% after the first cycle.

Michalska *et al* [43] prepared a new family of fluorine-free solid-polymer electrolytes, using poly (ethylene oxide) (PEO) as polymer and sodium salts with diffuse negative charges: sodium pentacyanopropenide (NaPCPI), sodium 2, 3, 4, 5-tetracyanopirolate (NaTCP) and sodium 2,4,5-tricyanoimidazolate (NaTIM). The electrochemical stability of electrolytes was lowest for NaPCPI (3 V vs. $Na^+/Na$), 5 V vs. $Na^+/Na$ was obtained for electrolytes with NaTCP and NaTIM salt shows stability window of 4.5 V (figure 11). The ionic conductivity was measured for PEO doped with $NaPF_6$, NaTCP and NaTIM, at O:Na molar ratios of 16:1, showed optimal σ values and for NaPCPI membrane was 20:1. $PEO_{16}NaPF_6$ exhibited σ of the order of 0.1 mS $cm^{-1}$ above 60 °C. The novel $PCPI^-$ and Huckel-type anions showed ionic conductivities on the order of 0.1 mS $cm^{-1}$ above 50 and is higher than $PEO_{16}NaPF_6$ based system. At 70 °C, $PEO_{16}NaTCP$ showed ionic conductivity values (so-called liquid-like) greater than 1 mS $cm^{-1}$. It was concluded that the ionic conductivity of the electrolyte depends on the cation-anion interactions and ion pair having lower cation-anion interaction energies dissociate more salt which increases the number of free charge Carrier. Also, the bulkier anion acts as a plasticizing agent, and reduces the crystallinity of the polymer matrix which results in increased polymer flexibility and mobility of the charge carriers. The increase in ionic conductivity may be attributed to a negative charge diffused over the entire anion via conjugated π bonds and electron-withdrawing cyano substituents. The calculated dissociation energies of the most stable ion pairs decreases as salts change from $NaPF_6$ > NaPCPI > NaTIM > NaTCP, and corresponding Ion pair dissociation energies are 485, 443, 420 and 407 kJ $mol^{-1}$, respectively. TGA analysis shows the thermal stabilities for polymer electrolyte with NaPCPI, NaTCP and NaTIM salts upto 600 °C, 540 °C and 570 °C respectively. DSC analysis evidences the correlation of change in $T_g$ with respect to salt concentration. Also, the aromatic nature of Huckel-type anions provides flexibility to PEO matrix which supports faster ion transport as evidenced by an enhancement in ionic conductivity.

Ma *et al* [44] reported the preparation of SPE based on perfluorinated sulfonimide salt-based SPEs, composed of lithium (trifluoromethanesulfonyl) (n-nonafluorobutanesulfonyl)imide ($Li[(CF_3SO_2)(n\text{-}C_4F_9SO_2)\text{-}N]$, LiTNFSI) and poly (ethylene oxide) (PEO) using solution cast technique. The obtained ionic conductivity value was $3.69 \times 10^{-4}$ S $cm^{-1}$ at 90 °C and an anodic electrochemical stability at 4.0 V vs $Li^+/Li$, and sufficient thermal stability (>350 °C). As shown in figure 12 a, the Li−S cell with LiTNFSI/PEO ($EO/Li^+$ = 20) blended polymer electrolyte affords the average discharge specific capacity of ∼450 mAh $g^{-1}$ at 0.2 C for more than 200 cycles, which is higher than that of ∼310 mAh $g^{-1}$ at 0.5 C at 60 °C. Fortunately, the cycling stability is superior with the negligible capacity loss for hundreds of cycles. However, this trend in consecutive capacity loss (i.e., poor cycling stability) during cycling (200 cycles) at 0.2C at 60 °C is observed in the LiTFSI-based electrolyte (figure 12 b). This indicates that the LiTNFSI-based SPEs would be potential alternatives for application in high-energy solid-state Li batteries.

Judez *et al* [45] reported the preparation of polymer electrolyte based on lithium bis(fluorosulfonyl)imide ($Li[N(SO_2F)_2]$,LiFSI) salt and poly(ethylene oxide) (PEO) as polymer host. The ionic conductivity obtained was $1.0 \times 10^{-4}$ S $cm^{-1}$ at 70 ºC. Also, the good electrochemical stabilities of Al current collector were seen with both SPEs as evidenced by the experiments of cyclic voltamogram and electrochemical impedance spectroscopy of Al electrode.

The cycling performance of Li-S polymer cells is studied and presented in Figure 30. For the first cycle almost same discharge capacity (900 $mAhg_{sulfur}^{-1}$ 30 wt. % S vs. 800 $mAhg_{sulfur}^{-1}$ for 40 wt. %) was observed with 30 wt. %, 40 wt % S. While, for 50 wt. % sulphar loading the discharge capacity was 600 $mAhg_{sulfur}^{-1}$ and may be due to the electronically insulating nature of S (figure 13 a). Further increase of thickness to 65 μm reduces the discharge capacity to 250 $mAhg_{sulfur}^{-1}$. The above decrease may be due to longer and more tortuous ion pathway leading to salt depletion because of low cation transference number. Also, the lack of back diffusion of polysulfide causes insulating layers and prevents utilization of the deeper-residing sulfur and insulating nature of sulphar also plays an active role with an increase of thickness. The specific discharge capacity of all the cells maintains 100−400 $mAhg_{sulfur}^{-1}$ at 0.5C and no capacity fading was observed after 50 cycles, irrespective of thickness or sulphar content (figure 13 c).

Ibrahim *et al* [46] studied the effect of $LiPF_6$ on the PEO as polymer host. The highest conductivity is obtained at $4.1\times10^{-5}$ S $cm^{-1}$ for 20 wt.% of $LiPF_6$ and beyond that decrease in conductivity is due to decrease of a number of free charge carriers which reduces the segmental motion of polymer chain. Both melting temperature and crystallinity shows decrease which evidence the enhancement of amorphous phase.

Thiam *et al* [47] prepared the solvent-free and oligomer-free 3D polymer electrolytes. The DSC analysis shows the decrease of $T_m$ and $\Delta H_m$ after cross-linking and it may be due to the constraint produced by the cross-linking which reduces the tendency of the chain re-organization. While, the hydrogenation does not affect the glass transition temperature. The highest ionic conductivity was $1.7\times10^{-4}$ S $cm^{-1}$ and is highest of the r network electrolytes. It was concluded that the network electrolyte is better than the other modified polymer electrolytes. Table 4 summarizes the some important results of Ionic conductivity, cation/ion transference number, thermal stability, voltage stability window and cyclic stability of polymer salt complex.

### ***4.2.*** *Ionic liquid based polymer electrolytes*

Ionic liquids are basically the low-temperature molten salts and are characterized by weak interactions, due to the combination of a large cation and a charge-delocalized anion. As the addition of salt to the pure polymer having insulating nature evidence the increase of ionic conductivity up to two or three order. So, the addition of an ionic liquid suppresses the crystallinity and promotes the release of more number of charge carriers. Another suitable approach is the incorporation of ionic liquid (EMITf, EMIM-TY,

BMPyTFSI $PYR_{13}FSI$ EMIMTFSI, BMITFSI) having a bulky cation and large anion win the polymer host. Also, the incorporation of IL overall increases the ionic conductivity of electrolyte system [22].

Das *et al* [49] prepared the polyethylene oxide/poly (vinylidene fluoride-hexafluoropropylene)-lithium bis(trifluoromethane)sulfonamide as salt and 1-propyl-3-methyleimidazoliuum bis(trifluromethylesulfonyl)- imide (PMIMTFSI) ionic liquid. DSC analysis concludes that decrease in glass transition temperature with the increase of IL content indicates the increase in polymer chain segmental motion and it leads to fast ion transport. The thermal stability of all samples was almost 350 $^{o}$C and is almost independent of the content of IL. Figs. 14 (a)–(d) shows the FE-SEM images for different IL contents and shows sponge like homogeneous structure with micropores is observed in the films and is attributed to the porous structure of PVDF-HFP polymer. The addition of IL in polymer salt system smoothens the surface evidence the decrease in crystallinity.

Chaurasia *et al* [50] investigated the modification in crystallization kinetics behavior of PEO+$LiClO_4$ with IL 1-ethyl-3-methylimidazolium hexafluorophosphate ($BMIMPF_6$) via solution cast technique. Polarizing optical microscope (POM) shows spherulites (order of microns and millimeters) in the form of spherical aggregates of lamellae which evidence the polymer crystallinity in the absence of pronounced stress or flow and the spherulites grow radially which is a measure of the mechanism of crystallization in the polymer. It is observed that the growth is very fast for 10 wt. % salt concentration and diameter increased from ~100 µm (within 10 s) to 310 µm ( with time 120 s). The addition of ILs in polymer salt system reduces the spherulites size an increased the number of nucleating sites. The spherulites growth rate ($G_s$) obtained from the spherulite size vs. time plot for polymer salt system is 1.77 µm/sec and decreases to 0.67 µm/sec for polymer salt with the ionic liquid. The decline in $G_s$ value indicates the incorporation of IL in polymer salt system which hinders the spherulites growth rate of polymer PEO.

Gupta *et al* [51] prepared polymer electrolyte using polymer poly (ethylene oxide) (PEO), lithium salt bis(trifluoromethylsulfonyl) imide (LiTFSI) and ionic liquid (IL) trihexyltetradeylphosphonium bis(trifluoromethylsulfonyl)imide via solution cast technique. XRD analysis evidences the simultaneous presence of halo region and crystalline peaks which indicate the semi crystalline nature of PEO and addition of IL enhances the amorphous phase (figure 15). SEM analysis shows the ion association at high IL content due to recrystallization. All samples were thermally stable up to 350 $^{o}$C. Cation transference number obtained was 0.37 and ion transference number was 0.99. The highest conductivity was $4.2\times10^{-5}$ S $cm^{-1}$ for 20 % IL and electrochemical stability window was 3.34 V.

Polu *et al* [52] investigated the effect of ionic liquid 1-ethyl-3-methyllimidazolium bis(trifluoromethylsulfonyl)imide (EMImTFSI) on poly(ethylene oxide) (PEO)+lithium difluoro(oxalato)borate (LiDFOB) based polymer salt matrix. XRD analysis indicates a decrease of crystallinity due to coordination interactions between $Li^+$ and $EMIm^+$ cations with ether oxygen atoms of PEO and it leads to the fast segmental motion of polymer chain. XRD analysis was also supported by DSC which shows a decrease of melting peak area and supports the idea of fast ion transport with the incorporation of IL. FTIR analysis confirmed the presence of interaction between ether group of PEO with the cation of salt and IL. The shift of C-O-C stretching peak of PEO towards lower wavenumber side evidences the cation interaction with ether group of polymer host (figure 16). The addition of IL increases the ionic conductivity and is $1.85 \times 10^{-4}$ S $cm^{-1}$ at room temperature and is attributed to increasing of a number of charge carriers with IL incorporation. Further, the increase of temperature increases the conductivity due to an increase of free volume due to polymer expansion. The decrease of activation energy is seen from 0.68 to 0.447 with 40 % IL content in polymer salt matrix. The electrochemical analysis shows the initial specific capacity 155 mAh $g^{-1}$ at the low current rate for 40 % IL system up to 50 cycles and after that capacity was 134.2 mAh $g^{-1}$.

Balo *et al* [53] reported the preparation of a gel polymer electrolyte (GPE) based on polymer polyethylene oxide (PEO)-lithium bis(trifluoromethylsulfonyl) imide (LiTFSI) with ionic liquid (IL) 1-ethyl-3-methylimidazolium bis(trifluoromethylsulfonyl)imide (EMIMTFSI). DSC study shows a decrease in $T_g$ with increasing amount of IL which evidences the enhancement of polymer flexibility and free volume which supports in the fast transport of ions in polymer electrolytes. Also, the variation of crystallinity from 80 %→24.4 %→4.3 % for pure PEO, PEO + 20 wt. % LiTFSI, and PEO + 20 wt. % LiTFSI + 10 wt. % EMIMTFSI respectively evidences the intrachain and interchain hopping of ions due to plasticization effect of ILs in GPE. TGA graph shows two decomposition temperature $T_{d1}$(related to uncomplexed polymer) and $T_{d2}$ (related to complexed polymer) and stability of pristine PEO (334 $^oC$) decreased to 309 $^oC$ on the addition of salt due to low lattice energy and more dissolution of salt. All GPEs shows thermal stability up to 310 $^oC$ with highest 363 $^oC$ for PEO + 20 wt. % LiTFSI + 5 wt. % EMIMTFSI. The conductivity increases with both increase of temperature and ionic liquid content may be attributed to increasing of polymer chain flexibility due to former and reduction of crystallinity by later. The room temperature conductivity of the sample $2.50\times10^{-6}$ S $cm^{-1}$ (PEO + 20 wt. % LiTFSI) increased to $1.43\times10^{-5}$ S $cm^{-1}$ (PEO +20 wt. % LiTFSI + 2.5 wt. % IL) with highest value $2.08\times10^{-4}$ S $cm^{-1}$ at 12.5 wt. % ILs (figure 17 a-c). The increase may be due to a decrease of activation energy with IL addition and temperature variation shows Arrhenius type behavior. The total contribution in all GPEs was due to ions as a high transport number ($t_{ion}$>0.99) was observed and cation transference number ($t_+$=0.39). The electrochemical stability window was around 4.6 V and is suitable for a battery application. The charge discharge of PEO + 20 wt. % LiTFSI + 12.5 wt. % EMIMTFSI was performed in the range 2.0-4.0 V with C/10 current rate. The discharge capacity at first cycle is around 56 mAh $g^{-1}$ due to the formation of solid electrolyte interface and 110 mAh $g^{-1}$ in 5th cycle, 120 mAh $g^{-1}$ in the $10^{th}$ cycle. The discharge efficiency (η) was more than 98 % after 100 cycles and cyclic performance was same in both flat and bends condition.

Cheng *et al* [54] reported the preparation of $P(EO)_{20}LiTFSI$ based electrolyte with 1-butyl-4-methylpyridinium bis(trifluoromethanesulfonyl)imide (BMPyTFSI) ionic liquid using solution casting method. The SEM analysis shows loose and porous morphology for $P(EO)_{20}LiTFSI$ electrolyte, and the addition of ILs in the polymer salt matrix changes the morphology resulting in an in compact and smooth morphology. DSC analysis shows that addition of BMPyTFSI in polymer sat matrix weakens the interaction among the polymer chains and prevents the polymer recrystallization. The ionic conductivity reported for polymer salt complex at 40 ◦C was $1.6 \times 10^{-6}$ S $cm^{-1}$ and increases to $6.9 \times 10^{-4}$ S $cm^{-1}$ with the addition of ILs. The low temperature increase in ionic conductivity is due to the high self-dissociating and ion-transporting abilities of the ionic liquid while the high temperature increase is associated with polymer segmental motion and salt dissociation. The temperature dependence variation of the ionic conductivity follows Vogel–Tamman–Fulcher (VTF) equation in the temperature range of 20–80 ◦C, and activation energy significantly decrease with from 21.3 kJ/mol for polymer salt matrix to 14.1 kJ/mol after addition of IL. The electrochemical stability window for polymer salt system was 4.8 V and increased to 5.3 V after addition of ILs and this enhancement may be attributed to due to the formation of a stable passive layer.

Chaurasia *et al* [55] prepared PEO:IL (1-ethyl-3-methylimidazolium tosylate, EMIM-TY) based polymer electrolyte films by solvent free hot-pressing technique. The XRD pattern shows absolute peaks of PEO in the range 2θ= 15°–30° riding over a small "halo" indicating a partially amorphous phase. The addition of ionic liquid in polymer shows a shift and splitting in PEO films and may be due to change in PEO-structure geometry due to interaction/complexation with the IL. A new peak growth is seen at 2θ~ 16° on the addition of ionic liquid starts growing in the (PEO+ EMIM-TY) membranes at the expense of the intensity of prominent peaks of pristine PEO at 18.95 and 23.35°. This additional peak is evidence of PEO+ EMIM-TY complex formation. The crystallinity calculated from XRD shows decrease form 32 % to 27 % with the addition of up to 0.3 mol of $EMIMBF_4$ ILs from 83 % to 64 % with 20 wt. % ionic liquid EMIM-TY. DSC thermo grams of pristine PEO shows an endothermic melting peak at 66.36 °C and the addition of ionic liquid (EMIMTY), results in the appearance of two endothermic peaks. The main peak $T_{m1}$ has been assigned to the melting temperature of crystalline (uncomplexed) pure PEO, whereas the second endothermic peak $T_{m2}$ is assigned to the crystalline complexed material. The crystallinity of the polymer electrolyte membranes decreased with increasing amount of ionic liquid (EMIM-TY) upto ~20 wt. % and after that it starts to increase.

FTIR analysis provides the presence of possible complexation between ether group of PEO and the end groups of the imidazolium cation ($EMIM^+$) of IL (EMIM-TY). The changes in the C–O–C vibrations of PEO and C-H stretching vibrations of IL cation aromatic ring clearly indicate that the polymer PEO complexes with the ionic liquid through the ether group of PEO is getting loosely attached to the imidazolium ring of the IL cation ($EMIM^+$). The ionic conductivity of the polymer PEO is very poor (~$5.8 \times 10^{-8}$ S $cm^{-1}$) and increases considerably with the increasing amount of ionic liquid in the PEO membrane. At 40 wt. %, the ionic conductivity of the polymer electrolyte PEO+ EMIM-TY membrane is ~$2.87\times 10^{-5}$ S $cm^{-1}$ at 30 °C. The increase in conductivity was due to increased number of charge carriers provided by the higher concentration of ionic liquid.

Yongxin *et al* [56] reported the preparation of new type of polymer electrolyte based on N-methyl-N-propylpiperidinium bis (trifluoromethanesulfonyl) imide (PP1.3TFSI), polyethylene oxide (PEO), and lithium bis (trifluoromethanesulfonyl) imide (LiTFSI) as salt. Thermal analysis was done by TGA and all films were thermally stable up to 200 °C. FTIR analysis concludes the decrease of crystallinity after incorporation of IL in the polymer salt matrix. The addition of IL improves the electrochemical stability window of polymer salt system from 4.5 V to 4.7 V (vs. $Li/Li^+$). The highest ionic conductivity obtained was ~$2.06\times 10^{-4}$ S $cm^{-1}$, and is much greater than the polymer salt system having the conductivity value only $3.95\times10^{-6}$ S $cm^{-1}$ at room temperature. The increase of temperature to 60 °C, increases the ionic conductivity to $8.68\times10^{-4}$ S $cm^{-1}$ and for the sample without IL was $3.26\times10^{-5}$ S $cm^{-1}$. The cation transference number of Li/$P(EO)_{20}$LiTFSI+ 1.27PP1.3TFSI/Li cell (10 mV signal) was 0.339 and is suitable for battery applications as shown in figure 18 a & b. Also the electrochemical analysis was done and capacity was 120 mAh/g and was stable up to 20 cycles with coulomb efficiency greater than 99%.

Singh *et al* [57] prepared polymer electrolyte using PEO as a host polymer, sodium methylsulfate (NaMS) as salt and 1-butyl-3-methylimidazolium methyalsulfate (BMIM-MS) as IL. SEM micrograph shows a change of pure PEO rough surface to smooth on the addition of IL which indicates faster ionic transport and absence of crystal domain. The increase in FWHM and absence of crystalline peaks corresponding to PEO indicates the enhancement in amorphous content of polymer system as evidence by XRD. DSC graph shows a decrease of melting enthalpy with the addition of IL which indicates the decrease of crystallinity. Also the decrease of glass transition temperature suggests the fast ion transport (figure 18 c). Thermal stability of all system was above 300 °C as measured by TGA. The highest ionic conductivity value was $1.05\times10^{-4}$ S $cm^{-1}$ for 60 wt. % of IL loading at room temperature. Cation transference number obtained was 0.46 and is a good ionic conductor with ion transference number 0.99. The electrochemical stability window was 4-5 V and shows the applicability of electrolyte for application in energy storage devices.

Vries *et al* [58] reported the ternary polymer electrolytes incorporating ionic liquids (Cation: $Pyr_{14}$ & $Pyr_{1201}$, Anion: $(FSO_2)_2N^-$, $(C_2F_5SO_2)_2N^-$, $(C_4F_9SO_2)(CF_3SO_2)N^-$) in poly(ethylene oxide) (PEO), lithium bis(trifluoromethanesulfonyl)imide (LiTFSI) based polymer salt system. TGA analysis shows that all samples are thermally stable up to 200 $^oC$ and $FSI^-$ containing samples own a lower thermal stability (figure 18 d). The ionic conductivity obtained was $10^{-4}$ S $cm^{-1}$ at room temperature and $10^{-3}$ S $cm^{-1}$ at 60 °C.

Simoneeti *et al* [59] investigated a quaternary, polyethylene oxide (PEO)-LiTFSI based electrolytes with N-methyl-N-propylpyrrolidinium bis(fluorosulfonyl)imide ($PYR_{13}FSI$) as an ionic liquid. DSC analysis evidence formation of hetrogenous phase and reorganization of the internal structure of polymer electrolyte evidenced by the shift of melting peak. Optical microscopy image for freshly prepared PEO-E electrolyte sample evidence only modest IL phase separation while upon two weeks aging presence of a large amount of segregated ionic liquid is confirmed. DSC analysis concludes that salt concentration increases the plasticizing effect and confirms the coexistence of heterogeneous phases. The analysis of impedance measurements shows conductivity value $3.4\times10^{-4}$ (-20 $^oC$), $2.43\times10^{-3}$ (20 $^oC$) S $cm^{-1}$, $9.1\times10^{-3}$ S $cm^{-1}$ (60 $^oC$) and is attributed to the formation of a 3-D network of highly conductive IL pathways with increasing IL content. Also, no effect of prolonged storage periods was observed on the conductivity and impedance response suggesting the presence of good ageing resistance. The thermal stability of all SPEs was up to 200 $^oC$ and was independent of ionic liquid concentration. The electrochemical stability window was 4.5 V (vs. the $Li/Li^+$ redox couple) and is attributed to release of the thin ionic liquid film during phase separation process onto the external electrolyte surface, which prevents the contact of the PEO host with carbon. The role of ILs was further studied by manufacturing Li/NMC polymer cells and upto 4.15 V a good profile was obtained and at high voltage rough profile was obtained (figure 19 a). For second panel no evidence of oxidation or degradation was noticed in the range 3-4.0 V with capacity value 107 $mAhg^{-1}$ (figure 19 b) and may be due to release of IL released by the PEO electrolyte bulk present in both the polymer separator and in the composite cathode. Another report by Simonnnet *et al* [60] reported the PEO-LiTFSI-$PYR_{13}$FSI-EC quaternary polymer electrolytes by hot press method. The ionic conductivity obtained was $1.5 \times 10^{-4}$ and $1.6 \times 10^{-3}$ S $cm^{-1}$ at -20 and 20 $^oC$.

Choi *et al* [61] investigated the effect of room temperature IL 1-butyl-3-methylimidazolium bis(trifluoromethanesulfonyl)imide (BMITFSI) (between 20 and 80 parts by weight (pbw)) on the poly(ethylene oxide)-(lithium bis(trifluoromethanesulfonyl) imide) [PEO-LiTFSI] based polymer salt matrix using ball milling and hot-pressing technique. They showed that the incorporation of ILs in the solid polymer electrolyte significantly reduces the electrolyte resistance, and the effect is more noticeable at lower temperatures. The addition of ILs increases the ionic conductivity of polymer salt complex $4.0\times 10^{-6}$ S $cm^{-1}$ (25 °C) to $3.9\times 10^{-3}$ $Scm^{-1}$ (20 °C). The highest conductivity was found to be $3.2\times 10^{-4}$ $Scm^{-1}$ (25 °C) and $3.2\times10^{-3}$ $Scm^{-1}$ (80 °C) for PEO-LiTFSI-BMITFSI (80 °C) system. The increase of conductivity was attributed to the reduction of coordinating interaction of $Li^+$ ions with the O atoms of PEO segments with the anion of RTIL which leads to a larger number of charge carriers supporting the fast ionic transport. CV data showed the enhancement of cathodic stability of RTIL below the plating potential of lithium on the addition in polymer salt complex and may be due to the formation of a stable passivation layer on the lithium electrode which conducts $Li^+$ ion effectively but prevents further reaction with RTIL. The charge/discharge shows an initial discharge capacity of 90 mAh/g and increases to 140 mAh $g^{-1}$ after 5 cycles.

The preparation of PEO–LiTFSI based polymer electrolytes by incorporating different N-alkyl-N-methylpyrrolidinium bis(trifluoromethanesulfonyl)imide, PYR1ATFSI ionic liquids was reported by Kim *et al* [62]. The ionic conductivity was $>10^{-4}$ S/cm for all systems and battery test of cell Li/$P(EO)_{10}$LiTFSI + 0.96 $PYR_{1A}$TFSI/$LiFePO_4$ shows a capacity of 125 mAh $g^{-1}$ and 100 mAh $g^{-1}$ at 30 °C and 25 °C, respectively. Joost *et al* [63] reported the polymer electrolyte based on $PEO_{20}$LiTFSI $[Pyr_{14}TFSI]_6$. The ionic conductivties obtained was $5.0\times10^{-4}$ at 20 °C and $2.5\times10^{-3}$ at 60 °c.

Pandey *et al* [64] reported the similar performance characteristics of electrical double layer capacitors (EDLCs) based on poly (ethylene oxide) (PEO)/ magnesium and lithium trifluoromethanesulfonate (triflate+ionic liquid 1-ethyl-3-methylimidazolium trifluoromethanesulfonate (EMITf) on the multiwalled carbon nanotube (MWCNT) electrodes. The PEO complexes with Mg and Li salts and added ionic liquid EMITf have been used to fabricate the EDLC cells with optimized compositions namely: (a) $PEO_{25}\cdot Mg(Tf)_2$+40 wt.% EMITf, and (b) $PEO_{25}\cdot$LiTf+40 wt.%. EMITf with an ionic conductivity of $\sim10^{-4}$ S $cm^{-1}$ at ambient temperatures. Also, the PEO-$Mg(Tf)_2$ and PEO-$Mg(Tf)_2$-EMITf electrolytes offer less ionic conductivity compared to $Li^+$ ion conducting polymer electrolytes. The electrochemical stability window was obtained by CV at two different scan rates of 1 and 5 mV $s^{-1}$ and is of the order of 4 V for the $Li^+$ ion conducting polymer electrolyte. While, the stability window for $Mg^{2+}$ ion conducting polymer electrolyte is upto 4.8 V. The capacitance values calculate for both cells are ~0.03 F $g^{-1}$ (for the Mg-system) and ~0.01 F $g^{-1}$ (for the Li-system). The comparative analysis of both cells concludes that divalent $Mg^{2+}$ ions also play an active role in the formation of a double layer and helps in the neutralization of more negative charges gathered at the electrode.

Figure 20 a & b shows the cyclic voltamogram of both EDLC cells recorded in the potential range of −1.0 to 1.0 V at different scan rates. The mirror image symmetry observed in both cells evidences the capacitive behavior of the cells with a double layer formation at the interfaces. But, at a higher value of scan rate slight deviation is due to the finite value of the equivalent series resistance (ESR) in polymer electrolyte based EDLCs. The capacitance value for $Li^{+}$ ion Cell-I is 3.1 F $g^{-1}$,for $Mg^{+2}$ cell-II 2.4 F $g^{-1}$ at a scan rate of 10 mV $s^{-1}$ and it confirms the superiority of Mg based electrolyte over Li ion based. The performance of cycle was tested for 50 cycles and confirms the proper formation of the interfaces during the initial cycling process. The galvanostatic charge–discharge was performed from 0 to 2.0 V at room temperature (~25 °C) as shown in Figure 20 c &d. The obtained capacitance value is in good agreement with CV studies. The high Columbic efficiency (η) ~90 %, obtained from charge–discharge experiment evidences the proper interfacial contacts between MWCNT electrode and both electrolytes.

The effect of EMImTFSI on the electrical, electrochemical and interfacial properties of $(PEO)_8$LiTFSI-10% NC based polymer electrolyte was reported by Karuppasamy *et al* [65]. XRD analysis suggests the complete dissolution of salt in the polymer matrix. Further addition of ionic liquid decreases the crystallinity of polymer and also disrupts the polymer chain reorganization tendency which supports fast ionic transport or conductivity. The addition of IL also enhances the polymer flexibility and segmental motion as evidenced by DSC analysis. To support the CRD and DSC data FTIR was done which provide strong evidence of a decrease of crystallinity in terms of shifting and decrease of intensity of peaks. Stress-strain behavior shows elongation and strain at break are 24 MPa and 8 % for electrolyte with 0 % IL. While after addition of IL the elongation and strain at break are 10 MPa and 12 %. This reduction in mechanical strength may be due to the higher plasticization and increased polymer chain flexibility (Figure 21). The highest ionic conductivity obtained was the order of $10^{-2}$ S $cm^{-1}$ at 343 K for 10 wt. % IL content and electrochemical stability window were 3.97 V.

### ***4.3.*** *Gel polymer electrolyte*

As the addition of salt to the pure polymer having insulating nature evidence the increase of ionic conductivity up to two or three order. So, the addition of a low molecular weight plasticizer/ionic liquid suppresses the crystallinity and promotes the release of more number of charge carriers. There are various plasticizers such as, EC, PC, DEC, PEGDME, PEG, SN and ionic liquid which are reported for enhancing the conductivity and battery performance. The main role of plasticizer is to increase the free volume which provides an easy path to cation for migration within the electrolyte system. The enhances amorphous phase content is also related to the enhanced ionic conductivity and may be due to more dissociation of salts to release more free charge carriers which are required for a highly efficient system.

The effect of poly ethylene glycol, propylene carbonate, ethylene carbonate and dimethyl carbonate on PEO-$LiClO_4$ based polymer electrolyte was reported by Das *et al* [66]. The simultaneous presence of crystalline and amorphous regions of the polymer was confirmed by SEM images. The smooth surface morphology after addition of salt is an indication of enhanced amorphous content via interaction between ether oxygen of PEO and $Li^{+}$ ions. Also the PEG plasticizer leads to the smoother surface as compared to other plasticizers (figure 22). Also the temperature dependent ionic conductivity for all plasticized polymer electrolytes follows well known VTF relation which evidences the confirming coupled motion of $Li^+$ ions and segment of the polymer chain.

Kumar *et al* [67] reported the preparation of lithium ion conducting polymer electrolytes based on polyethylene oxide (PEO) - lithium trifluoromethanesulfonate ($LiCF_3SO_3$ or LiTf) with an ionic liquid 1-ethyl 3-methyl imidazolium trifluoromethanesulfonate (EMITf). The optical micrographs of pure PEO film, shows the spherulitic texture of PEO along with dark boundaries indicating the amorphous content in the polymer. Further addition of salt results in a larger proportion of dark boundaries/regions which evidences the increase of amorphous content. While, for the large content of ionic liquid, a mud-like dense liquid wrapped around the spherulites is observed. A broad hump observed between 15° and 30° suggests the presence of partial amorphous nature of the polymer and addition of salt increases the intensity (figure 23). Further addition of ILs increases the intensity of the hump and shows a decrease in the degree of crystallinity of the polymer electrolyte. The presence of crystalline peaks even after the addition of high content of ILs evidences the persistence of crystalline nature of the plasticized PEO-complex and is a correlation with optical micrograph. FTIR spectra indicate a change in the band corresponding to polymer host which indicates the presence of interactions of ionic liquid component ions with the host polymer PEO. The presence of two bands at 749 and 757 $cm^{-1}$ are due to pure PEO ($CH_2$ rocking mode) and $PEO_{25}$-LiTf complex ($\delta_{s\text{-}CF3}$ mode of vibration of free triflate ions) predicts that some proportion of pure PEO is left un-complexed in PS complex and further addition of ILs in PS complex dis-appears the un-complexed PEO peak which confirms the interaction of ILs cation with ether group of PEO.

Raman analysis evidences the interaction of PEO with ionic liquid EMITf. Then the addition of salt in pure polymer reduces the peak at 1062 $cm^{-1}$ and another peak appears at 1033 $cm^{-1}$ corresponding to the $SO_3$ stretching mode of free triflate anions. The additional peaks observed in Raman (1033 $cm^{-1}$) and FTIR (1016 $cm^{-1}$) on the addition of EMITf are due to the vis-à-vis reduction of intensity of PEO peak at 1062 $cm^{-1}$ and confirm the interaction of PEO with the ionic liquid EMITf.
The thermal stability of pure PEO was ~340 °C and increases to ~375 °C on the addition of salt. Further addition of ILs in polymer salt complexes reduces thermal stability to ~350 °C due to the volatility of ILs. DSC analysis shows a decrease of melting peak corresponding to pure PEO from ~69 °C to ~62 °C on the addition of salt. Further addition of ILs decreases the endothermic peak indicates the disruption of the polymer chain. The crystallinity of pure PEO film decreases from ~89 % to 71 % on addition of salt. While, at high ILs content crystallinity was 22 %. (Addition of salt in pure polymer host increases the $T_g$ and it indicates the reduction in flexibility of PEO chains due to crosslinking of the polymer through the salt. Further, the addition of ILs decreases the $T_g$ suggesting an increase in polymer flexibility.

The addition of salt increases the ionic conductivity and highest conductivity was ~$3 \times 10^{-4}$ S cm$^{-1}$ for 40 wt. % ILs content. The σ vs. 1/T variation shows a sudden jump at ~60–65 °C for all the polymer electrolyte films, which is attributed to jumping from semi-crystalline to amorphous (i.e. solid to liquid) phase transition at $T_m$ of PEO. The decease of height of this jump with increasing IL content is associated with enhancement of amorphous content and shows Arrhenius-type behavior. The lowest value of activation energy obtained was for 5 wt. % IL content and a further increase of IL content almost saturate the activation energy value.

The total ionic transference number obtained was found to be >0.99, and evidence the dominant of ionic conduction in the polymer electrolyte. The effect of temperature and ionic liquid ion mobility was studded by using $^7$Li NMR. The $^7$Li NMR absorption spectra of polymer salt system show narrow line width with an increase of temperature to 100 °C. While, a decrease in the line width with a gradual increase in temperature evidences the increase in Li mobility with increasing temperature. A narrow line width for ILs doped polymer salt matrix reveals the increase in ion mobility as well as the role of IL as a plasticizer, hence the increase in polymer flexibility and a decrease of viscosity. The electrochemical stability window (ESW) was studied by the cyclic voltammetry and is in the range of −1.4 to 1.6 V (i.e. ~3.0 V) for polymer salt complex. Further, the addition of ILs increases the voltage window up to −2.6 to 2.3 V (i.e. ~4.9 V).

Li *et al* [68] reported the preparation of GPE using PEO as host polymer and $LiPF_6$ as liquid electrolyte solution via in situ polymerization method. SEM image of GPE evidence that unifrom structure is an indication of favorable transport for $Li^+$. The electrochemical window of GPE was measured by LSV and is 4.9 V. The ionic conductivity obtained was $3.3\times10^{-3}$ S cm$^{-1}$ and was independent of temperature. The high cation transference number 0.76 indicates its applicability for the application (figure 24). The electrochemical profile shows the reversible capacity of the $LiFePO_4$ at 0.5C for GPE is 133 mAhg$^{-1}$. While, the discharge capacity is 103 mAhg$^{-1}$ after 500 cycles and capacity retention obtained was 81%.

Zhu *et al* [69] reported preparation of two types of gel polymer electrolyte by incorporating 1-ethyl-3-methylimidazolium bistrifluoromethanesulfonylimide (EMI-TFSI)and N-methyl-N-propylpiperidinium bistrifluoromethanesulfonylimide ($PP_{13}$-TFSI) in PEO as polymer host by solution cast technique. DSC analysis concludes that both IL weakens the mutual interaction between polymer chains and disrupts the polymer crystallization. The ionic conductivity achieved was 10-5 S/cm at room temperature and was the maximum for $PEO_{20}$LiTFSI+1.0 EMI-TFSI of the order of $2.67\times10^{-4}$ S cm$^{-1}$ at 40 °C. After incorporation of ILs, the decomposition voltage reached was upto 5.2 V. The discharge profiles for the cell $LiFePO_4/PEO_{20}LiTFSI$ + 1.0EMI-TFSI/Li shows the specific capacity of 132 $mAhg^{-1}$ at a 0.05 C rate, with a voltage of about 3.4 V vs $Li/Li^+$. After an increase in C-rate upto 0.2 the capacity obtained was 117 mAh $g^{-1}$.

Egashira *et al* [70] reported the preparation of gel polymer electrolyte using the two kinds of an ionic liquid having different cations (EMITFSI and HTMATFSI), and poly(ethylene oxide) branched poly(methyl methacrylate) (PEO-PMA) using ion-exchange reaction. The DSC graph of the gel electrolyte is shown in the figure 25 a & b. The peak corresponding to the glass transition temperature shows the noticeable shift depending on both polymer matrix, ionic liquid component and the existence of LiTFSI. The highest ionic conductivity obtained was $2.3 \times 10^{-4}$ S $cm^{-1}$ and lowest activation energy 30 kJ/mol for the Li-EMI(9). It was concluded that the Quaternary ammonium-based ionic liquid is better than the imidazolium-based ionic liquid for enhancing the ion mobility, hence the ionic conductivity.

Natarajan *et al* [71] reported the composite gel polymer electrolyte using magnesium aluminate ($MgAl_2O_4$) and LiTFSI as salt with a combination of 1,3-dioxolane (DOL) and tetraethylene glycol dimethyl ether (TEGDME) as a plasticizer by a simple solution casting technique. The smoother surface seen by FESEM shows better ionic conductivity in case of composite gel polymer electrolyte. The thermogravimetric curves show improved thermal stability as compared to the pure PEO (figure 25 c). The addition of plasticizer shows almost no effect on the glass transition temperature and may be due to the formation of amorphous phase by addition of salt in the polymer matrix. The highest ionic conductivity was obtained for 20 wt. % of plasticizer and is of the order of $3.4 \times 10^{-4}$ S $cm^{-1}$ at 30 °C.

Jayasekara *et al* [72] synthesized interpenetrating network structure where rods of polymer electrolyte is placed in the nanoscale channels of anodic aluminum oxide (AAO) membranes using PEO as polymer and lithium triflate ($LiSO_3CF_3$) as salt. The highest ionic conductivity obtained was $5 \times 10^{-4}$ S $cm^{-1}$ and is four order higher than the pure polymer electrolyte. The decrease of melting point observed from the DSC analysis suggests the decrease of crystallinity. The crystalline thickness plays an important role in ion conduction and is 4 nm for the polymer electrolyte with AAO membrane as compared to 20 nm for the pure PEO and is an indication of enhanced amorphous content.

Kim *et al* [73] investigated the effect of EC, PC, PG on PEO-$LiN(CF_3SO_2)_2$ (LiTFSI) based polymer matrix. The cation transference number with PG was 0.516 and is greater than the plasticizer-free electrolyte (0.487). While for EC and PC was 0.381 and 0.262, respectively. The ionic conductivity and diffusion coefficient was also higher for electrolyte doped with PC and may be due to the interaction of PC with host polymer and salt [74]. The interaction of PC with PEO disrupts the coordination interaction between cation and PEO which enhances the ion mobility. Later one reduces the ion pairing effect and more free charge carriers are released. The electrochemical stability window for all samples was ~4 V at 80 °C (figure 25 d).

Ma *et al* [75] synthesized $TiO_2$-grafted NHPE comprising of ion-conducting poly(ethyleneglycol) methyl ether methacrylate (PEGMEM) and non-polar stearyl methacrylate (SMA) using in-situ polymerization/crystallization method. The successful synthesis of $TiO_2$-grafted PEGMEM/SMA nanohybrid was confirmed by $^1$H-NMR spectroscopy and FTIR spectroscopy. It is noted that 5 wt. % $TiO_2$-grafted NHPE offers the highest ionic conductivity of $0.74\times10^{-4}$ S cm$^{-1}$ at 25 °C, of the PEGMEM/SMA polymer electrolyte and the conductivity increases to $1.10\times10^{-4}$ S cm$^{-1}$ at 30 °C. Also, the decrease in activation energy from 5.51→4.86 kJ mol$^{-1}$ confirms that lithium ions have more mobility in $TiO_2$-grafted NHPE. The charge-discharge study shows that the cell delivers an initial discharge capacity of 153.5 mAhg$^{-1}$ at 1C with a capacity retention ratio of 96 % after 300 cycles (figure 26).

The addition of plasticizer in $(PEO)_9$-LiTf + 10 wt. % $TiO_2$ polymer matrix shows that meting point ($T_m$) decreases from 60 °C to 50 °C and glass transition temperature ($T_g$) from -46 °C to -50 °C on the addition of 50 wt. % plasticizer [76]. EC was chosen as plasticizer due to its low viscosity as compared to other plasticizers. This reduction in $T_g$ weakens the complexation between $Li^+$ ion and the polymer chain enhancing the cation migration. Conductivity was also enhanced to $1.6\times10^{-4}$ S cm$^{-1}$ on the incorporation of 50 wt. % EC plasticizer. The addition of plasticizer enhances the amorphous phase as evidenced by a decrease in $T_g$.

Bandara *et al* [77] synthesized composite polymer electrolyte using the polyethylene oxide, alumina nano-fillers and tetrapropylammonium iodide as salt. The highest conductivity was obtained for the 5 wt. % alumina nanofiller and was $2.4 \times 10^{-4}$, $3.3 \times 10^{-4}$ and $4.2 \times 10^{-4}$ S cm$^{-1}$ at 0, 12 and 24 °C. The enhancement in conductivity may be due to the formation of conducting paths by the nanofiller surface. Also the activation energy was smaller for the 5 wt. % alumina and was 0.20 eV. This reduction in activation energy promotes fast ion transport or the high ionic conductivity.

Chen *et al* [78] investigated the effect of succinonitrile (SN) plasticizer on PEO-LiTFSI-1% LGPS based polymer electrolyte prepared by solution cast technique. DSC analysis indicates that SN plays an active role in increasing the flexibility of polymer chain and more amorphous phase is formed. Also, the addition of SN reduces the interaction of cation with ether group of polymer which supports faster migration of ions. Initially up to 10 % SN crystallinity decreases but at higher concentration increase in crystallinity is due to the dominance of insulating nature of SN which blocks conducting pathways. The maximum ionic conductivity was $1.58\times10^{-3}$ Scm$^{-1}$ at 80 °C and $9.10\times10^{-5}$ S cm$^{-1}$ at 25 °C for 10 % SN. The increase in conductivity was due to increase of the fraction of an amorphous phase and is supported by DSC data. The decrease of conductivity at the higher content of SN was due to non-ionic plastic crystal nature, which leads to insulating nature of SN and blocks the free ions. The SPE was thermally stable up to 255 $^{o}C$ and electrochemically upto 5.5 V. The cycling and rate performance of solid-state lithium battery Li/$PEO_{18}$-LiTFSI-1% LGPS-10 % SN/$LiFePO_4$ is presented in fig. 27 (a-d) and specific capacities are 158.1, 144.0, 133.3 and 86.2 $mAhg^{-1}$ at 0.1 C, 0.2 C, 0.5 C and 1 C. The discharge specific capacity still maintains 152.1 $mAhg^{-1}$ after 60 cycles, and was 94.7% of the maximum specific capacity with 99.5 % coulombic efficiency. At high rate (0.5 C) the reversible capacity increases due to the activation process. The discharge capacity after 100 cycles was 120 $mAhg^{-1}$ and is an excellent performance for the prepared SPE.

Pitawala *et al* [79] reported the combined effect of nanofiller ($Al_2O_3$) and plasticizer (EC) on $(PEO)_9$-LiTf ($O/Li^+=9$) based solid polymer electrolyte. The maximum conductivity was $2.1\times10^{-5}$ S $cm^{-1}$ obtained for 15 wt. % nanofiller. Also, the conductivity enhancement was higher for nanofiller based polymer electrolyte than plasticized polymer electrolyte. Loss of mechanical property by the addition of plasticizer was regained by the addition of nanofiller. Lewis acid base type interactions provide a favorable additional site for cation migration, while plasticizer enhances the amorphous phase as evident from DSC data. The addition of DOP plasticizer provided the highest ionic conductivity value of $7.60\times10^{-4}$ S $cm^{-1}$ for PEO-15 wt. % ($LiCF_3SO_3$) + 20 wt. % DOP as reported by Klongkan [76]. The addition of DOP plasticizer in a PEO-15 wt. % ($LiCF_3SO_3$) solid polymer electrolyte shows drop in strength, elongation and Young's due to greater mobility of the polymer chains provided by polymer–plasticizer interactions which make sliding of chains easier by reducing the friction (figure 28). Some important results in terms of ionic conductivity, cation/ion transference number, thermal stability, voltage stability window and cyclic stability of ionic liquid and gel polymer electrolytes are summarized in Table 5.

### ***4.4.*** *Solid polymer electrolyte*

As discussed above a lot of work was done by addition of ILs and plasticizer in the polymer salt sytems. Although they have sufficient ionic conductivity and stability window, poor mechanical strength and thermal stability lead the researchers to make a polymer electrolyte system with improved electrochemical and mechanical properties. The ultimate solution for the above is the preparation of solid polymer electrolytes by the addition of nanofiller and nanoclay. The addition of nanofiller will result in improved properties by the presence of various interactions between the surface group of nanofiller and the polymer salt system [76, 80-81]. The increase of amorphous content by the addition of nanofiller will lead to enhanced conductivity also along with mechanical strength. Inoporation of nanoclay is also an imoprtant approach for the enhancement of overall properties. The negative surface charge present in the clay galleries provides the space for the intercalation of polymer chains. Aslo the thermal stability gets improved due to presenv of polymer chains between the clay galleries. So, both the approches and their effect on the PEO prperties is discussed in the following section.

### ***4.4.1.*** *Dispersed Type Polymer Electrolyte*

As, in previous we have discussed the polymer salt system and ionic liquid based system. The addition of ionic liquid/plasticizer imrpoves the ionic conductvity but mechnail properties get disturbed which affects the performance of abettery ssytem. So, the alternative of this draback is the use of nanofiller in the polymer salt ssytem and is one of the best approaches for improving the electrochemical properties as well as mechnaical properties. A lot of of work has been done towards enhancement of electrical conductvity, electrochemical stability window, thermals atbility and mechanical stability.

Nanofillers are classified into two types on the basis of support to transport of ions in polymer matrix; one that is invlved in ionic transort is called active nanofiller ($Li_{10}GeP_2S_{12}$ (LGPS), $Li_{1.3}Al_{0.z}Ti_{1.7}(PO_4)_3$ (LATP), $LiN_2O_3$) and another which supports the polymer staructure only and no active role in ion transport is passive filler ($Al_2O_3$, $CeO_2$, $TiO_2$, $BaTiO_3$).

The first report on the addition of nanofiller (α-alumina) in PEO-$LiClO_4$ polymer electrolyte was given by Weston and Steel in 1982 [82]. It was concluded that mechanical stability of film was improved by the addition of 10 wt. % nanofiller. Another report by Wieczorek in 1996 reported the enhancement in conductivity by Lewis acid base approach with α-$Al_2O_3$ as nanofiller and PEO as the electrolyte. The presence of various interactions in the polymer matrix between Lewis base centers on ether oxygen and the strong acid $Li^+$ cation modifies the polymer chain structure which supports fast ionic transport [83]. The effect of nano $SiO_2$ and $TiO_2$ fillers on the thermal, mechanical and electrochemical properties of PVA:PVdF:$LiCF_3SO_3$ have been investigated by Hema et al., [84]. The addition of nanofiller improves the thermal stability as evidenced by TGA and is attributed to due to the bond strength of Si–O, and Ti–O in the nanofiller. The conductivity was enhanced by the addition of nanofiller and is $3.7 \times 10^{-3}$ S $cm^{-1}$ compared to filler free system at 303 K. The electrochemical stability window was 5.2 V and is suitable for the application.

Wang *et al* [85] reported the preparation of solid polymer electrolyte based on PEO-$LiClO_4$+ $Li_{1.3}Al_{0.z}Ti_{1.7}(PO_4)_3$ (LATP), $TiO_2$, $SiO_2$. LATP is active nanofiller while $TiO_2$ and $SiO_2$ are passive nanofiller. FTIR spectra of pure PEO and LATP show no change/shift w.r.t polymer nanocomposite spectra and it evidences the mechanical dispersion of LATP in polymer matrix but is not forming a complex. Complexation of polymer and salt ($LiClO_4$) is confirmed by the FTIR spectra. The prepared films were free-standing and translucent and translucency decreases with increasing LATP loading. SEM micrograph shows the uniform complete distribution of for 5 and 10 wt. % of LATP loading and at high content agglomerates of particles start to appear (figure 29). The XRD spectra of PEO shows its fundamental peaks at 19.3° and 23.4° which confirms the crystalline structure while, a broad feature is obtained for LATP at 24°confirming its amorphous nature. The addition of LATP in polymer salt matrix suppresses the crystallinity. DSC results show that the addition of LATP nanoparticles gradually reduces the crystallinity from 45 % to 37 % as nanofiller prevent the polymer chain from regular alignment. The highest ionic conductivity obtained was 1.71 × $10^{-4}$ S $cm^{-1}$ at 20 °C is achieved for 10 wt. % LATP nanoparticles and EO/Li = 5 and the activation energy was 2.10 kJ $mol^{-1}$ for this which is lowest among all samples.

Zhao *et al* [86] reported preparation of SPEs using PEO as a host polymer, LiTF as salt and $Li_{10}GeP_2S_{12}$ (LGPS) sulfide electrolyte as nanofiller. DSC results show that LGPS suppresses the polymer crystallization which directly suggests an increase in free volume and hence, more amorphous phase is confirmed which supports fast ionic transport. XRD spectra show that cubic single phase (space group 141/acd) of LGPS provide a suitable tunnel for cation migration (figure 30 a). SPEs film is semi-transparent and flexible as in given figure 30 d. The maximum ionic conductivity was 1.21×$10^{-3}$ S $cm^{-1}$ at 80 °C which is higher than the ionic conductivity of PEO-only membrane is only 7.98×$10^{-4}$ S $cm^{-1}$ at 80 °C (figure 30 b). The enhancement in conductivity is attributed to single ion conductor nature and unity transport number for LGPS. Also, the presence of weak interaction between lithium and sulfide ions than that of lithium and oxygen ions concludes that Li ion can easily migrate upon addition of LGPS in the polymer matrix [87]. The electrochemical stability window was 5.2 V for LGPS added SPE which is higher than that of pure PEO 4.2 V and this may be due to the wide potential window of LGPS itself. Charge/discharge curve for cell $LiFePO_4$/$PEO_{18}$-LiTFSI-1% LGPS/Li shows high capacities of 158, 148, 138 and 99 mAh $g^{-1}$ with current rate 0.1 C, 0.2 C, 0.5 C and 1 C at 60 °C, respectively, which makes them a promising candidate for application in Lithium ion battery.

Ali *et al* [88] reported preparation of polymer electrolyte based on PEO/PEG-$LiClO_4$ and $CeO_2$ as a nanofiller using standard solution cast technique. XRD analysis shows the presence of coordination interactions between the $Li^+$ ion and ether oxygen atoms of polymer chain which reduces the crystallinity (figure 31). The decrease in crystallinity is due to disorder produced in polymer chain with the addition of nanofiller which supports amorphous phase formation and hence, faster ion transport. The maximum ionic conductivity was 1.18×$10^{-4}$ S $cm^{-1}$ for 1 wt. % nanofiller dispersion in the polymer matrix and may be due to interchain and intrachain hopping of ion movements (Activation Energy=0.41 eV). Further, at high nanofiller content conductivity reduction may be attributed to blockage effect provided by nanofiller particles.

Yoon *et al* [89] reported the effect of silica aerogel on PEO-PVDF-$LiClO_4$ based solid polymer electrolyte. XRD analysis concludes that polymer blend with silica aerogel contains highly amorphous phase. The complex formation was confirmed by FTIR and various bands corresponding to PEO and PVDF are observed (figure 32). FTIR analysis revels that at high salt concentration ion association occurs which decrease the number of free charge carriers. The highest ionic conductivity was $1.70\times10^{-4}$ S $cm^{-1}$ at 30°C for PEO:PVDF ratio 3:1.

Jung *et al* [90] prepared hybrid polymer electrolyte using LAGP ($Li_{1.5}Al_{0.5}Ge_{1.5}(PO_4)_3$) as nanofiller in PEO-$LiClO_4$ matrix. FESEM micrograph shows heterogeneous particle size distribution of LAGP powders in polymer salt matrix. EDX mapping image suggests complete dissolution of salt and even distribution of nanofiller. XRD spectra show disorder in the structure of the solid polymer electrolyte as crystalline peaks corresponding to PEO were disappeared on the addition of LAGP. Also LAGP was in original for as no degradation was observed in the polymer film. The highest ionic conductivity was $1.0 \times 10^{-5}$ S $cm^{-1}$ for 70 wt. % LAGP and is higher than polymer salt system at room temperature and equivalent circuit in inset represents a solid electrolyte sandwiched between two blocking electrodes. Also, conductivity increases with temperature to $2.6 \times 10^{-4}$ S $cm^{-1}$ at 55°C. The electrochemical stability window was enhanced from 4.5 V to 4.75 V on the addition of LAGP in the polymer matrix.

The electrochemical properties were studied by assembling the solid-state Li/$LiFePO_4$ cells with hybrid solid electrolytes (LAGP-60, LAGP-70 and LAGP-80) in the voltage range of 2.6–4.0 V at a constant current rate of 0.2C and 55°C. The initial discharge capacity was 137.6 mAh $g^{-1}$ and 142.2 mAh $g^{-1}$ for the 20th cycle. The discharge capacities of the Li/HPE/$LiFePO_4$ cells as a function of the cycle number and discharge capacities of the cells decreased from their initial capacities of 133.0~138.5 mAh $g^{-1}$ to 113.4~121.5 mAh $g^{-1}$ at the 100th cycle, corresponding to 81.9~91.3% of the initial values. Also, the capacity retention was improved with increasing the content of LAGP, which can be ascribed to the high electrochemical stability and good interfacial properties as LAGP may suppress harmful interfacial side reactions between the electrodes and the electrolyte, which results in good cycling stability.

Mohanta *et al* [91] prepared the PEO based solid polymer electrolyte with lithium trifluromethanesulphonate ($LiCF_3SO_3$) as salt and porous silica as nanofiller. TEM analysis of the electrolyte shows porous nature of morphology. FTIR analysis confirmed the epoxy capping on the surface of calcined porous silica nanostructure and successful immobilization of the surface group on the silica surface. The Nyquist plot for the chem-$SiO_2$ nanofiller and epoxy-$SiO_2$ nanofiller based electrolyte is shown in figure 33. The highest conductivity was $1.03\times10^{-4}$ S $cm^{-1}$ for 1% epoxy-$SiO_2$ and may be due to high amorphous content as compared to other samples.

Dam *et al* [92] investigated the conduction mechanism of PNC based on polyethylene oxide as polymer host, lithium trifluoromethanesulfonate as salt and nano-crystalline zirconia ($ZrO_2$) as filler using broadband dielectric spectroscopy. The maximum value of dc conductivity ($\sigma_{dc}$) was $2.04\times10^{-5}$ S $cm^{-1}$ for $PEO_{20}$-$LiCF_3SO_3$-8 wt. % $ZrO_2$. The experimental data have been fitted using Jonscher power law (JPL) and it shows two mechanisms of conductivity (i) ion jumps from one site to its neighboring favorable vacant site, contributing towards the dc conductivity and (ii) short time period available the correlated forward backward hopping of mobile ions give rise to ac conductivity at high frequencies.

Moremo *et al* [93] reported the preparation of poly (ethylene oxide) (PEO) /sodium bis(trifluoromethanesulfonate) imide (NaTFSI)+$SiO_2$ based electrolyte using the solvent-free hot-pressing technique with different EO:Na molar ratio. DSC analysis shows $T_g$ at -60.7 °C and melting peak at 65 °C ($\Delta H_m$=155.7 $Jg^{-1}$) for pure PEO. The $T_g$ variation shows an irregular trend with varying salt concentration. Initially, the $T_g$ increases as NaTFSI concentration increases by promoting a loss of rigidity of the polymer backbones and may be due to stronger interaction and slowing of polymer segmental motion. While, at the highest concentration (i.e. 6:1) plasticization effect comes into picture due to the flexibility of the S-N-S bond in TFSI. Two peaks are observed in melting process in the 12:1-8:1 composition range and are assigned to the onset of small crystallites melting, allowing the growth of the larger ones which melt at a higher temperature. The $T_m$ peak, the crystallinity decreases due to plasticization action of salt observed for the second endothermic peak. Further addition of $SiO_2$ in polymer salt complex decreases the melting temperature which directly enhances the amorphous content and may be due to the interaction of nanofiller with polymer hoist via Lewis acid base interactions.

The ionic conductivity decreases from 6:1 down to 20:1 for PEO:NaTFSI, and $PEO_{20}$:NaTFSI and $PEO_{12}$:NaTFSI system shows value 1.1 mS $cm^{-1}$ and 1.3 mS $cm^{-1}$ at 80 °C, respectively. A nonlinear effect was observed in conductivity data due to an increase of NaTFSI concentration and may be due to a compromise between amorphization effect of the salt and charge carrier number. The addition of nanofiller at 5 wt. % content shows no change in conductivity value and may be due to masking of the effect of nanofiller by the plasticizing action of $TFSI^-$ anion. At 10 wt. % nanofiller decrease in ionic conductivity may be due to blockage of ion conducting pathways. The sodium transport number was measured by applying a fixed DC voltage and following the DC current as a function of time. The sodium ion transference number increases up to the addition of 5 wt. % $SiO_2$ and then slightly decreases for higher filler concentration due to the interaction of nanofiller with sodium ion (figure 34).

Ganapatibhotla *et al* [94] reported the influence of nanofiller surface chemistry and ion content on the conductivity of PEO-$LiClO_4$+$Al_2O_3$ based solid polymer electrolytes. Alumina nanofiller was chosen due to the existence of polymorphs: the α-form has acidic or electron pair-accepting surface sites, β has basic surface sites or electron withdrawing sites, and γ has a combination of acidic and basic surface sites resulting in neutral surface chemistry.

Here, hydrogen atoms in acidic and neutral particle filled electrolytes participate in three processes: vibration, segmental relaxation of PEO, and a rotational process consistent with a liquid analogue of the $PEO_6$:$LiClO_4$ structure. On the basis of data. enhancement in conductivity was due to the formation of the hydrogen bonds between −OH groups on the acidic particles with the anions. These interaction leads to release of more free charge carriers which contribute in conduction while neutral particles are having basic groups form a hydrogen bond with cations through the ether group of polymer host, and thus either decrease or have no effect on conductivity. It is proposed in the mechanism that that Li ions hop from one coordination site to another, enabled by energy barriers that are accessible at room temperature. While, the basic sites in the neutral surface act as Li traps, interrupting the movement of Li. At eutectic composition, there is no surface chemistry dependence which is due to the stability of multiple layers of $PEO_6$:$LiClO_4$ and PEO by acidic particles and neutral particles (which have both acidic and basic sites). At a composition of EO/Li = 8:1, acidic fillers enhance conductivity, Li transference number, and Li ion diffusivity more than neutral fillers at all temperatures with highest enhancement at 30 °C. At a composition of 14:1, fillers only enhance conductivity above the melting temperature of PEO, where acidic particles impact conductivity to a larger extent than neutral fillers. It is notable that the influence of surface chemistry nearly vanishes at the eutectic composition (EO/Li = 10:1). Above the eutectic temperature (50 °C), there is no statistically relevant difference, and below the eutectic temperature, the neutral particles are more efficient. These findings indicate that the mechanism of conductivity enhancement at the eutectic composition differs from other compositions.

Vignarooban *et al* [95] reported enhancement in conductivity based on poly (ethylene oxide)–lithium trifluoromethanesulfonate ($LiCF_3SO_3$ or LiTf) based polymer electrolyte with $TiO_2$ nano-filler. The highest conductivity was of $4.9\times10^{-5}$ S $cm^{-1}$ for 10 wt. % nanofiller and is attributed to the formation of transient conducting pathways due to the Lewis acid–base type interactions of caion with surface groups of the nanofiller. The dual functionality of the nanofiller was pointed out, one is to promote the ion migration by reducing the reorganization tendency of polymer and second is a decrease in ionic coupling due to Lewis acid base interactions. DSC spectra show that meting point ($T_m$) decreases from 64 °C to 60 °C and glass transition temperature ($T_g$) from -39 °C to -46 °C on the addition of 10 wt. % nanofiller and supports increase in ionic conductivity value (figure 35 a)

Lee *et al* [96] reported the effect of Barium Titanate nanofiller on structural modification and electrochemical properties of PEO-PVdF based system. XRD and DSC analysis indicated that change of the crystallinity value means the crystal growth of PEO was weaken by the molecular interaction of the polymer matrix by addition of nano-sized $BaTiO_3$. It was concluded that the presence of $BaTiO_3$ improved the ionic conductivity of the CPE with highest at 15 wt. % nanofiller ($1.2\times10^{-4}$ S $cm^{-1}$) by slightly decreasing the CPE crystallinity. At higher nanofiller content blocking the effect of nanofiller comes into the picture due to nanofiller aggregation which reduces the conductivity. TGA analysis confirms the thermal stability up to 330 $^oC$ on the addition of nanofiller and degradation of polymer matrix occurs at higher temperature. The electrochemical stability window was enhanced by the addition of nanofiller and is −2.3 to 2.4 V which is sufficient for energy storage devices (figure 35 b).

Zhang *et al* [97] prepared the $PEO_{20}$–LiBOB based solid polymer electrolyte with nano-sized MgO as nanofiller. The electrochemical stability is measured by cyclic voltammetry and addition of nano-MgO improves the anodic stability (figure 36 a & b). The two anodic peaks are appeared at at 0.97V and 2.05V, while two cathodic peaks at 1.53V and 0.68V. The peaks at 0.97V and 0.68V are possibly due to reduction and oxidization of ion species formed by dissociation of the lithium salt in PEO segments. While, peaks at 1.53 V and 2.05 V are associated with trace water molecule that exists in the SPE. The addition of nanofiller increases the initial oxidative decomposition up to 4.2V and may be due to absorption of trace water and unstable compounds present in polymer salt system. For polymer salt system, an average anodic current density reduces from initial 0.01 $mAcm^{-2}$ to 0.003 $mAcm^{-2}$, at 4.5V and remains constant at 3.0V that is around 0.003 $mAcm^{-2}$ during 20 full CV. The compatibility of $PEO_{20}$–LiBOB and $PEO_{20}$–LiBOB–5 wt. % MgO toward prepared $LiCoO_2$ and $LiNi_{1/3}Co_{1/3}Mn_{1/3}O_2$ is studied. The polymer salt system shows charge capacity and discharge capacity 168.8 and 156.8 mAh $g^{-1}$, respectively, and discharge capacity degrades to 142.5 mAh $g^{-1}$ after 20 cycles. The addition of 5 wt. % nanofiller gives discharges capacity value 157 mAh $g^{-1}$. The application of LiBOB and nano-sized metal oxides is a convenient and economical way to develop PEO based polymer electrolytes for 4V class cathodes.

Zhu *et al* [98] reported the preparation of PEO-LiTFSI based electrolyte with aluminum 1,4-benzenedicarboxylate (MIL-53(Al)) as nanofiller. The ionic conductivity of the electrolyte is $3.39\times10^{-3}$ S $cm^{-1}$ at 120 $^{o}$C EO: Li ratio of 15:1. The addition of filler disrupts the polymer host crystallization nature and dissociates the salt to release number of charge carriers. The decrease of phase transition temperature is seen with the addition of nanofiller and suggests that amorphous phase formed supports the ionic transport to achieve high ionic conductivity. The cation transport number obtained by AC impedance spectroscopy was 0.343 and was higher as compared to 0.252 for polymer salt system. This increase in cation transport is attributed to the formation of the metal organic framework. Also, the electrochemical stability window was improved from 5.13 V (4.99 at 120 $^{o}$C) to 5.31 V (5.10 V at 120 $^{o}$C) on the addition of filler in polymer salt system at 80 $^{o}$C (figure 36 c). All electrolyte system was thermally stable up to 200 $^{o}$C. The dynamical properties studied by stress-strain curve shows an increase of stress after addition of filler (figure 36 d). The cycling performance of battery shows initial discharge capacity 127.1 mA h $g^{-1}$ (at 5 C) at 80 $^{o}$C and 136.4 mA h $g^{-1}$ at 120 $^{o}$C. After 300 cycles, the discharge capacity was 116.0 mA h $g^{-1}$ at 80 $^{o}$C and 129.2 mA h $g^{-1}$ at 120 $^{o}$C.

Lin *et al* [99] reported the preparation of poly (ethylene oxide) (PEO)-$LiClO_4$+monodispersed ultrafine $SiO_2$ ($MUSiO_2$) composite polymer electrolyte via in situ synthesis. The in situ hydrolysis was used as it offers the opportunities to form the much stronger interaction between polymer chains of PEO and $SiO_2$ nanospheres, which is crucial for decreasing polymer crystallinity. Two interaction mechanisms were discussed here which affects the polymer chin density and reduce the crystallinity. In former one under hydrolysis condition the hydroxyl groups at the ends of PEO chains can chemically bind with those on $SiO_2$ surface and another deals with the mechanical wrapping of polymer chains when the $SiO_2$ spheres grow. The addition of salt in PEO shows a decrease in intensity in XRD spectra which directly evidence the decline of crystallinity. The differential scanning calorimetry (DSC) analysis confirms a higher fraction of amorphous phase for in situ PEO-$MUSiO_2$ composite which may be due to the presence of strong interactions between $MUSiO_2$ spheres and polymer chains also the uniform distribution of $MUSiO_2$ spheres from in situ hydrolysis improves the effective surface area of $SiO_2$. The dissociation or separation of $Li^+$ and $ClO_4^-$ was studied by FTIR using Gaussian-Lorentz fit and for the in situ CPE, 98.1% of dissociation ratio was observed which is higher than ceramic-free SPE (85.0%), PEO-fumed $SiO_2$ CPE (87.4%), and ex situ CPE (92.8%) counterparts. So, in situ hydrolysis technique results in more dissociation of salt and may be attributed to the maximum surface area of $SiO_2$ with the small size and interactions of $ClO_4^-$ with $SiO_2$ surface. The decrease of crystallinity from XRD reveals the improved segmental motion of polymer chains and enhances the coordinating interaction of Li+ with ether groups of PEO chains. The ionic conductivity obtained from in situ hydrolysis was in the range of $10^{-4}$ $-10^{-5}$ S $cm^{-1}$ and is ten times higher value than the ex situ PEO-$MUSiO_2$ CPE counterparts. At 60 °C the ionic conductivity value was $1.2 \times 10^{-3}$ S $cm^{-1}$ and is comparable to the liquid electrolyte. The electrochemical stability window was 4.7 V for ex-situ CPE and is 5.5 V for in-situ CPE and is attributed to strong adsorption effect on anion for in situ CPE, which suppresses its anodic decomposition at high potential.

The electrochemical properties were tested between 2.5 and 4.1 V with a similar charge and discharge rates. Figure 37 a shows the voltage profiles of all-solid-state batteries with in situ CPE under different rates at elevated temperature (90 °C), and for rate below 1 C (170 mAh $g^{-1}$ based on the weight of $LiFePO_4$), the cell shows a high capacity retention of 120 mAh $g^{-1}$ while, the flat plateau with low over potential indicates the fast kinetics of $Li^+$ transport in CPE. Figure 37 b shows excellent rate capability with >100 mAh $g^{-1}$ specific capacity retention at a lower operating temperature (60 °C). The high value of over potential at 60 °C than 90 °C, may be due to the decreased ionic conductivity and slower $Li^+$ transport kinetics at 60 °C. Figure 37 c shows the comparison on rate capability of ceramic-free SPE, ex situ CPE, and in situ CPE at 1 C rate with 90 °C temperature. As capacity was improved and almost double capacity retention (~120 mAh $g^{-1}$) was observed for in situ CPE than ex situ CPE (~65 mAh $g^{-1}$). The ceramic-free CPE shows much lower capacity retention (~50 mAh $g^{-1}$) and much higher over potential indicating the slow $Li^+$ transport. For in-situ CPE cycle stability was more and after 80 cycles, >105 mAh $g^{-1}$ of specific capacity can still be retained (figure 37 d), and absence of any growth of dendrites supports the excellent electrochemical stability of our CPE within the voltage window.

Nnacy *et al* [100] reported the effect of $Al_2O_3$ nanofiller on a blend of polyethylene oxide (PEO) and polypropylene glycol (PPG) with zinc trifluoromethanesulfonate [$Zn(CF_3SO_3)_2$] as salt. XRD analysis shows the decrease of intensity on the addition of salt and nanofiller which indicates the altering on the structure of polymer chain due to the feasibility of coordination interactions occurring between zinc ions and ether group of PEO. Also crystallinity decreases with the addition of nanofiller but doe $Al_2O_3$ content (>3 wt. %) an increase in crystallinity was observed due to the possible reorientation of PEO polymer chains. It may be attributed to the reduction of dipole-dipole interaction or hydrogen bonding of the polar groups near the surface in of nano-$Al_2O_3$ and free volume of the polymer matrix decreases. SEM analysis shows remarkable rough structure for pyre PEO indicating its crystalline nature and addition of salt changes the surface homogeneity. Further addition of nanofiller shows a relatively smooth surface through whole polymer electrolyte with irregular sized small pores due to a decrease of spherulite structure. No phase separation occurrence was seen in polymer nanocomposite film. DSC analysis shows a decrease of area and shift in melting peak on the addition of nanofiller and is attributed to enhanced amorphous phase and polymer flexibility. The increase of ionic conductivity on the addition of nanofiller indicates the presence of strong Lewis acid base interaction between polymer/salt and nanofiller. Also, the activation energy decreases upto 3 wt. % nanofiller content and is lowest value is 0.44 eV. The electrochemical stability window was 3.6 V.

Pandey *et al* [101] investigated a comparative study of preparation of nanocomposite polymer electrolyte based on PEO: $NH_4HSO_4$ + $SiO_2$ by hot press technique (HP) and solution cast (SC) technique. Hot press technique was preferred over the solution cast technique as it is rapid, least expensive and dry procedure to prepare solvent-free polymer electrolyte films. The SEM micrograph of polymer salt complex film prepared by SC technique shows distinct spherulites separated by dark boundaries while, samles prepared by HP shows the absence of spherulitic texture. The presence of spherulites indicates the lamellar structure of the crystalline phase, and dark boundaries are associated with polymer amorphous nature. The addition of nanofiller in polymer salt matrix disturbs the spherulitic texture. The X-ray diffraction (XRD) patterns of pure PEO was affected by preparation techniques and observed in terms of peak broadening while, the Scherrer length "*l*" (a measure of crystallite size of PEO) was shortened for samples prepared by HP technique. The formation of lamellar structure observed in SEM for samples prepared by solution cast technique was confirmed by XRD. The presence of a broad region in the 2θ range ~15°–30° for $SiO_2$ dispersed NCPE films evidences the increase of amorphous content. Differential scanning calorimetry (DSC) analysis confirms the presence of a dominant endothermic peak at ~58–62 °C associated with semi crystalline-amorphous phase transition and/or melting Temperature ($T_m$) of PEO as well as to two-step changes owing to glass transition temperature. Both $T_g$ and $T_m$ decrease with the addition of nanofiller as compared to the pure polymer and an additional endothermic broad peak observed at ~85 °C is attributed to amorphous nature of polymer nanocomposites. FTIR confirmed the formation of composites on the addition of nanofiller in polymer salt system. The conduction was dominated by ions in polymer nanocomposites as a high ion transference number was observed which is $t_{ion}$~0.97–0.98. The ionic conductivity was highest for samples prepared by HP technique over SC technique and is also confirmed by less.

Klongkan prepared a solid polymer electrolyte (SPE) using PEO-$LiCF_3SO_3$ as host matrix, $Al_2O_3$ as nanofiller and dioxyphthalate (DOP) and polyethylene glycol (PEG) as plasticizers by a ball milling method followed by a hot pressing process [76]. The highest ionic conductivity was 1.00 $\times10^{-6}$ S cm$^{-1}$ for 15 wt. % $LiCF_3SO_3$ and is due to increase in a number of free cations. Decrease at high concentration is due to ion-ion interaction which traps the cation and reduces conducting pathway [81]. DSC spectra show reduction in $T_m$, $T_g$ and $X_c$ values to 54.34 °C, 64.37 °C and 37.31%, respectively for 15 wt. % $LiCF_3SO_3$ and suggest the increase in segmental motion of polymer chain. Capiglia *et al* [102] prepared SPE comprising of PEO as polymer host, $LiClO_4$ & $LiN(CF_3SO_2)_2$ as the doping salts with $SiO_2$ as nanofiller. The highest ionic conductivity was 1.4 $\times10^{-4}$ S/cm for PEO –$LiN(CF_3SO_2)_2$ –5 wt. % $SiO_2$ and the activation energy was 0.54 V. It was concluded that calcination of nanofiller also affects the ionic conductivity.

Sun *et al* [103] reported the effect of ferroelectric filler ($BaTiO_3$) on the PEO-$LiClO_4$ based polymer matrix. The highest ionic conductivity achieved was $1\times10^{-5}$ S/cm for1.4 wt. % of filler at 25 °C and $1.2\times10^{-3}$ S/cm at 70 °C. The cation transference number was 0.37 and is much higher than polymer electrolyte without filler. The electrochemical stability window was more than 4 V for all samples. Appetecchi *et al* [104] prepared the solvent free polymer electrolyte by a hot-press technique using PEO as polymer host, $LiCF_3SO_3$ as salt and $SiO_2$ as nanofiller. The cell prepared Li/PEO:$LiCF_3SO_3$:$SiO_2$/$LiFePO_4$ shows 100 % efficiency up to 400 cycles.

Choudhary *et al* [105] synthesized the (PEO-PMMA)-$LiClO_4$ based polymer nanocomposite electrolyte with various nanofiller of different particle size and dielectric permittivity ($Al_2O_3$, $SiO_2$, $SnO_2$ or ZnO) using the solution cast method followed by a melt-press technique. XRD concludes that the salt gets completely dissociated via the formation of ion-dipolar coordination in the polymer matrix and blend polymer is less crystalline than the pure PEO. From impedance study it was found that the conductivity variation follows the order No filler (NF) > ZnO > $SnO_2$ > $SiO_2$ > $Al_2O_3$ and all PNCs shows conductivity of the order of $10^{-5}$ S cm$^{-1}$. The decrease of the ionic conductivity was attributed to the slowdown of the cooperative polymers chain segmental motion on the dispersion of nanofiller.

Keller *et al* [106] synthesized the hybrid ceramic-polymer electrolytes, with $Li_7La_3Zr_2O_{12}$ (LLZO) as ceramic and $P(EO)_{15}LiTFSI$ as polymer salt matrix using the hot-press technique. The salt (LiTFSI) was chosen due to superior charge delocalization of the anion resulting in less coordinated $Li^+$ cations. The crystallinity was decreased on addition of salt while the addition of ceramic has no effect of the onset of the melting point and the glass transition temperature. Although the change in the shape of the glass transition curve suggests the disruption of the crystallization which evidences the increase in the amorphous content. Thermal stability of all the films was upto 370 $^oC$.

Scroasti *et al* [107] reported the impedance spectroscopy study of two different types of nanocomposite polymer electrolyte membranes with the addition of $TiO_2$ and $SiO_2$ nanofiller in the PEO-$LiClO_4$ salt matrix. The electrical conductivity, transport number and mechanical properties were improved. The increase in conductivity may be due to the presence of cross-linking centers for the PEO segments provided by a surface group of nanofiller which decrease the polymer chain reorganization tendency. The modification in polymer chain leads to the formation of conducting paths via interaction by surface groups on nanofiller.

### *4.4.2. Intercalated Type solid Polymer Electrolyte*

Dam *et al* [108] reported the structural and electric properties of polymer nanocomposite electrolytes consisting of PEO as polymer host, lithium hexafluoroarsenate ($LiAsF_6$) as salt, and dodecylamine modified montmorillonite (DMMT) clay with a surfactant (organic dodecyl amine). XRD spectra of prepared PNCs shows (001) reflection at an angle $2\theta = 7.51°$ as for MMT ($d_{001}$=11.77 Å) and modification of clay peak shifts toward lower angle $2\theta = 5.82°$ ($d_{001}$=15.78 Å). The increase of interlayer spacing may be due to cation exchange between the clay layers and the surfactant Figure 38 shows that addition of DMMT in polymer salt complex shifts the DMMT (001) plane towards the lower angle side up to 15 wt. % of DMMT concentrations. The increase in basal interlayer spacing evidences the successful intercalation of polymer chains in the clay layers. At high clay content (>30 wt. %) the basal interlayer spacing decreases as now, intercalation of the polymer chain is restricted in DMMT clay layers due to the increased clay content. The absence of XRD peak corresponding to $LiAsF_6$ in PNC indicates complete dissolution of salt in the polymer matrix. The almost negligible effect was observed on characteristic peaks of PEO on the addition of clay.

The conductivity variation with temperature follows VTF behavior which suggests that polymer segmental motion affects the dc conductivity. The increase of temperature increases the diffusion rate and segmental motion of polymer chains. The highest conductivity was $4.0\times10^{-5}$ S $cm^{-1}$ at 273 K for 2 wt. % DMMT. The increase in conductivity may be due to Lewis base nature of silicate layers of the DMMT clay and interact with lithium cations results in the increase of the fraction of free ions (number density of charge carriers $n_i$). The optimum concentration of clay was better for the conductivity than the lower or higher content.

Das *et al* [109] prepared PNC using PEO-PDMS+$LiCF_3SO_3$ and Hectorite clay. The presence of ion-polymer, and ion– clay interaction was confirmed by Raman spectroscopy. The polymer clay interaction was confirmed and observed in terms of the shift in the peak position of 842 $cm^{-1}$ and 857 $cm^{-1}$, and peak profile of 1061 $cm^{-1}$ and 1232 $cm^{-1}$ (figure 39). And, increase in COC stretching at 1122 $cm^{-1}$ corresponding to PEO is evidence of ion polymer interaction due to clay intercalation inside the polymer matrix. Further analysis was done to study the ion dissociation effect by deconvolution the anion peak and for PNC with 1 % clay content higher free area suggest the number of free charge carriers for conduction. The highest conductivity was $5.2\times10^{-5}$ $Scm^{-1}$ for 1 wt. PNC.

As layered silicates disrupt the crystallization of polymer so, regarding this Sunitha *et al* [110] reported the effect of MMT on the $(PEO)_4$LiCBSM based polymer electrolyte. DSC and XRD analysis evidence increase of amorphous phase and the highest ionic conductivity was $2.15 \times 10^{-4}$ $Scm^{-1}$ at 323 K (figure 40 a & b)

Mohapatra *et al* [111] reported the preparation of PEO-$LiClO_4$ based solid polymer electrolyte with Na-montmorillonite (MMT) as clay. XRD analysis evidences the decrease of PEO crystallinity as observed in terms of decreased peak intensity and increased FWHM. The decrease of crystallinity is attributed to the interaction of clay with polymer salt complex. Also, the Basal spacing increases on modification of clay and strong intercalation of polymer salt complex are observed in clay galleries due to dipolar interaction. TEM analysis further evidences the intercalation/exfoliation in the clay galleries. DSC analysis shows a decrease in glass transition temperature with addition of clay and is an indication of enhancement in polymer flexibility. The thermal stability of prepared PNCs was upto 300 ºC and may be due to barrier effect in which clay layers prevents decomposing of polymer salt complex. But at high content clay layers may accumulate the heat on the surface and decrease the degradation temperature. The highest conductivity obtained was $\sim9.43\times10^{-4}$ $Scm^{-1}$ for a 1 wt. % clay loading. The voltage stability was increased for 7.5 % clay loading and is 3 V. The ionic transference number achieved was 0.99 while, cation transference number was 0.50 much higher than polymer salt system which has 0.25-0.30 (figure 40 d).

Ratna *et al* [112] prepared the Poly (ethylene oxide) (PEO)/clay (Nanocore I30E)) nanocomposites with $LiBF_4$ as salt. The increase of Basal spacing indicates the formation of intercalated nanocomposites. Crystallinity is reduced after addition of salt in pure polymer and is due to the interaction of cation with polymer ether group as concluded by DSC (figure 40 c). The PEO/$Li^+$/clay composites and PEO/clay composts show different behavior due to the interaction of negatively charged clay layers with the cation.

It was concluded that two effects are present in PEO/$Li^+$/clay composite electrolytes: one reduces the crystallinity and the other favors the crystallinity. Former one dominates at low clay loading due to steric hindrance produced by the huge surface area of randomly oriented clay in the polymer matrix. At high clay loading later one dominates due to the expansion of the silicate layers and clay content which reduces the PEO---$Li^+$ interaction and crystallinity increases. The highest conductivity was obtained for 7.5 % clay loading due to the interaction of lithium cations with silicate layers which disrupts the strong intra-association between $Li^+$----$BF_4^-$ and supports the fast migration of lithium ions.

Recently Lin *et al* [113] reported the preparation of SPE comprising of PEO-LiTFSI and natural halloysite nano-clay (HNT). The highest ionic conductivities values for PEO+LiTFSI (EO:Li=15:1)+HNT(10 %) are $1.11\times10^{-4}$ , $1.34\times10^{-3}$ and $2.14\times10^{-3}$ S $cm^{-1}$ at 25 °C, 60 °C, 100 °C respectively. Also the optical microscope study evidences the increase in conductivity as a reduction of phase transition temperature and the crystal size is observed. The cation transport no. ($t^+$) was 0.40 and is much higher as compared to of pure PEO ($t^+$=0.10). Also, the electrochemical stability window (ESW) was much higher as measured by LSV and confirms the stability up to 6.35 and 4.78 V respectively at 25 °C and 100 °C. TGA analysis confirms thermal stability up to 430 °C (figure 41 b). The addition of HNT also improves the mechanical strength of SPE and for 400 % strain the stress for HNT based SPE is 2.25 MPa which is large as compared to 1.25 MPa for PEO-LiTFSI electrolyte (figure 41 c).

The enhancement in conductivity was due to the presence of two face surfaces; the outer surface contains a -Si-O-Si- silica tetrahedral sheet, while the inner surface consists of -Al-OH groups from the octahedral sheet. HNT supports salt dissociation and cation gets coordinated with an outer layer having negative surface charge layer while anion gets adsorbed on the inner surface of HNT. The Lewis acid base interactions between the PEO---LiTFSI----HNT shorten the path for $Li^+$ via formation of 3D channels and a high speed pathway enhances the ionic conductivity. Also the battery was fabricated using SPE as electrolyte and stable discharge capacities were observed with an averaged value of 745 ± 21 $mAhg^{-1}$ in the 100 discharge/charge cycles. The 87 % retention was observed compared to the second discharge capacity, and is close to 100% efficiency for each cycle.

Zhang *et al* [114] prepared solid polymer electrolyte comprising of polyethylene oxide (PEO), lithium bis (Trifluromethanesulfonyl)imide (LiTFSI), and montmorillonite (MMT). XRD spectra show a strong (001) reflection at 2θ=5.0° corresponding to MMT with interlayer spacing 1.76 nm. The addition of MMT in polymer salt composites indicates an increase in interlayer spacing to 1.96 nm and is due to polymer chain intercalation in nanoclay galleries. Also, the decrease in peak intensity with the addition of nanoclay indicates an increase of amorphous content. The maximum conductivity was $2.75\times10^{-5}$ S $cm^{-1}$ at 25 °C for the composite containing 10 wt. % MMT and $3.22\times10^{-4}$ S $cm^{-1}$ at 60 °C. Further at higher clay content insulating nature of clay dominates which reduces the ionic conductivity. The $Li^+$ transference number was increased from 0.17 to 0.45 for PEO/LiTFSI/10 % MMT based polymer electrolyte and is attributed to the interaction of Lewis acid sites on the anionic surface of MMT with ether group of PEO. The electrochemical stability window is 4 V and is useful for application purpose. The electrochemical performance of Li/S cells was investigated using galvanostatic charge/discharge at 60 °C. The specific capacity of 998 mAh $g^{-1}$ was obtained in the first discharge at 0.1 C, and a reversible capacity of 591 mAh $g^{-1}$ is achieved in the second cycle (figure 44 b). Figure 42 c shows the cycling performance of the cells with PEO/MMT solid polymer electrolyte and initially capacity increase is seen for the first 20 cycles at 0.1 C. After 100 cycles, the reversible specific discharge capacity of 634 mAh $g^{-1}$ was obtained and is equivalent to 63.5 % capacity retention of initial discharge capacity. Figure 44 d shows rate capability of cells for different C rates, namely 0.1, 0.2, and 0.5 C, it shows a general trend of decrease of capacity with an increase of discharge rate.

Thakur *et al* [115] reported the preparation of ($PEO_{25}$–$NaClO_4$) and an organically modified sodium montmorillonite (NaMMT) based polymer nanocomposite by tape casting technique. The XRD patterns confirm the intercalation of clay in polymer salt complex and increase in basal interlayer spacing is observed due to an ion exchange on clay modification which confirms the strong interaction between polymer chain and clay. The clay peak $d_{001}$ reflection shifts towards lower angle side are evidencing the successful intercalation of polymer in to the clay nanometric channels which increase its width and are observed at only low clay content. At high clay content width decreases due to a decrease of intercalation of polymer in clay channels. The FTIR spectrum evidences the presence of strong interactions on intercalation of clay in polymer salt system. Also, the stretching band of $ClO_4^-$ ion shows enhanced symmetry which evidences the strong ion–clay interaction AFM study shows the formation of new nanostructure on the addition of clay and presence of humps suggest atomic corrugation at the nanoscale while the multiphase composite behavior of the polymer matrix on intercalation is confirmed by color pattern. Also, the conductivity increases with the addition of clay with a maximum at 5% clay concentration at room temperature with value ~$2.20 \times 10^{-7}$ S $cm^{-1}$. The increase in conductivity may be due to the interaction of negative changes in nanoclay surface and sodium cations as confirmed by XRD dominating at low clay content. While, at high clay content inverse effect is seen which shows a decrease of mobility due to the formation of ion pairs and is in good agreement with FTIR results.

Chaudhary *et al* [116] prepared the solid polymer electrolyte based on poly (ethylene oxide) (PEO), lithium perchlorate ($LiClO_4$) with montmorillonite (MMT) nanoplatelets by direct melt compounded hot press technique. The dc conductivity variation follows the Jonscher power law with exponent value 0.87 to 0.91. The highest conductivity was $4.69 \times 10^{-7}$ S $cm^{-1}$ for the 2 wt. clay content. Also the synthesis methods used in this study affect the arrangements of intercalated/exfoliated MMT nano-platelets and aligns them parallel to the film surface, which affected the ion conduction paths, and may favor the enhancement in ionic conductivity. Table 6 represents the some significant results of ionic conductivity, cation/ion transference number, thermal stability, voltage stability window and cyclic stability in dispersed and intercalated type solid polymer electrolytes.

## 5. Proposed Ion Transport Mechanisms

For the complete understanding of improvement in properties of polymer electrolyte system it becomes crucial to know that, how the transport of ions occurs in the polymer matrix. For ion transport one important property possessed by polymer host is that it must have an electron rich donor group such as; ether group (—Ö—) as in the case of PEO and high dielectric constant (~4-5). This ether group will support the dissociation of salt via coordinating interaction to cation and making anion immobile which is supposed to attach with the polymer chain backbone. Further for complete dissociation of salt in a polymer matrix, it must have low lattice energy which will release number of free charge carriers for transport. As the conductivity is proportional to a number of free charge carriers, electric charge and charge carrier mobility ($\sigma = ne\mu$). So, a number of free charge carriers is directly linked with the dissociation of salt and which is further affected by the lattice energy of salt and dielectric constant of the host polymer [117-118].

Now, the second factor which is important is 'ion mobility'. In the polymer electrolyte since, polymer flexibility plays an important role in ion transport. Polymer flexibility is observed by the glass transition temperature of the polymer host ($T_g$) and at this temperature a rigid phase changes into the rubbery or viscous phase. Lower will be the $T_g$, it will provide more flexible phase easily and this will increase the polymer chain flexibility which plays a crucial role in ion transport. Above the glass transition temperature more free volume is created due to the fast segmental motion of polymer chains and provides enough sites for conduction. The increased polymer flexibility will increase the disorder in the polymer chain and will enhance the ion mobility by providing suitable coordinating sites for transport via interaction with electron rich sites.

### *5.1. Polymer salt system*

The addition of salt in the polymer host provides free ions for transport. Graham S. MacGlashan *et al* [119] reported a structure of the PEO:$LiAsF_6$ complex with a 6:1 composition and proposed a conduction mechanism and in which the $Li^+$ cations were arranged in rows, with each row located inside a cylindrical surface formed by two PEO chains (figure 43 a & b). Each polymer chain adopts a non-helical conformation that defines a half-cylinder; the two half-cylinders, which are related by a center of inversion, interlock on both sides.

Six ethylene oxide, starting from a C-O bond. Each cation is units are required to represent the chain conformation, which coordinated simultaneously by both chains, involving three ether oxygen's from one and two from the other, giving a total coordination around $Li^+$ of 5 and forming a somewhat distorted square pyramid (figure 43 b). The cation is coordinated with the ether group while, anions are located outside the dimensions of the cylinder in the interchain space. The crystal structure possesses highly aligned cylindrical tunnels, and the tunnels promote easy $Li^+$ ion transport along the chains, involving the sixth ether oxygen, which is not part of the static coordination environment around $Li^+$. The highly aligned tunnels provide a path for cation transport form one tunnel to next which results in high ionic conductivity (figure 44) and the anions are unable to penetrate in the tunnels due to less separation between PEO cylinders which prevents anion passage in between [120].

### ***5.2.*** *Ionic liquid polymer electrolyte*

Basically, the addition of ILs in the polymer salt matrix reduces the intermolecular interaction between the polymer chains and hence increases the mobility and flexibility of polymer chain segment which results in increased amorphous content. The enhancement in the number density of charge carriers (n) on the incorporation of IL content evidences the enhancement in ionic conductivity. The Ionic liquid may get incorporated in the PEO matrix and two possibilities are: (i) liquid drop retaining its separate identity. They may be as "coalesced drops" giving continuous pathway or "uncoalasced drops". (ii) The IL cation ($EMIM^+$) complexed with ether oxygen C–O–C group of PEO is spread out uniformly throughout the membrane [46]. As the pure polymer is of crystalline nature (Figure 45 a) and the addition of salt makes polymer chain flexible (figure 45 b). Further addition of ionic liquid in the polymer salt system increases the polymer segmental motion or flexibility which results in faster ionic transport or ionic conductivity (figure 47 c) [121].

### ***5.3.*** *Gel polymer electrolyte*

A plasticizer is a low molecular weight organic solvent that may be added to increase polymer segmental motion and flexibility and thereby improve the ionic conductivity of salts dissolved in the system (figure 46). Plasticizer penetrates into the polymer matrix and presence of interaction between polymer and plasticizer disrupts the arrangement of the polymer chain. The addition of plasticizer enhances the flexibility of the polymer electrolytes as reported by many researchers [122-123]. Plasticizer reduces the cohesive force between polymer chains and disordered arrangement of polymer chain now provides more coordinating sites to the ion for transport. This disruption in crystallization of polymer host reduces the glass transition temperature also and increased segmental motion of polymer chain leads to high ionic conductivity due to the availability of more free volume.

### 5.4. *Solid polymer electrolyte*

#### 5.4.1. *Dispersed type SPE*

The ability of nanofiller to promote fast ion transport is due to using of its structure for ion conduction via surface interaction. More importantly, nanofiller also enhance a mechanical property which is an important property to prevent dendrite growth formation and is related to the modulus of the polymer electrolyte. As ceramic nanofiller supports fast ion transport in polymer electrolytes, so it becomes essential to know the role played by nanofiller in conduction. A report by Croce *et al* [124] reported the above with support for a model. The significant interaction between surface groups of the ceramic nanofiller, host polymer and anion of salt was remarkable regarding this enhancement in conductivity. It was concluded that competition occurs between a surface group of nanofiller and cation for making a complex with polymer chain and the anion of salt. Structural modification of polymer chain occurs due to these types of interactions, (i) Nanofiller surface group interactions with polymer and hinder polymer reorganization tendency which directly supports the fast migration of cation via ceramic surface, and (ii) salt dissociation is improved due to a decrease in ionic coupling by surface interactions of nanofiller.

Another model was reported by Sun Ji *et al* [125] based on dielectric properties of nanofiller and dipole orientation of polymer by their ability to align dipole moments. In polymer salt system, two dipole moment are arranged in opposite directions to minimize its energy and form a stable state, and the addition of nanofiller changes dipole moment of the system which directly increases the dielectric constant value. Also, another possibility is that the nanofiller can penetrate the space between the polymer chains and large surface area of nanofiller prevents or retards crystallization of polymer host. It was concluded that observed change in dipole moment creates a vibration which gives activation energy to polymer matrix and facilitates the ion transport which increases the conductivity.

Croce *et al* proposed a model addressing the role of the different surface group of filler in enhancing the transport properties and specific interaction of PEO+$LiCF_3SO_3$ –$Al_2O_3$ (basic, acidic and neutral) based composite polymer electrolytes. The addition of nanofiller helps in controlling the PEO chain crystallization kinetics, as well as leads to specific surface interactions with the electrolyte components. The Lewis acid groups of the added ceramics (e.g. the OH groups on the $Al_2O_3$ surface) may compete with the Lewis–acid lithium cations for the formation of complexes with the PEO chains, as well as with the anions of the added LiX salt [124, 126].

The structural modifications occurring at the ceramics surface, due to the specific actions of the polar surface groups of the inorganic filler, which may act as cross-linking centers for the PEO segments and for the $X^-$ anions and a Lewis acid–base interaction centers with the electrolyte ionic species which together increases the number of free ions. The former promotes the structure modifications, and later one enhances the salt dissociation by lowering the ionic coupling. The acidic surface group of filler favors the hydrogen bonding with anion and polymer chains segments. This results in an enhancement in salt dissociation and amorphous phase formation as observed in ionic conductivity measurement (figure 47 a). While in the filler with the neutral surface the weak interactions with both polymer and anion of salt affect the ionic conductivity a little (figure 47 b). In the case of filler with a basic surface group (figure 47 c) confirms that, although no obvious changes occur, structural modification scan be in order $Al_2O_3$ acidic>$Al_2O_3$ neutral>$Al_2O_3$ basic.

Jayathilaka *et al* [127] proposed a model based on Lewis acid/base type interactions between the filler with different surface groups (basic, neutral, weakly acidic and acidic) and the ionic species comprising of $(PEO)_9$LiTFSI/10 wt.% $Al_2O_3$. As in a PEO based polymer electrolyte, a $Li^+$ ion is coordinated to about four ether oxygen atoms of the same or different PEO chains and ion movement occurs by the continuous breaking of old coordinating sites and formation of new sites supports the fast segmental motion of polymer chain. Now the dispersion of nanofiller $Al_2O_3$ (particle size 104 μm and pore size 5.8 nm) with a different surface group (a) filler free; (b) acidic; (c) basic; (d) neutral; and (e) weakly acidic. The different surface group provides a path for ionic transport by interactions with polymer and salt.

For the polymer electrolyte system without nanofiller anion is attached weakly with polymer segmental motion and provides no direct support for ion transport. For, the $Al_2O_3$ filler with the acidic surface, the $TFSI^-$ anion is attracted strongly with the filler surface having OH polar groups via hydrogen bonding and get separated from the salt while free lithium ion now coordinates with polymer ether group and migrates 128]. In a PEG–$LiClO_4$–$Al_2O_3$ system with basic surface groups a strong interaction between Li cation and oxygen at a surface group of filler is confirmed by FTIR spectra [129]. The weak bond formation occurs between lithium cation and oxygen surface group of filler which provides additional sites for lithium ion migration along with coordinating sites of polymer host. This dual path formation for ion transport one enhances the overall mobility or the conductivity. It was concluded form the above two that the degree of conductivity enhancement is highest for acidic alumina grains than basic alumina grains as the acidic alumina increases the anionic contribution to the conductivity.

Now, in the case of filler with neutral surface acidic and basic sites are supposed to be same and former ones are meant to promote the anion migration and the later one the cation migration. Now the anions may get interacted with both acidic and basic sites due to the random distribution of sites on filler surface and cation–anion interactions reduces the mobility as compared to the pure acidic and basic surface group. While, for the weakly acidic surface groups, alumina grain is expected to be of more acidic nature than basic groups which redces the number of additional sites for the cation and a decrease in conductivity is seen for weakly acidic alumina. The above model concludes that the enhancement in ionic mobility is due to Lewis acid–base surface group's interaction with cations and anions which support the creation of additional sites resulting in favorable conduction pathways for the migration of ions.

Thakur *et al* [130] reported substantial improvement in the mechanical stability, thermal stability, and conductivity of four series of ion-conducting dispersed phase composite polymer electrolytes (PEO + $NaClO_4$ - $Na_2SiO_3$, $SnO_2$) and proposed a mechanism based on polymer–filler interaction among the composite components. FTIR spectroscopy concludes that addition of filler is effective in loosening/stiffening of the polymeric segments. As cations coordinate with four ether oxygen in PEO–salt complexes and an ion–polymer interaction results in two types of transient cross-links (inter- or intra-molecular) of polyether chains [131-132]. The filler $Na_2SiO_3$ is basically an inert oxide with a net dipolar shift in the dielectric medium and $SnO_2$ is the oxide of basically acidic nature. In former the electropositive end of the dipoles may interact with the ether group of polymer host or $ClO_4^-$ ions, while later has a natural tendency to interact with the cross-link formation increases the viscosity and reduces the polymer chain flexibility. Two types of transient cross-linking between two nearby polymers are possible:

(i) Directly through the cation

(ii) Through cation–anion interaction or a combination of both

At low filler concentration, two possibilities arise, one is that filler may interact with the polymer directly or second is that there filler may compete with cations ($Na^+$) coordinated with ether oxygen or with anions ($ClO_4^-$). Both types of interactions reduce the chances of transient cross-linking of polymeric chains and increase the flexibility of the polymer matrixes also evidenced by FTIR spectra (downshifting of C–O–C stretching band). At high filler content polymer filler interaction leads to bridging effect between two polymeric chains via acidic or polar filler particles and enhances polymer stiffens as observed in FTIR spectra (upward shifting of the maximum of C–O–C stretching band). So polymer backbone with improved mechanical strength is obtained which can be improved with high filler concentration.

### ***5.4.2.*** *Intercalated/exfoliated type SPE*

Polymer-layered silicate nanocomposite has aroused much interest since Toyota's first successful formulation of Nylon 6-clay nanocomposite. The controleed parametres like conductvity and mobility can be achieved by incorporation of clay in the polymer which also prevents the mgration of the anion in the matrix. This also resulrts in the large intefacial area which hinders the polymer chain arrangement and increases amorphous content with mainintaing the mechanical properties. The incorporation of clays is a promising approach for enhancing ionic conductivity due to two properties (i) high cationic exchange capacity, (ii) ability for participating in intercalation and swelling processes, (iii) provides sigle ion conduction. The single ion conductionobtained in PNC with clay is due to layered starucre of clay. Then main advantage of the addition of clay is that cation gets easily intercalated in the clay galleries. One common approach to enhance the conduction is a modification of clay from hydrophilic to hydrophobic nature which increases the interlayer spacing expands. The growth of interlayer spacing is a major factor that provides the soft intercalation of polymer-salt complex in to the clay layers. The addition of clay in polymer electrolytes favors the cationic movement as cations will be allowed to move freely through the clay layers whereas the bulky and heavy anions are not authorized to pass through the clay layers. Another significant role palyed by clay is restricing the movement of anions through clay layers but also by increasing the viscosity, and catiomnic contribution dominates in the polymer matrix [131].

The exfoliated nanoplatelets of MMT in the PNCE films minimize the ion pairing effect while intercalation impedes the polymer crystallization and enhance the nanometric channels for cations mobility [132-136]. Modification of clay is necessary for better dispersion in a solvent via intercalation with a polymer [137]. On the basis of the presence of a type of interaction between the polymer matrix and organically modified layered silicate (OMLS) three type of polymer nanocomposites are acheived: intercalated and exfoliated [138].

- *Intercalated nanocomposites*: it is the result of penetration of polymer matrix in the galliers between silicate layers and forms an alternate arrangement of polymer chains and silicate layer in a regular fashion. The above arrangement is independent of polymeer to clay ratio and the distance between each other is in the range of few nanometer (1-4 nm). XRD is used to identify intercalated structures due to the presence of alternate arrangement of polymer chains and clay.
- *Flocculated nanocomposites*: they are the same as intercalated nanocomposites, except for the fact that some silicate layers are, sometimes, flocculated due to the edge-edge interactions between hydroxyl groups of the silicate.
- *Exfoliated nanocomposites*: it is the result of the complete separation of individual layers of silicate layers and radnom dipserison with an average distance (>6 nm) in polymer matrix depanding on clay charge. Generally, the clay content in an exfoliated nanocomposite is much lower than in an intercalated nanocomposite and is preferred over intercalated as they provide the best property improvements. Also it depends on clay loading unlike the intercalated one. XRD patterns show no peak for this due to the absence of ordered arrangement and large spacing.

Jing *et al* [139] investigated the free volume, structural transition and ionic conduction for PEO/$LiClO_4$/clay (OREC) nanocomposite electrolytes. It was concluded that OREC content increases the ionic conductivity of PNC and is highest for 3 wt. % clay content. At this clay content increase in free volume was the maximum. XRD analysis showed an increase in the d-spacing by organic modification of the clay and this evidences polymer intercalation in clay. Also, the polymer chain is disturbed, and interaction of nano-rectorite with polymer matrix enhances the free volume. Conductivity mechanism is reported and explained on the basis of free volume theory. The addition of PEO hinders the interaction between cation and polymer leads to the release of more free ions. Also, the addition of OREC can create a higher interfacial area between the polymer matrix and the clay which provides a path for cation migration enhances the conductivity of nanocomposite. But, as OREC exceeds 3 % then ion movement is restricted due to the formation of two type of complexes and free volume decreases.

Another model was proposed by Choudhary *et al* [116] on the basis of dielectric parameters to explain the effect of various interactions on the PEO local chain dynamics and the mechanism of PEO coordinated lithium cation motion. It is generally agreed that the $Li^+$ ion has four to six sites of coordination with PEO and MMT nano-platelets [2, 10]. In figure 48 (complex I), the $Li^+$ coordination with four neighbour etheric oxygens of the PEO segments in PEO–$Li^+$ domain is depicted and after incorporation of clay in polymer salt matrix complex formation of the exfoliated MMT and ether group of PEO with $Li^+$ is observed as depicted in complex II and complex III. Another complex formation is observed between MMT siloxane groups and lithium cation as shown in complex IV of figure 48. Complex V depicts the presence of supramolecular transient cross-linked structure between PEO and exfoliated MMT nano-platelet while, polymer chain intercalation with the cation in MMT gallery is sketched in complex VI. Complex VI also evidences that anions ($ClO_4^-$) are immobile due to bulky size and for MMT loaded system anions stay outside the clay galleries. This supports the fast ionic conduction for cation via polymer chain segmental motion [140]. But, the interaction of exfoliated MMT structures prevents cation migration due to less number of conduction paths. It was concluded that cation mobility is enhanced by two ways; (i) faster segmental motion of polymer chains and (ii) number of available ion conduction paths, as well as the height of potential barrier that must be least for successful jump from one site to another.

## 6. Summary and prospects

The electrolyte is a critical component of any energy storage/conversion device. The electrolyte plays the dual role by providing physical separation to electrodes and medium for to and fro swimming of ions enabling the transport. An ideal separator must possess some characteristic properties such as, minimum thickness, low internal resistance, shape flexibility, broad thermal/electrochemical stability window, mechanical stability and structural stability. However it is difficult to uphold all properties together but polyethylene oxide separators provide balanced properties amongst all polymers due to the presence of ether group and low glass transition temperature which enables us to build a high performance system.

In this review, we have introduced the recent progress in the prospective of PEO based polymer electrolytes with a sharp focus on dispersed/intercalated type and ionic liquid/plasticizer incorporated polymer electrolyte for energy storage devices. The desirable properties of polymer host, salt, solvent, nanofiller, ionic liquid, plasticizer and nanoclay required for synthesization and good performance of electrolyte in the battery are summarized. In addition to this the preparation methods for the polymer electrolytes by various researchers are described and it also includes the required characterization for studying the surface properties, electrochemical properties, thermal properties, stability properties done in the area of polymer electrolytes. The key role of the ionic liquid, nanofiller, plasticizer and clay in the enhancement of electrochemical properties of polymer electrolyte in energy storage devices is reported. The reliable transport mechanism proposed by various researchers to explain the increase in transport properties are concluded with a sharp eye on the presence of various types of interaction like; polymer----salt, polymer---nanofiller, salt---nanofiller, polymer---ionic liquid, polymer----clay.

The property of solvent, salt, nanofiller and plasticizer also affects the performance of polymer electrolyte. Salts with bulky anion size, high room temperature stability and low lattice energy are a beneficial candidate for achieving the high ionic conductivity. Salts with broader electrochemical stability window and environmentally friendly nature will fulfill the need for ideal salt for the polymer electrolyte. Also the role of solvent can't be neglected in the preparation of polymer electrolyte. The solvent must be of nonflammable nature, low viscosity and must be capable of complete dissolution of polymer chains inside which results in increased interaction of polymer chain with cations by swelling properly. Plasticizer with high dielectric constant and low viscosity is an alternative candidate for enhancing the properties of polymer electrolytes. Passive filler only supports indirectly the ion migration by providing addition al conducing pathways via Lewis acid base interactions and also dissociates the more salt. Instead of passive fillers, active fillers can be an alternative for the promotion of ions due to their supportive nature in ion transport and increased number of charge carriers.

At present, ionic liquid based polymer electrolytes are interesting candidates for a separator in energy storage device. They provide improved conductivity and electrochemical properties due to the release of more free charge carriers and a decrease of crystallinity. Although ionic conductivity is higher for gel polymer electrolytes but poor mechanical properties bound them for practical applications in the high performance energy storage device.

Low molecular weight Plasticizer incorporated polymer electrolytes are also used from last decade as a separator. They provide hindrance in polymer chain arrangement by penetrating inside the chain which increases the polymer chain flexibility and supports fast migration of ions which is characteristic of a high performance device. Also, the free volume by plasticizer provides an easy path to cation for migration. The addition of plasticizer also reduces the interaction of cation with polymer ether group which evidences the enhanced conductivity. But disturbance of mechanical properties affects the performance of the device and is one of the drawbacks of these types of systems.

So, best alternative having balanced ionic conductivity and mechanical properties is solid polymer electrolyte which can be classified into two types based on the addition of nanofiller and nanoclay known as dispersed and intercalated type polymer electrolytes. Further nanofiller are two types on the basis of support to transport of ions in a polymer matrix; one that is involved in ionic transport is called active nanofiller and another which supports the polymer structure only and no active role in ion transport is passive filler. The high dielectric constant of nanofiller dissociates more salt and release more number of free charge carriers. Nanofiller mainly play a dual role here; one is to promote the ion migration by reducing the reorganization tendency of polymer and second is a decrease in ionic coupling due to Lewis acid base interactions. Improved conductivity values are obtained for the dispersed type polymer electrolyte along with suitable thermal and mechanical properties which make them as an alternative to the energy storage device.

Clay intercalated polymer electrolyte is promising for a high energy storage system. The intercalation of clay in polymer chain increases the basal spacing which supports fast ion transport. Another advantage of the intercalated type polymer electrolyte is that single ion conductor can be obtained which is desirable in the case of polymer electrolytes, as anion is supposed to get trapped within the polymer chains. Clay blocks the anion outside the clay galleries and the only cation is transported inside.

So, for the successful development of advanced polymer electrolyte with overall balancing properties solid polymer electrolytes are attractive and further optimizing the properties of solid polymer electrolyte can fulfill the dream of an ideal energy storage device using polymer electrolytes. In context to above one important thing we have to keep in mind is that cost and properties as well as safety must be in the relaxed zone of human beings. In the future, polymer electrolyte/separators with shape flexibility are needed due to lack of time and space by a human being. Even an interesting one is the development of a device which can be folded easily and can sustain for long in remote areas during the journey. There is also need to search the new method of battery fabrication including layer by layer deposition and some different physical structure which can provide advanced energy storage system.

**List of abbreviations and acronyms**

| | |
|---|---|
| BMITFSI | 1-Butyl-3-Methylimidazolium Bis(Trifluoromethanesulfonyl)Imide |
| $BMIMPF_6$ | 1-Ethyl-3-Methylimidazolium Hexafluorophosphate |
| BMPyTFSI | 1-Butyl-4-Methylpyridinium Bis(Trifluoromethanesulfonyl)Imide |
| CPE | Composite Polymer Electrolyte |
| DMMT | Dodecylamine Modified Montmorillonite |
| DOP | Dioctyl Phthalate |
| EC | ethylene carbonate |
| EDLCs | Electrical Double Layer Capacitors |
| EMIMTFSI | 1-Ethyl-3-Methylimidazolium Bis(Trifluoromethylsulfonyl)Imide |
| EMIM-TY | (1-Ethyl-3-Methylimidazolium Tosylate, |
| EMITf | 1-Ethyl-3-Methylimidazolium Trifluoromethanesulfonate |
| ESD | Energy Storage Device |
| ESW | Electrochemical Stability Window |
| GPE | Gel Polymer Electrolyte |
| LAGP | $Li_{1.5}Al_{0.5}Ge_{1.5}(PO_4)_3$ |
| LATP | $Li_{1.3}Al_{0.z}Ti_{1.7}(PO_4)_3$ |
| LGPS | $Li_{10}GeP_2S_{12}$ |
| $LiAsF_6$ | Lithium Hexafluoroarsenate |
| LIB | Lithium Ion Battery |
| LiBOB | Lithium bis-(oxalato)borate |
| $LiCoO_2$ | Lithium Cobalate Oxide |
| $LiNO_3$ | Lithium Nitrate |
| LiTNFSI | Lithium (Trifluoromethanesulfonyl) (N-Nonafluorobutanesulfonyl)Imide |
| MMT | Montmorillonite |
| $MUSiO_2$ | Monodispersed Ultrafine $Sio_2$ |
| MWCNT | Multiwalled Carbon Nanotube |
| NaMMT | Sodium Montmorillonite |
| NaPCPI | Sodium Pentacyanopropenide |
| NaTCP | Sodium 2, 3, 4, 5-Tetracyanopirolate |
| NaTFSI | Sodium Bis(Trifluoromethanesulfonate) Imide |
| NaTIM | Sodium 2,4,5-Tricyanoimidazolate (Natim). |
| OMLS | Organically Modified Layered Silicate |
| PC | propylene carbonate |
| PE | Polymer Electrolyte |
| PEG | Polyethylene Glycol |
| PEGMEM | Poly(Ethyleneglycol) Methyl Ether Methacrylate |
| PEO | Polyethylene Oxide |
| PG | poly(ethylene glycol dimethyl) ether |
| PNC | Polymer Nanocomposites |
| PVP | Poly (Vinyl Pyrrolidone) |
| $PYR_{13}FSI$ | N-Methyl-N-Propylpyrrolidinium Bis(Fluorosulfonyl)Imide |
| SPE | Solid Polymer Electrolyte |
| TEGDME | 1,3 Dioxolane (DIOX) / Tetraethyl Eneglycol Dimethylether |
| VTF | Vogel-Tammann-Fulcher |
| $Zn(CF_3SO_3)_2$ | Zinc Trifluoromethanesulfonate |
| $ZrO_2$ | Zirconia Oxide |

**Acknowledgement**

One of the authors acknowledges with thanks for financial support from CUPB and partial funding from UGC Startup Grant (GP-41).

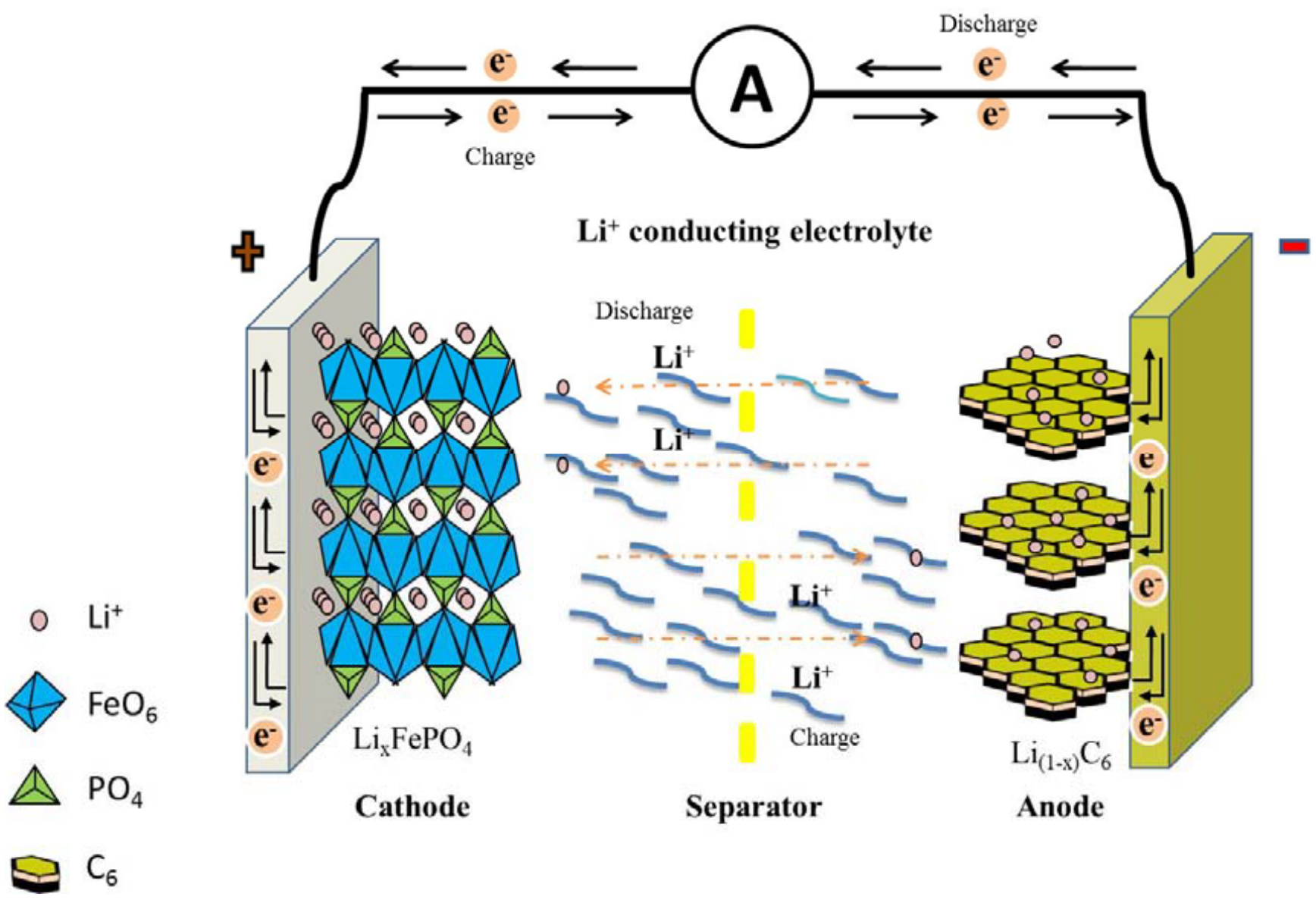

**Figure 1**. Schematic illustration of a typical lithium-ion battery [Reproduced from Ref. 4, Copyright (2014), with permission from The Royal Society of Chemistry].

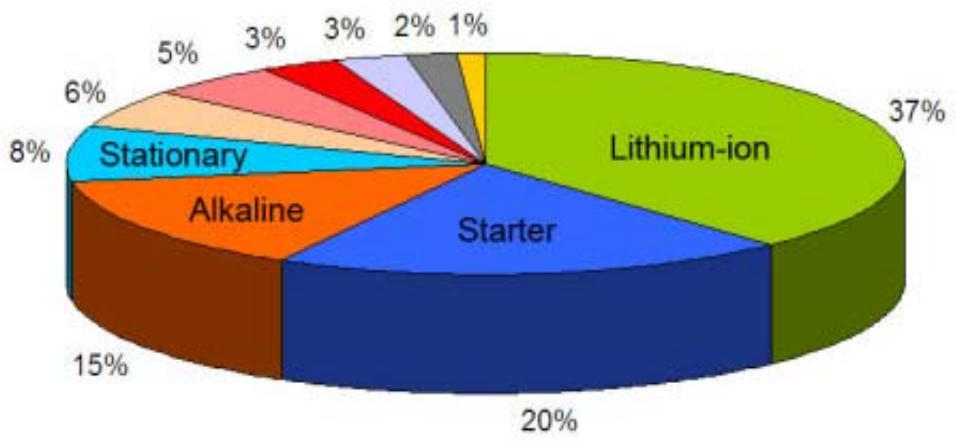

**Figure 2.** Revenue contributions by different battery chemistries **[6].**

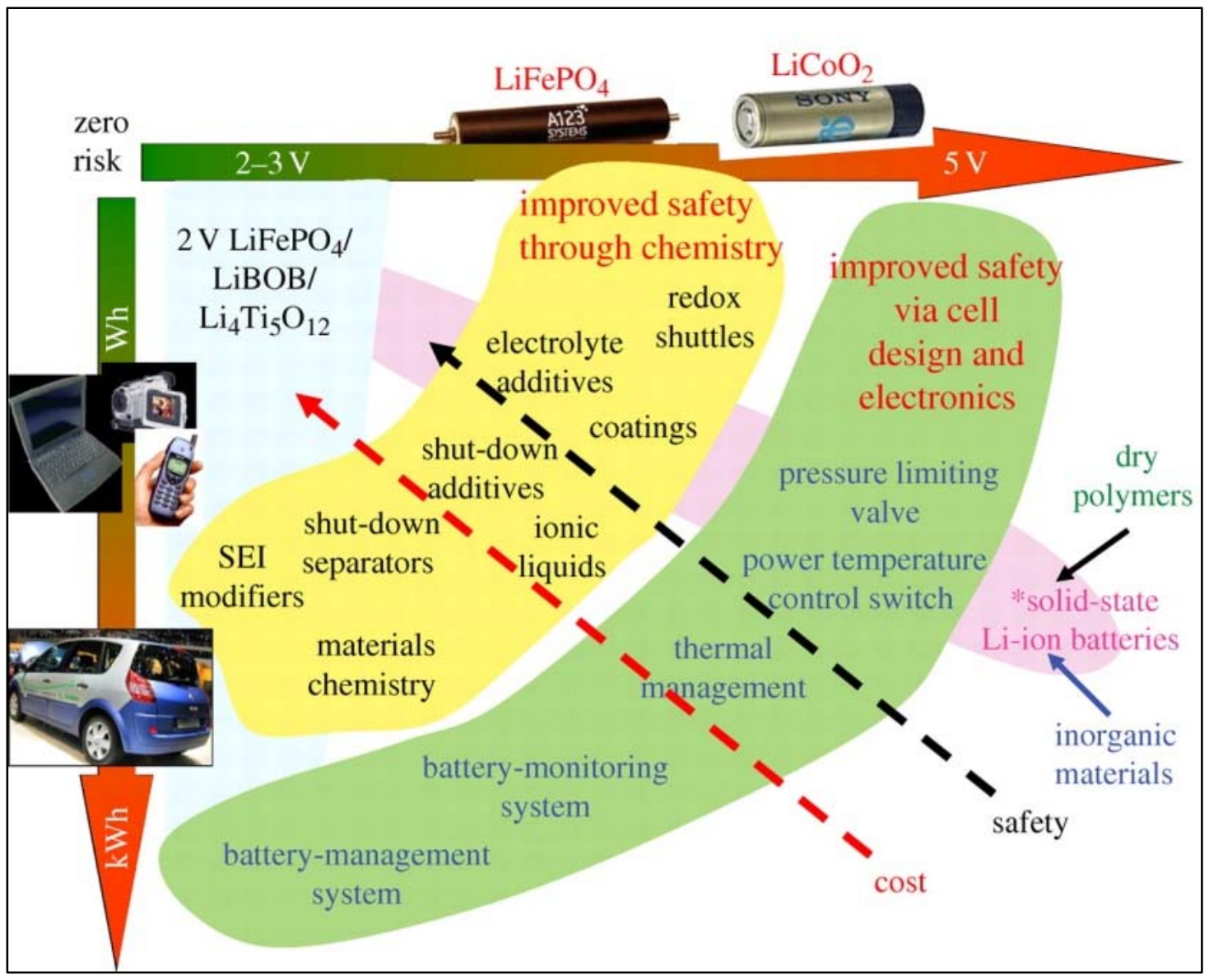

**Figure 3**. Safety aspects in Li batteries. The various chemical and engineering/devices approach aimed towards increasing Li-ion battery safety are mentioned. The distinction is made as a function of both (i) the type of applications considered (y-axis) and (ii) the type of redox chemistry involved (x-axis). Designing low-voltage systems, $Li_4Ti_5O_{12}/LiFePO_4$, increases safety at the expense of cost, conveying the general message that changes in safety and cost always track each other. While long far-reaching, the development of solid-state batteries is mentioned as the ultimate solution to reach zero risks [Adapted from ALISTORE-ERI white paper on safety; M. Armand, P. G. Bruce, L. Croguennec, M. Morcrette & J.-M. Tarascon 2008, internal communication) **[1].**

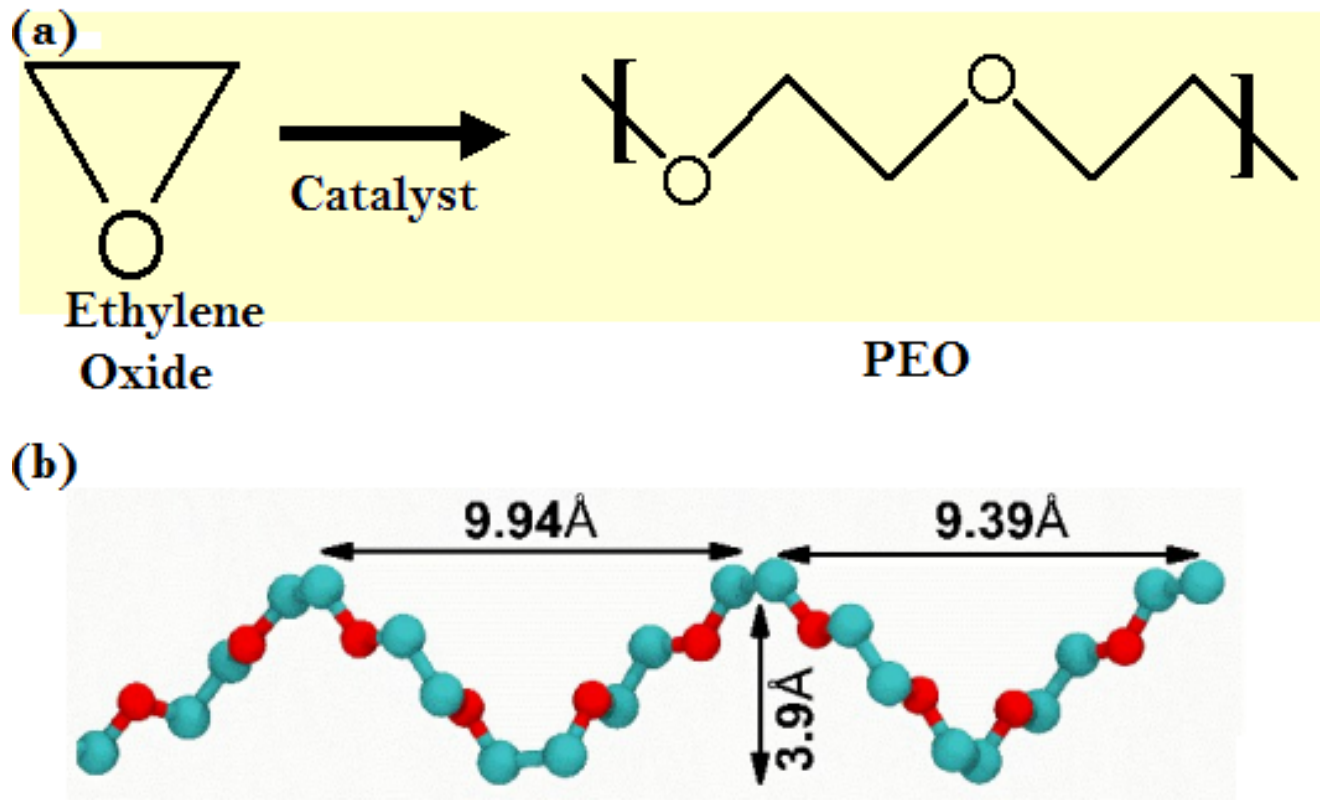

**Figure 4. (a)** Synthesis of polyethylene oxide **[17] (b)** helical portion of the PEO chain (hydrogen's are omitted for clarity) [Reproduced from Ref. 18, Copyright (2017), with permission from The Royal Society of Chemistry].

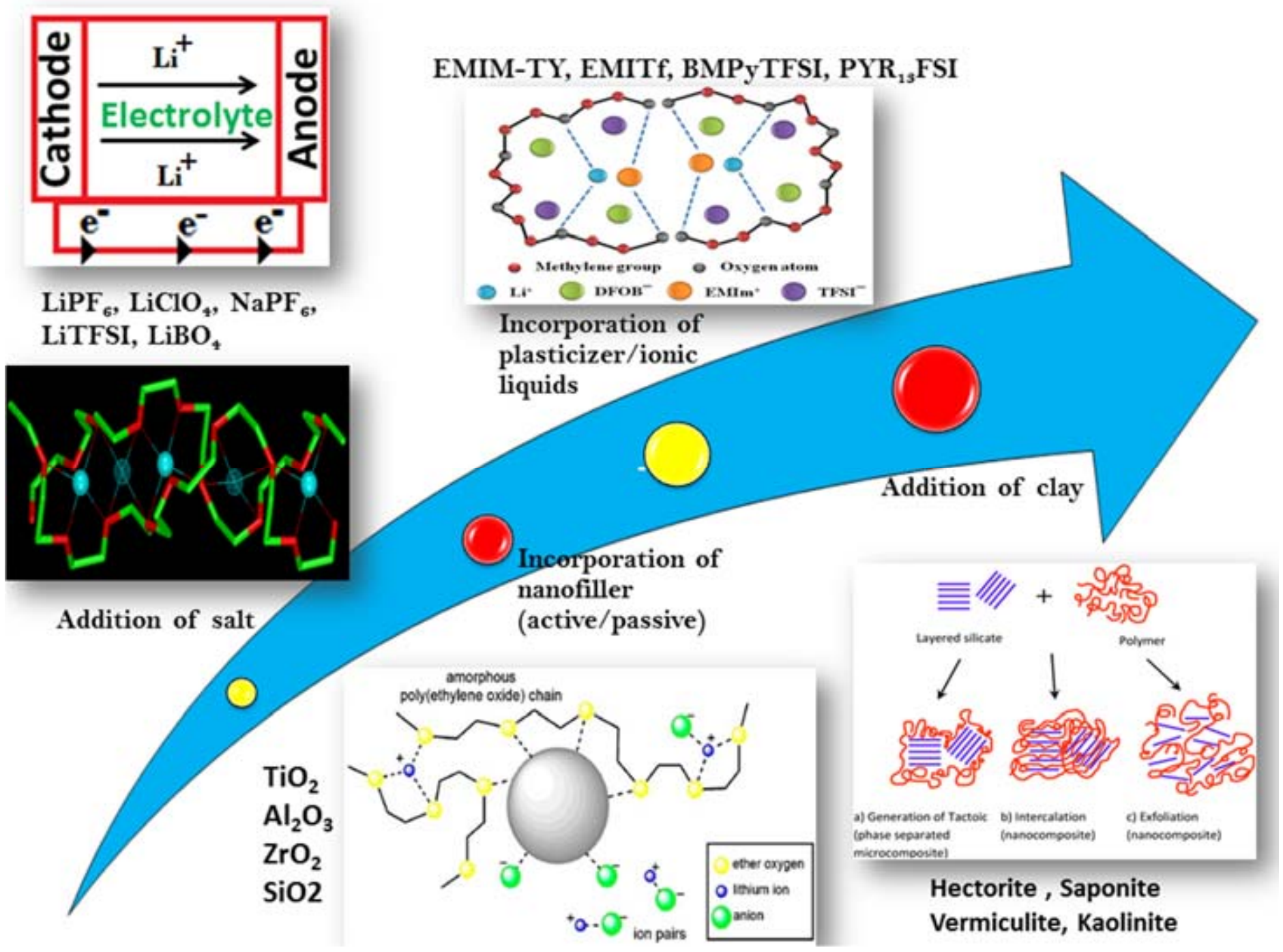

**Figure 5.** Approaches for enhancement of properties of polymer electrolytes

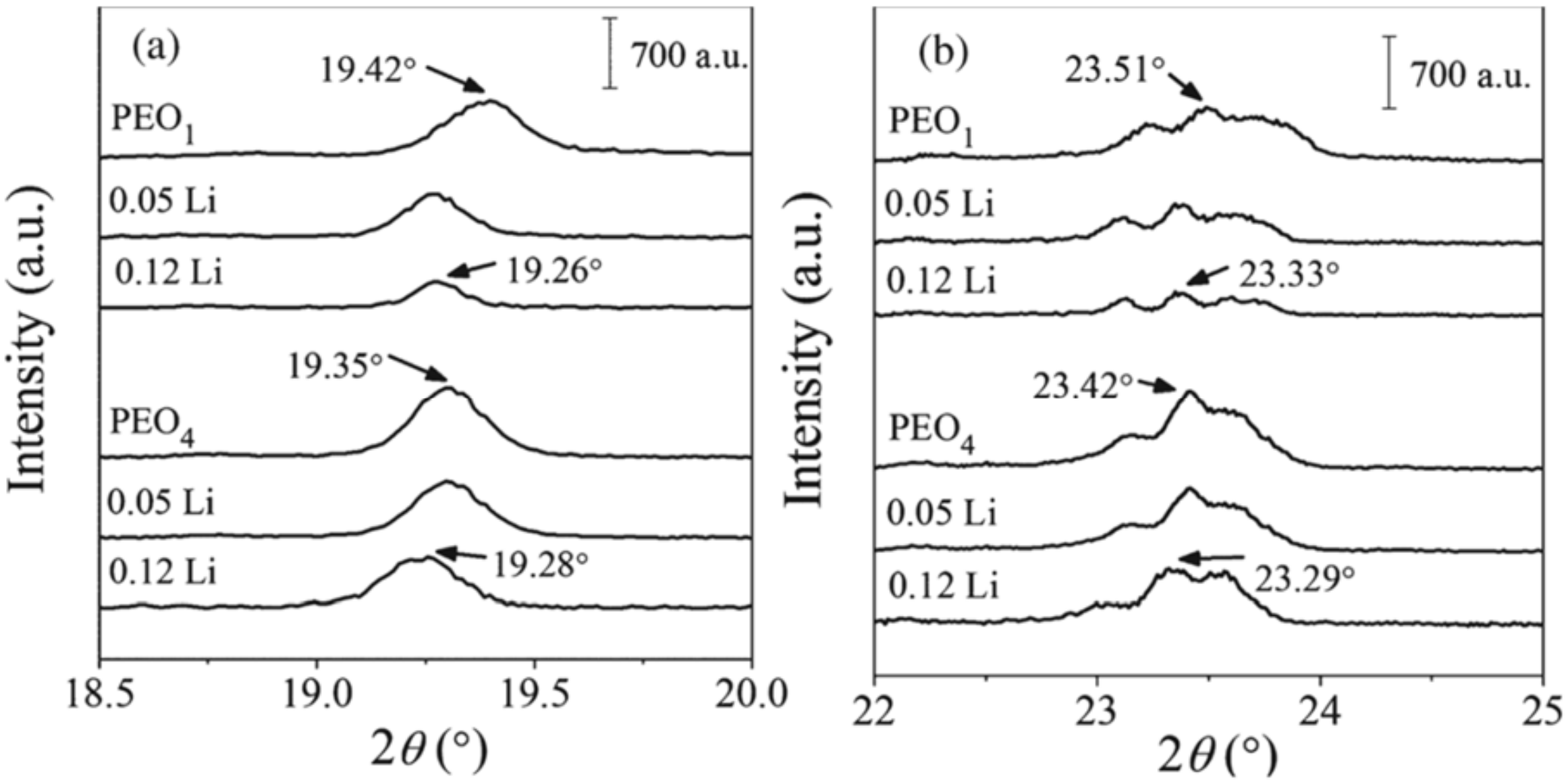

**Figure 6**. XRD diffractograms at (a) 2 theta =19$^o$ and (b) 2 theta =24$^o$ for $PEO_1$ and $PEO_4$ at salt concentrations $Y_S$ =0, 0.05 and 0.12 [Reproduced from Ref. 35, Copyright (2017), with permission from John Wiley and Sons].

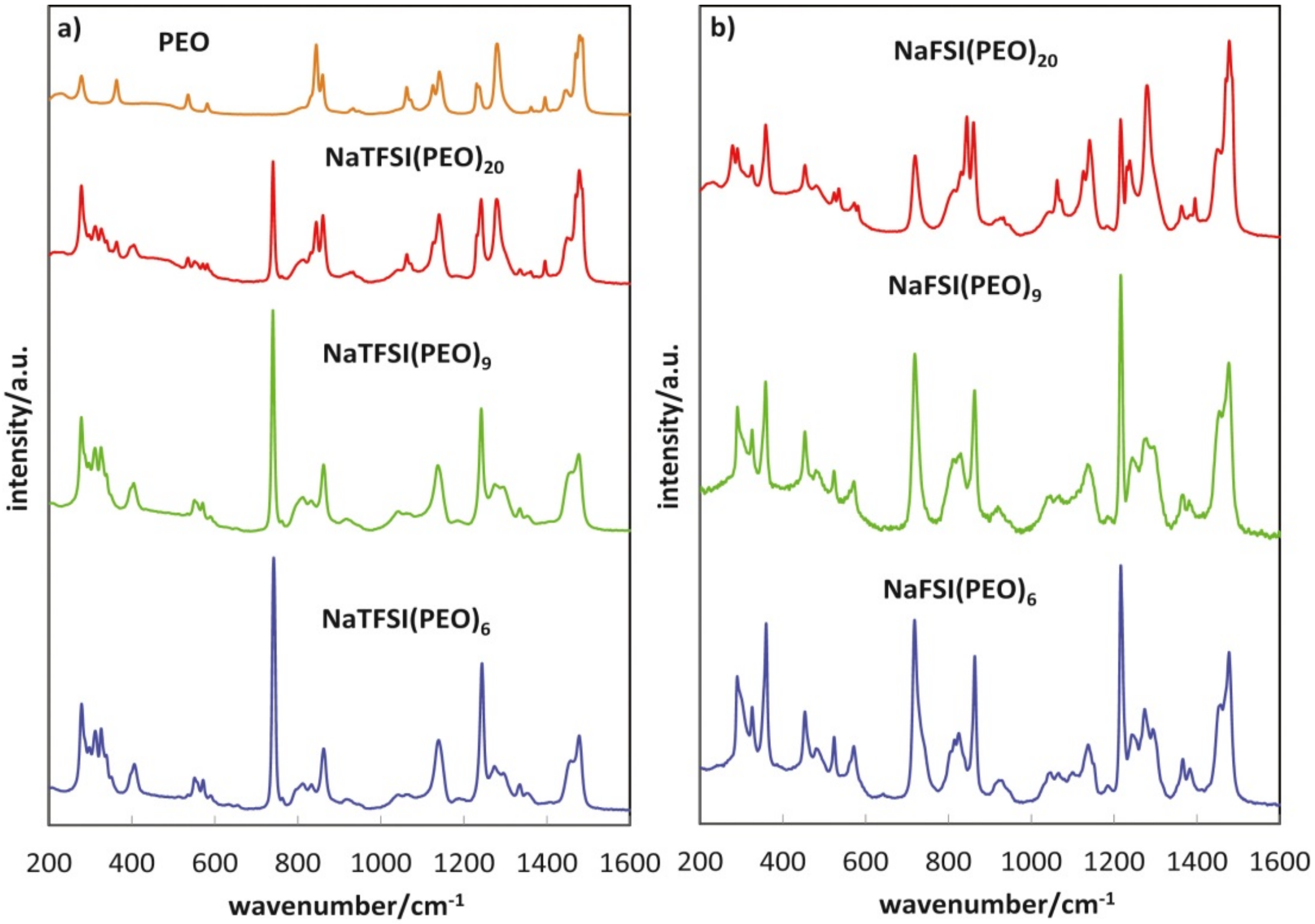

**Figure 7**. Raman spectra of: a) neat PEO and NaTFSI(PEO)$_n$, and b) NaFSI(PEO)$_n$, all in the range 200-1600 cm$^{-1}$ [Reproduced from Ref. 36, Copyright (2015), with permission from Elsevier].

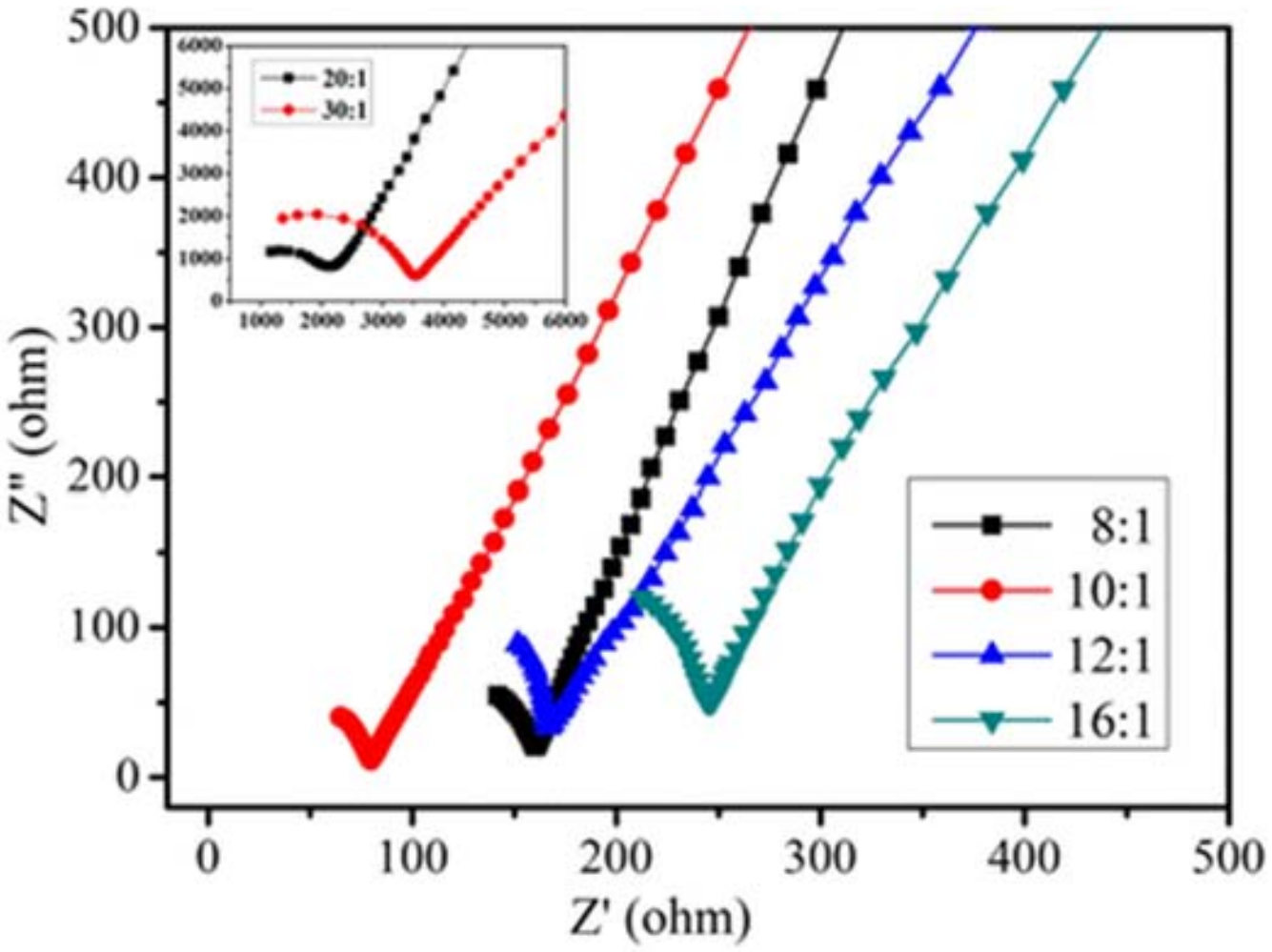

**Figure 8**. Nyquist impedance plots for the PEO–LiDFOB solid polymer electrolyte with different EO:Li molar ratios at 23 °C [Reproduced form Ref. 37, Copyright (2015), with permission from Springer].

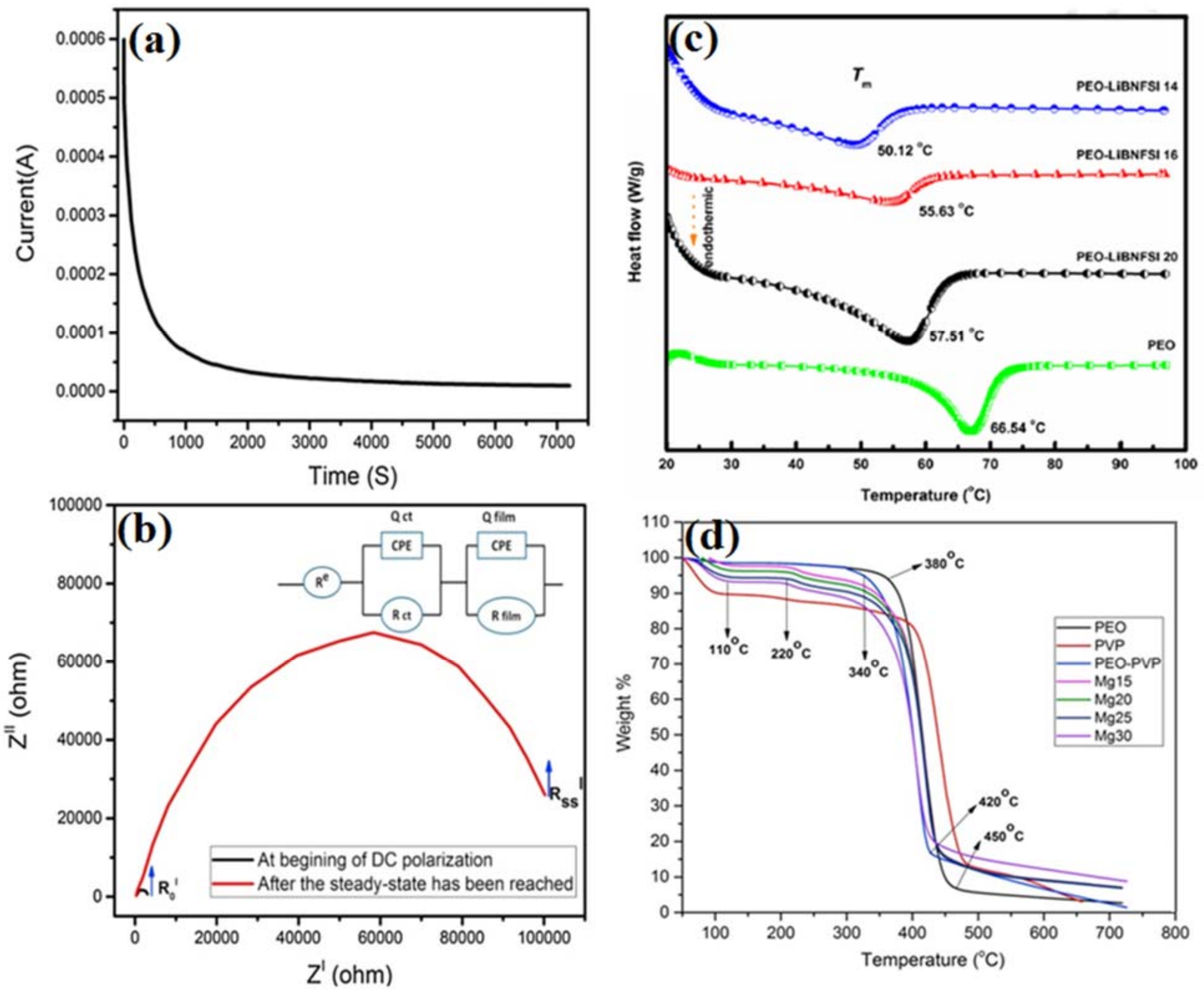

**Figure 9**. **(a)** Polarization curve of the cell, **(b)** Impedance plots of the cell before polarization and after the steady state has been reached. Inset of (b) represents the equivalent circuit for deconvolution of electrochemical impedance plot [Reproduced form Ref. 32, Copyright (2017),with permission from Elsevier], **(c)** DSC thermogram of pure PEO, PEO-LiBNFSI 20, PEO-LiBNFSI 16 and PEOLiBNFSI 14 [Reproduced form Ref. 38, Copyright (2017), with permission from Elsevier] and **(d)** TGA curves of PEO, PVP, PEO-PVP blend and SPE membranes [Reproduced form Ref. 39, Copyright (2017), with permission from Elsevier].

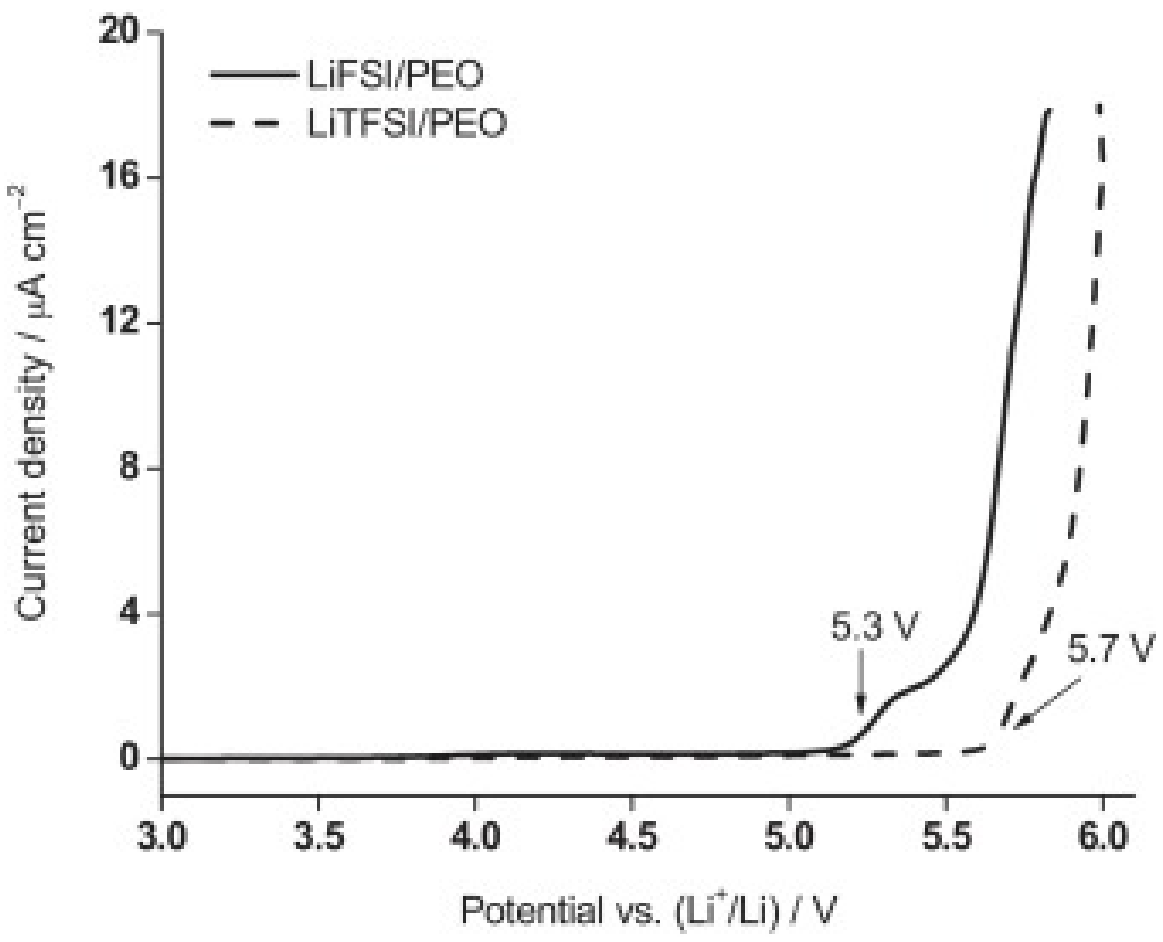

**Figure 10.** Linear sweep voltammograms of polymer electrolyte of LiX/PEO (X = FSI and TFSI) with a molar ratio EO/$Li^+$ = 20 at 80 ◦C; scan rate: 0.5 mV $s^{-1}$ [Reproduced form Ref. 42, Copyright (2014), with permission from Elsevier].

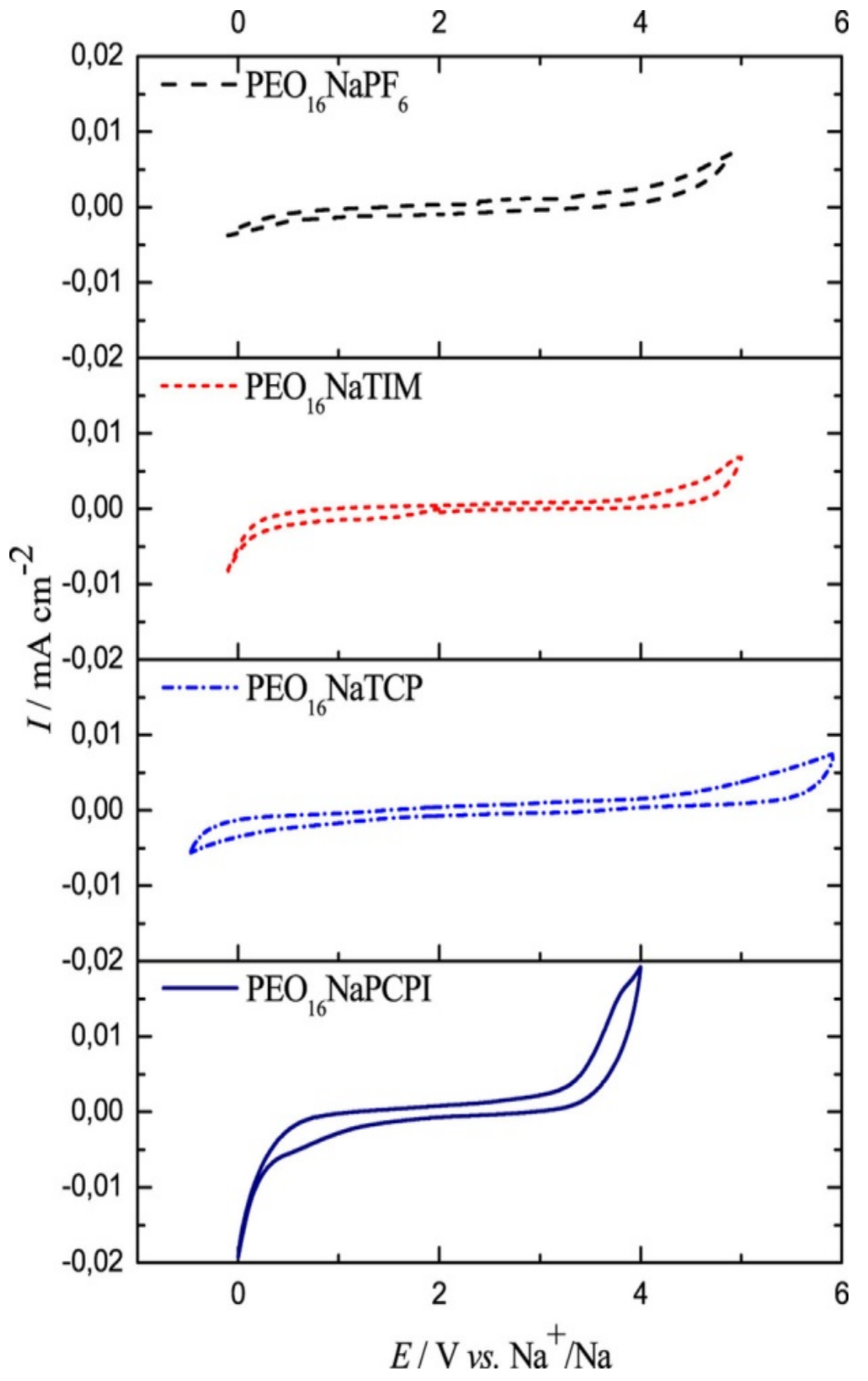

**Figure 11.** Cyclic voltammetry of electrolytes of PEO-based, solid-polymer electrolytes with salts of $NaPF_6$, NaTIM, NaTCP and NaPCPI at 50 °C [Adapted from Ref. 43, Copyright (2017), Nature Publishing Group].

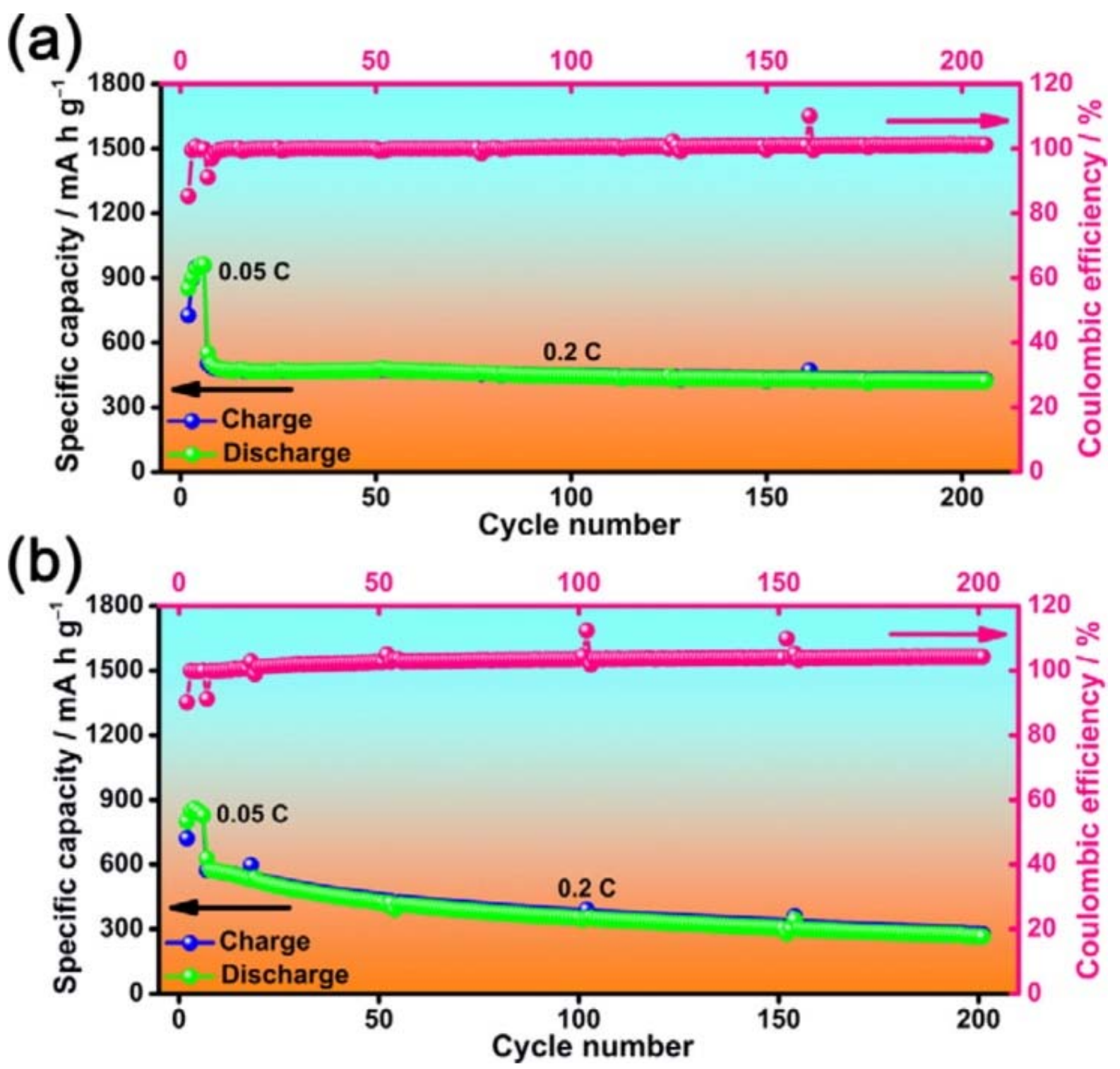

**Figure 12.** Cycling performances of the Li−S cells at 0.2C (after 5 cycles for activation at 0.05C) at 60 °C. (a) LiTNFSI/PEO (EO/$Li^+$ =20) blended polymer electrolyte; (b) LiTFSI/PEO (EO/Li+ = 20) blended polymer electrolyte. [Reprinted with permission from Ref. 44, copyright (2016) American Chemical Society]

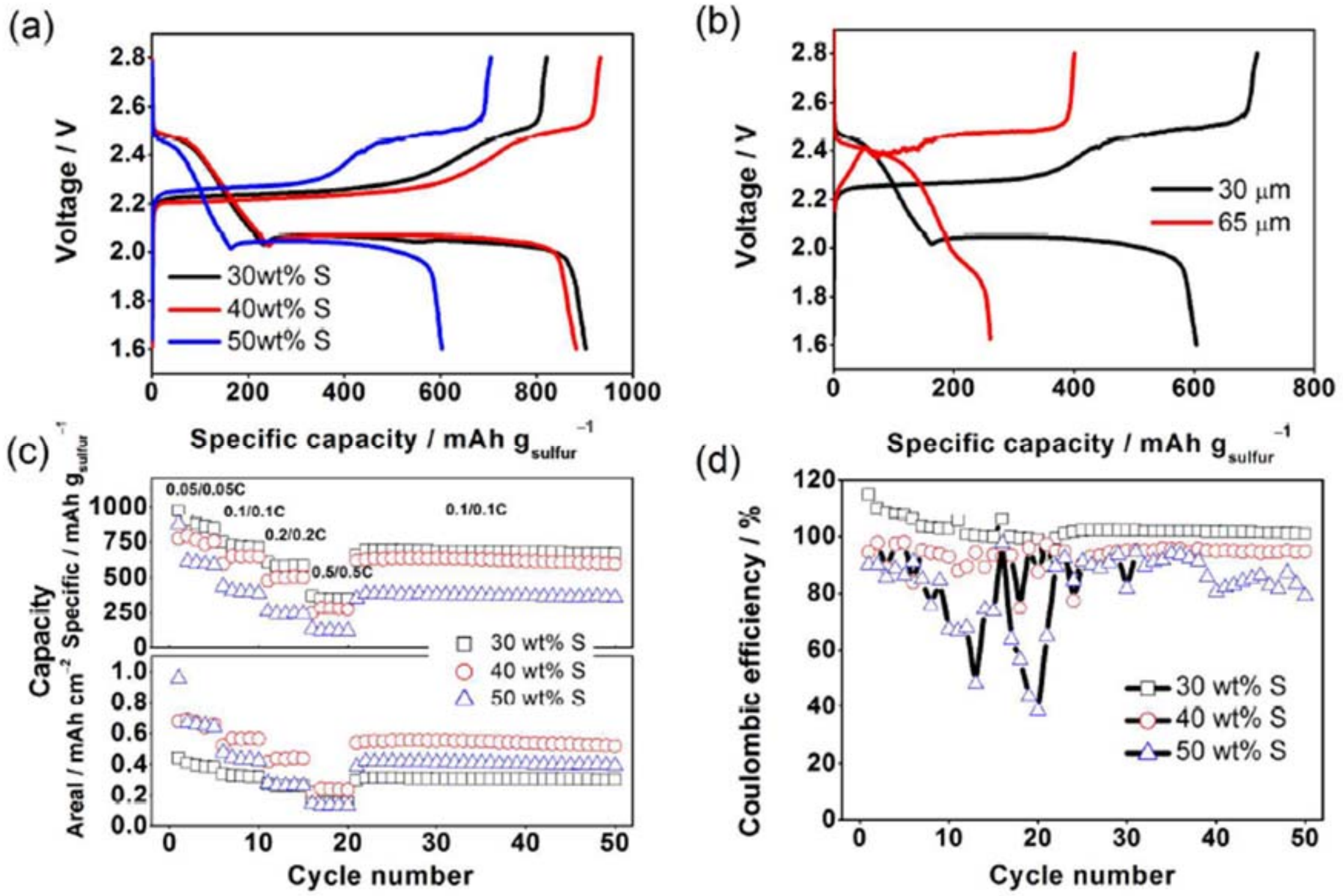

**Figure 13.** Discharge/charge profiles of the Li-S cells using the LiFSI/PEO electrolyte in the first cycle at a discharge/charge rate of 0.05/0.05C at 70 °C, (a) S cathodes with the thicknesses around 30 μm but different S contents, (b) S cathodes with the same S content of 50 wt% but different electrode thicknesses. Discharge capacity and areal capacity (c), and Coulombic efficiency (d) vs. cycle number for the LiFSI-based Li-S cells at 70 °C [Reprinted with permission from Ref. 45, Copyright (2017) American Chemical Society**].**

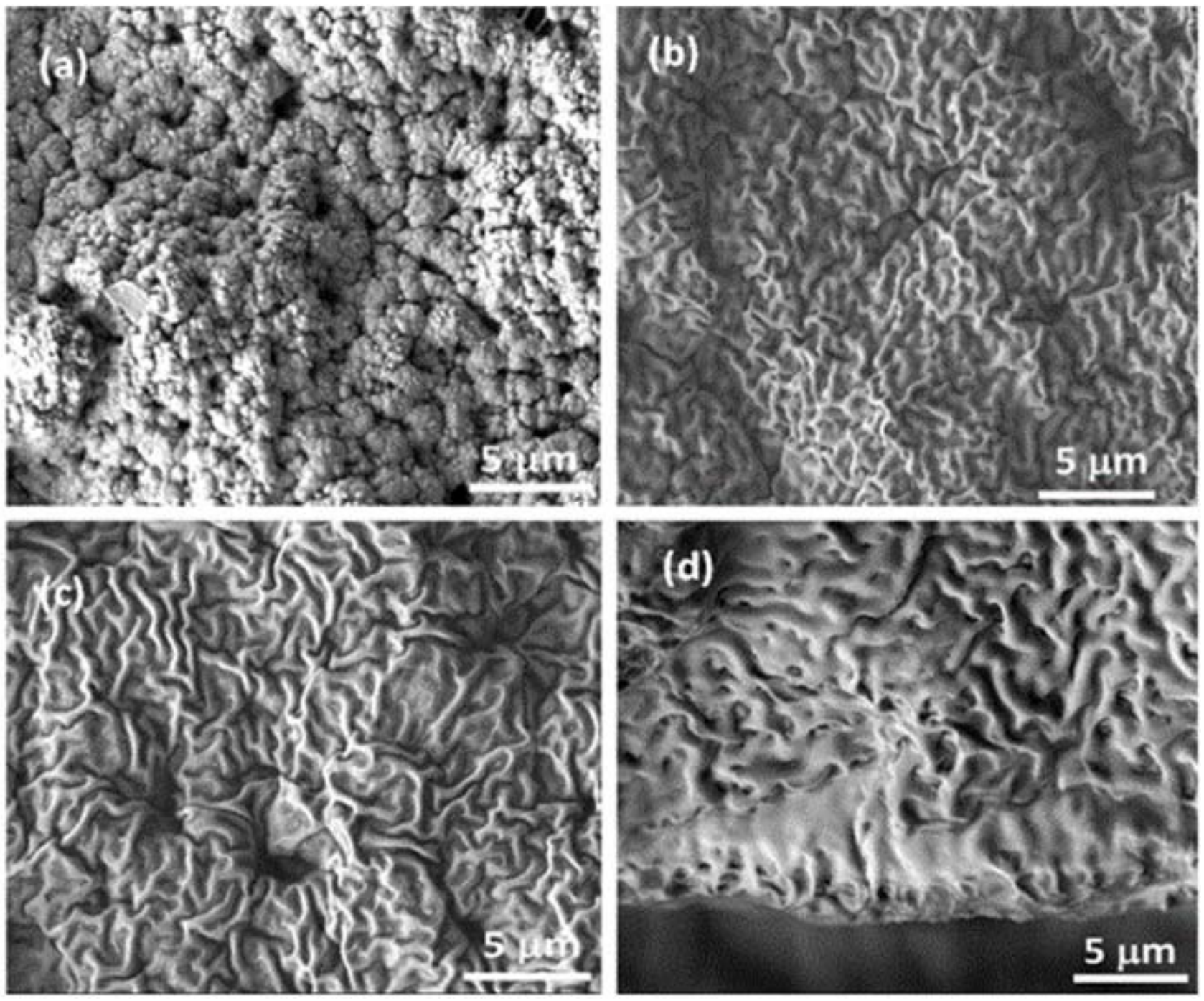

**Figure 14.** FE-SEM micrographs of PEO/(PVDF-HFP)-LITFSI-X wt. % PMIMTFSI solid polymer electrolytes: (a) X=0, (b) X=20, (c) X=50, and (d) X=50 (cross-section) [Reproduced from Ref. 49, Copyright (2016), with permission form AIP Publishing LLC].

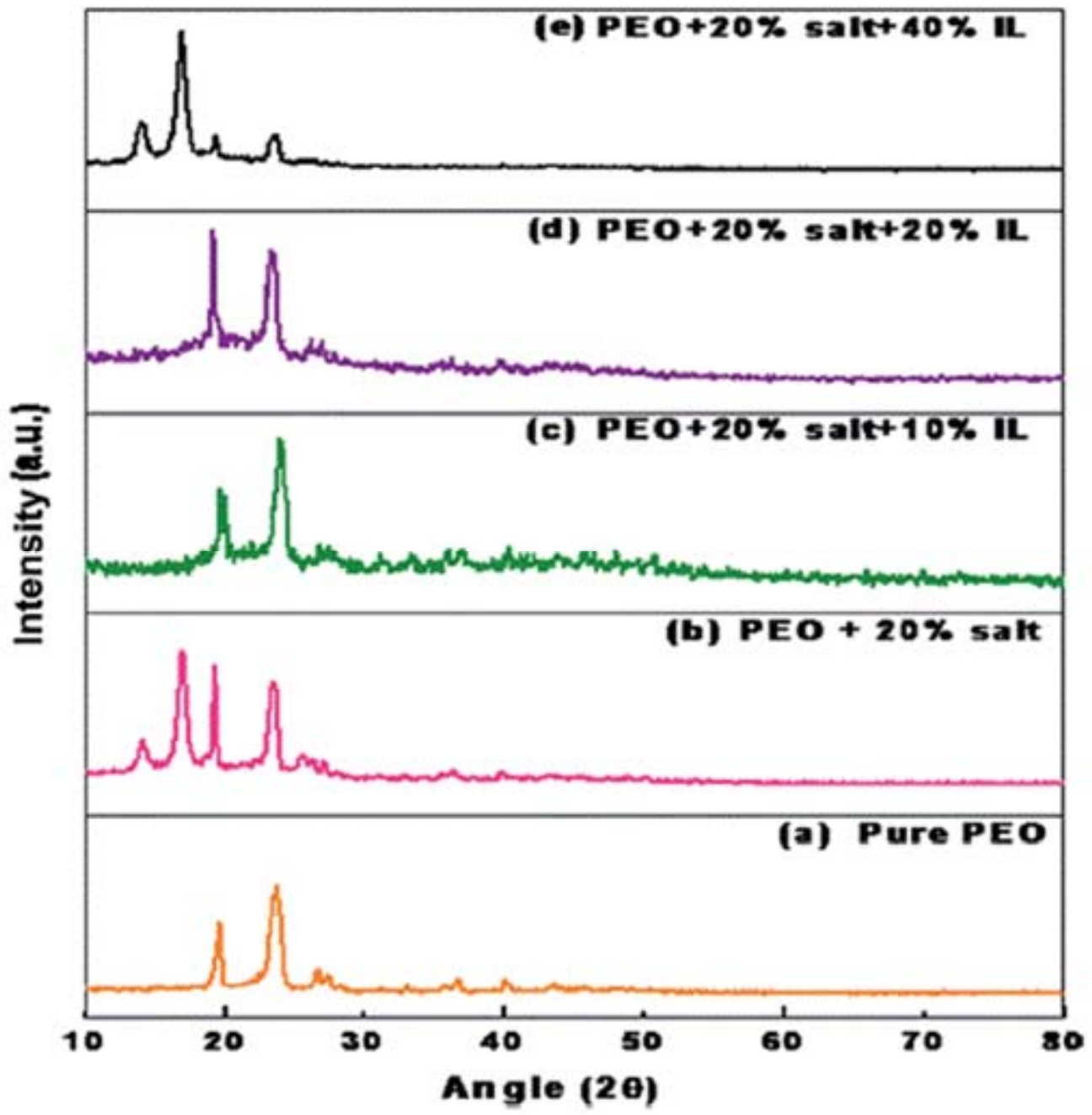

**Figure 15.** XRD pattern for (a) pure PEO and polymer electrolyte PEO + 20 wt % LiTFSI + X% IL (b) X=0, (c) X = 10, (d) X = 20, (e) X = 40 at room temperature [Reproduced from Ref. 51, Copyright (2016), with permission from Royal Society of Chemistry].

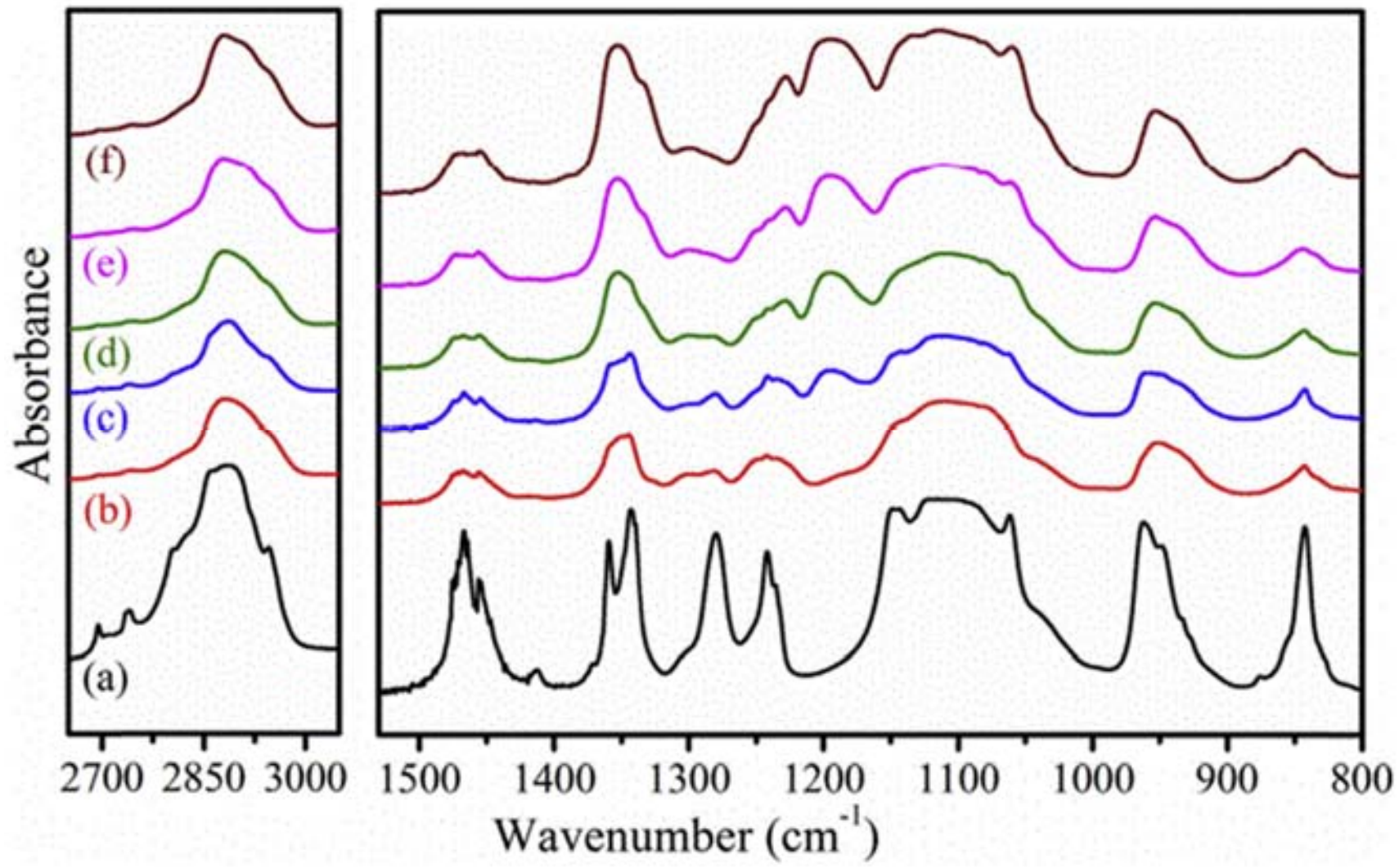

**Figure 16**. FT-IR spectra of a) pure PEO and PEO20-LiDFOB-IL electrolyte membranes with different IL contents b) 0%, c) 10%, d) 20%, e) 30% and f) 40% [Reproduced form Ref. 52, Copyright (2017), with permission from Elsevier].

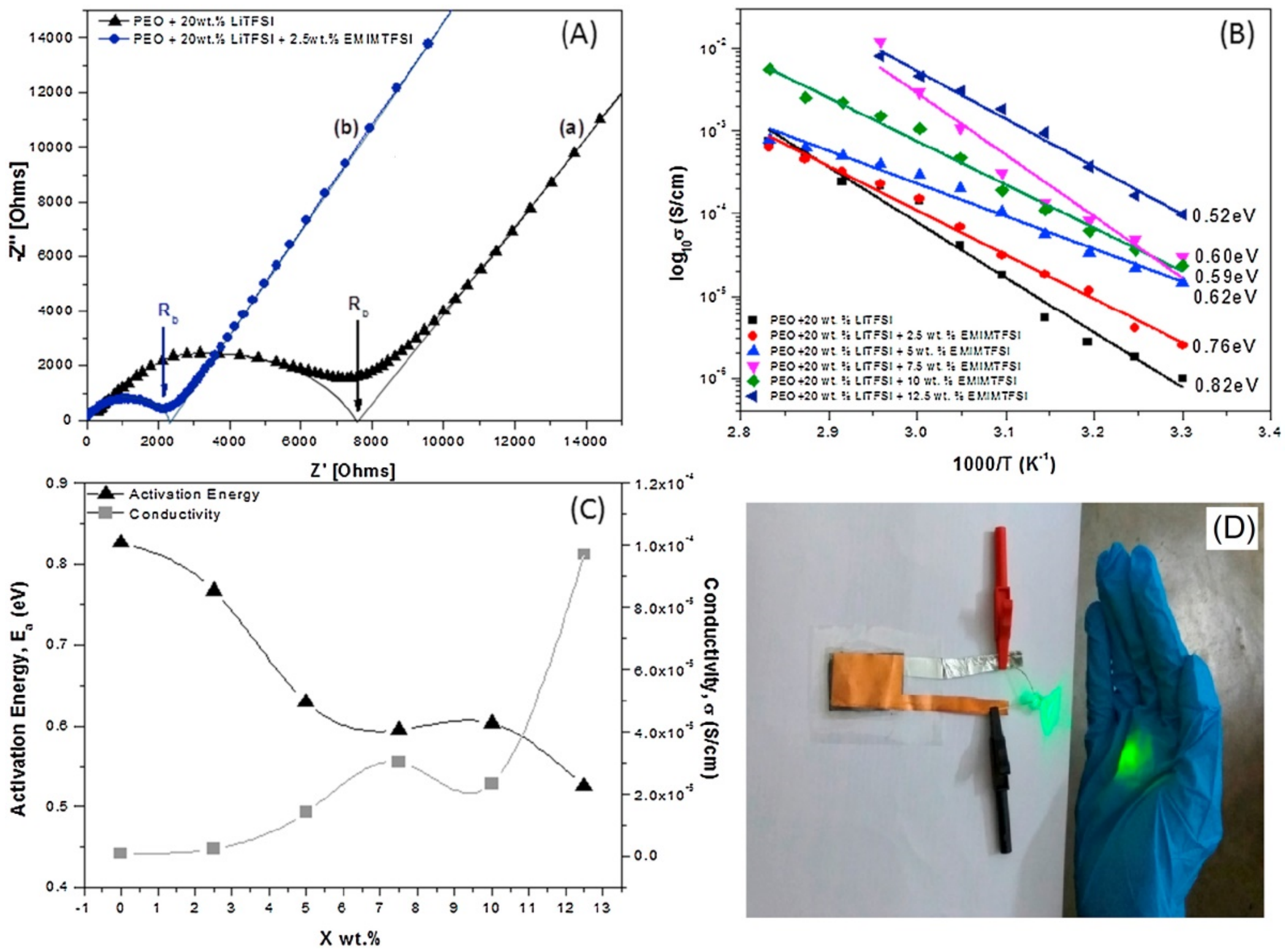

**Figure 17.** (A) Typical Nyquist plots of (a) PEO + 20 wt. % LiTFSI (b) PEO + 20 wt. % LiTFSI + 2.5 wt. % EMIMTFSI; (B) Temperature dependent conductivity; (C) Activation energy and room temperature conductivity of the GPE membranes PEO + 20 wt. % LiTFSI + x wt. % EMIMTFSI (x = 0, 2.5, 5, 7.5, 10 and 12.5) and (d) A prepared LPB cell (Li/PEO + 20 wt. % LiTFSI + 12.5 wt. % EMIMTFSI/$LiMn_2O_4$) in working condition **[**Reproduced form Ref. 53, Copyright (2017), with permission from Elsevier**].**

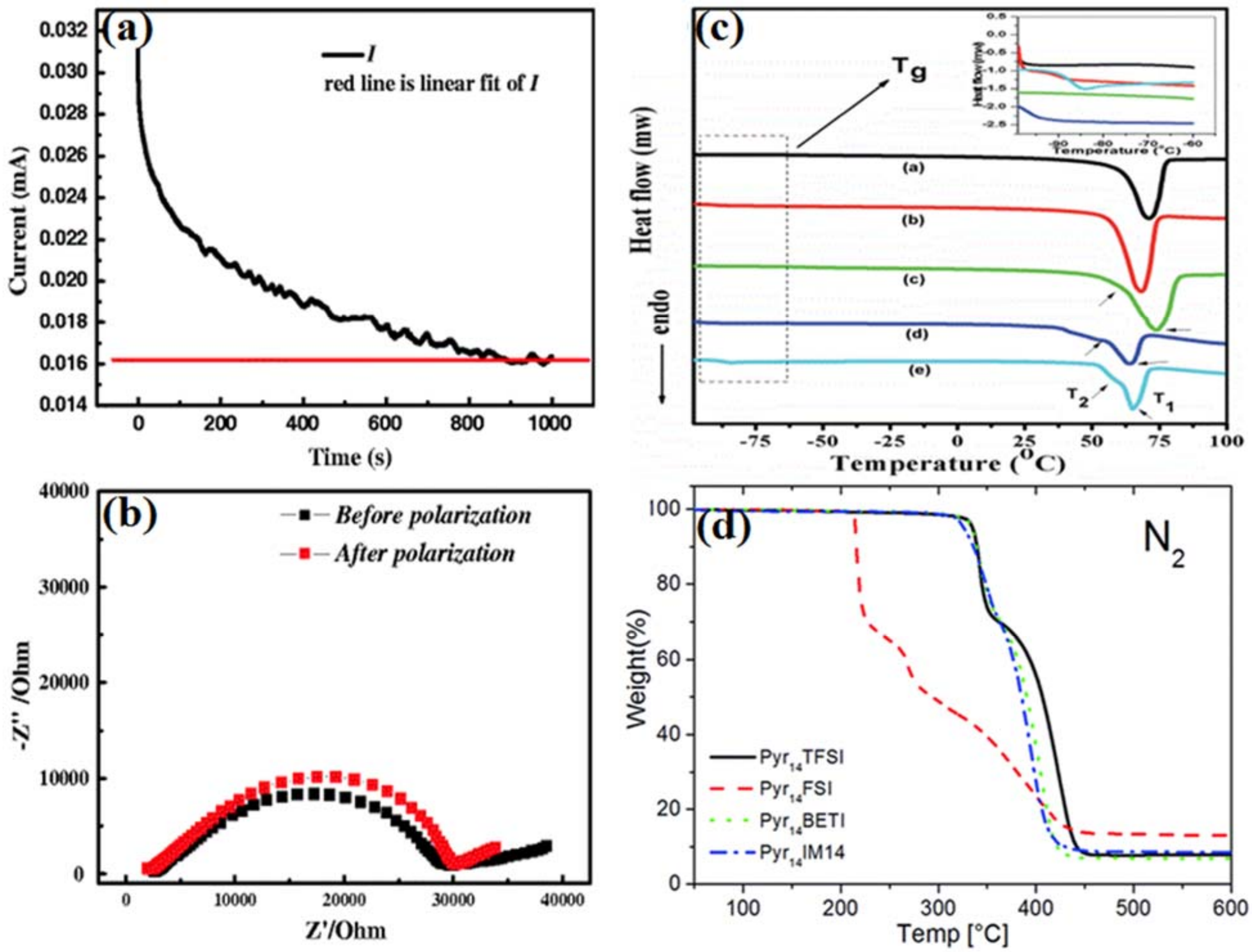

**Figure 18. (a)** Chronoamperometry of the cell with $P(EO)_{20}LiTFSI$+ PP1.3TFSI at 20 °C ΔV=0.010 $V_{dc}$ pulse, **(b)** Impedance response of the same cell before and after the dc polarization [Reproduced form Ref. 56, Copyright (2012), with permission from Springer], **(c)** DSC thermograms of (a) pristine PEO (b) PEO + 20 wt. % BMIMMS (c) PEO + 10 wt. % NaMS (d) (PEO + 10 wt. % NaMS) + 20 wt. % BMIMMS and (e) (PEO + 10 wt. % NaMS) + 60 wt% BMIM-MS [Reproduced from Ref. 57, Copyright (2016), with permission from Royal Society of Chemistry] & **(d)** TGA of polymer electrolytes in inert atmosphere. Scan rate 5 °C $min^{-1}$ [Reproduced from Ref. 58, Copyright (2015), with permission from Royal Society of Chemistry].

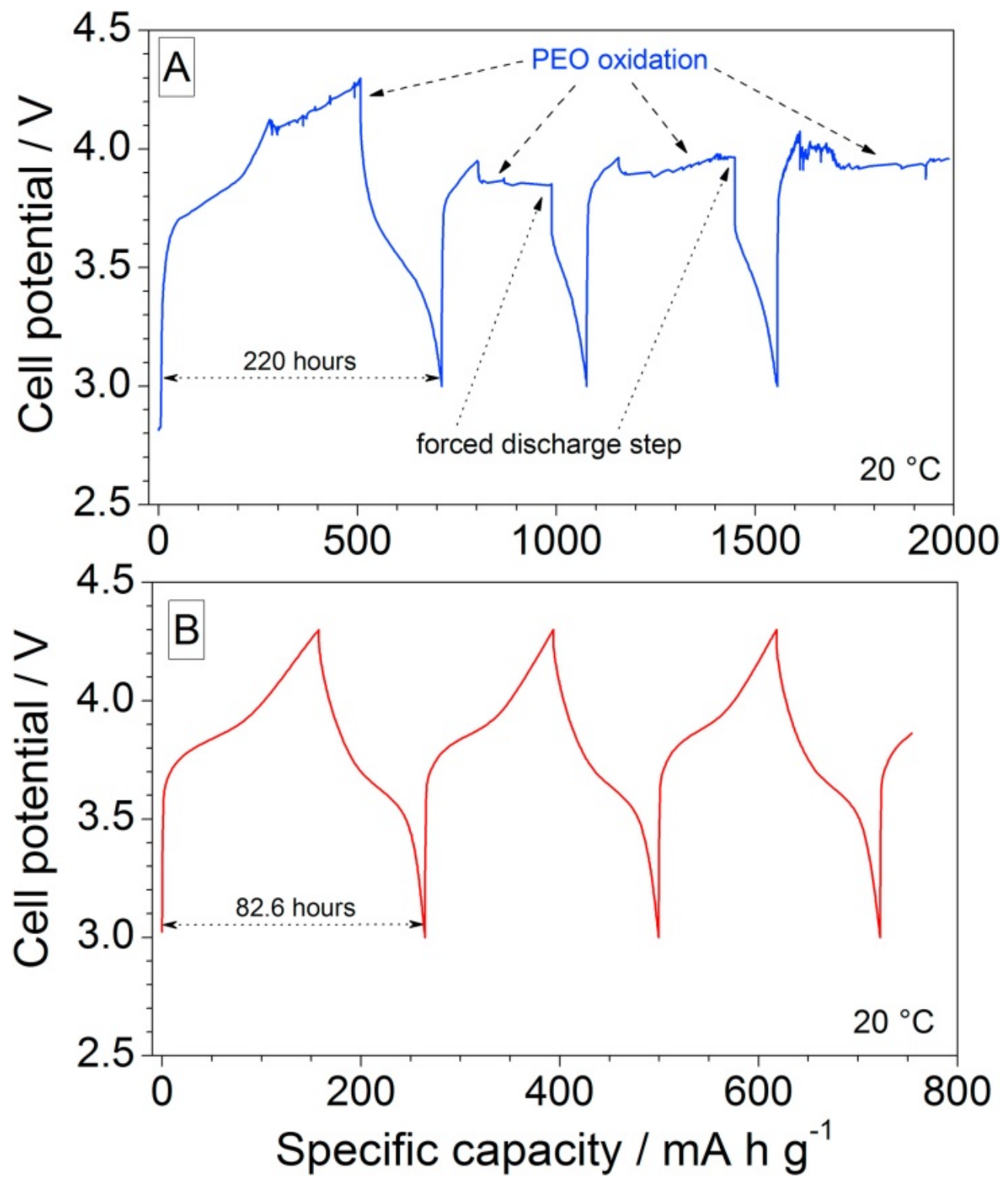

**Figure 19**. Voltage vs. capacity profiles, obtained at 0.02C and 20 °C, of Li/NMC half-cells with the PEO-E polymer electrolyte. **Panel A**: cell tested as manufactured. **Panel B**: cell kept in rest condition for more than one week before cycling tests [Reproduced form Ref. 59, Copyright (2017), with permission from Elsevier].

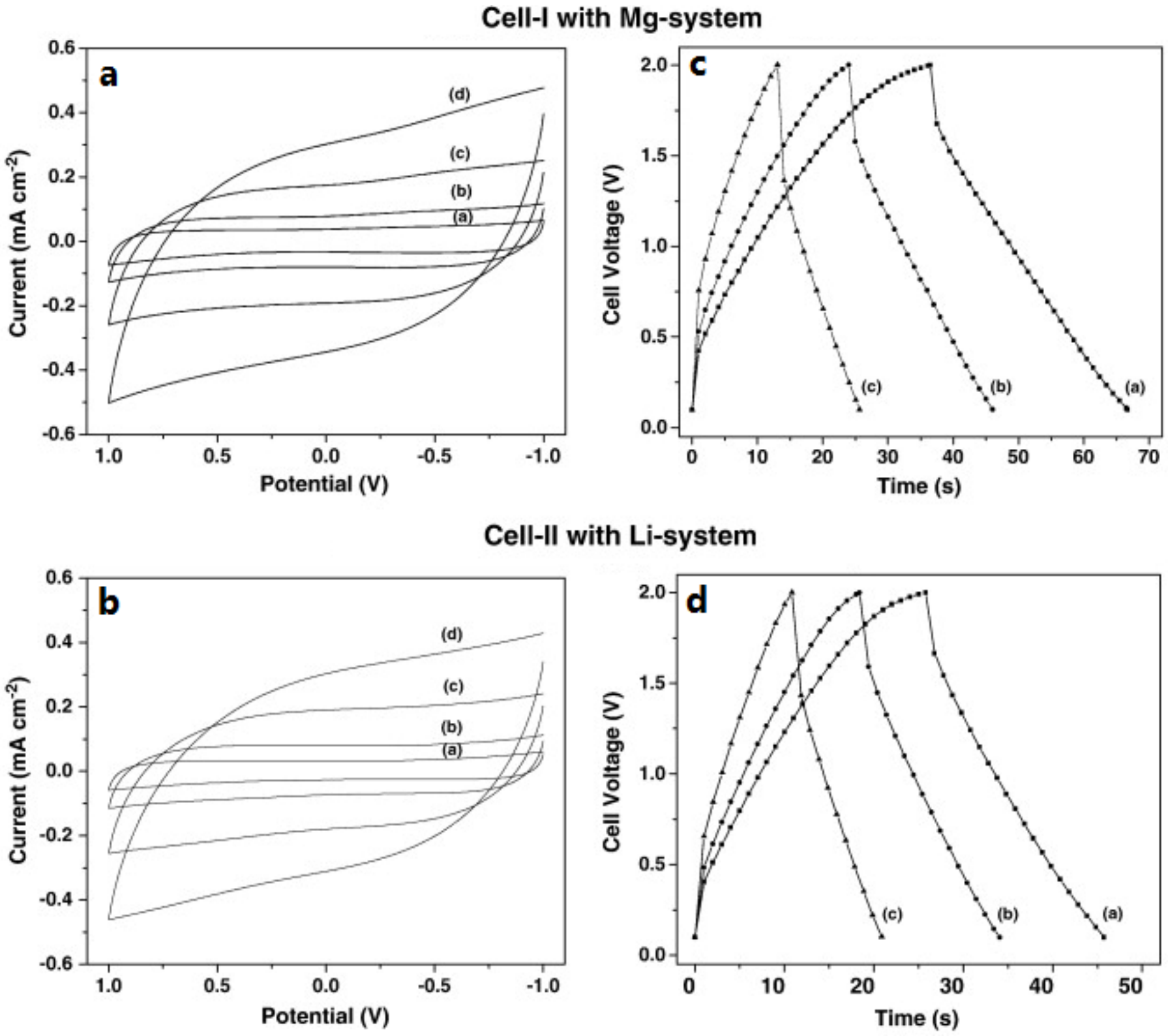

**Figure 20**. Cyclic voltammograms of (a) Cell-I, (b) Cell-II at different scan rates: (a) 10, (b) 20, (c) 50 and (d) 100 mV $s^{-1}$ and Charge–discharge characteristics of (a) Cell-I, (b) Cell-II at different current densities of (a) 150, (b) 200, and (c) 300 μA $cm^{-2}$ **[**Reproduced form Ref. 64, Copyright (2011), with permission from Elsevier**].**

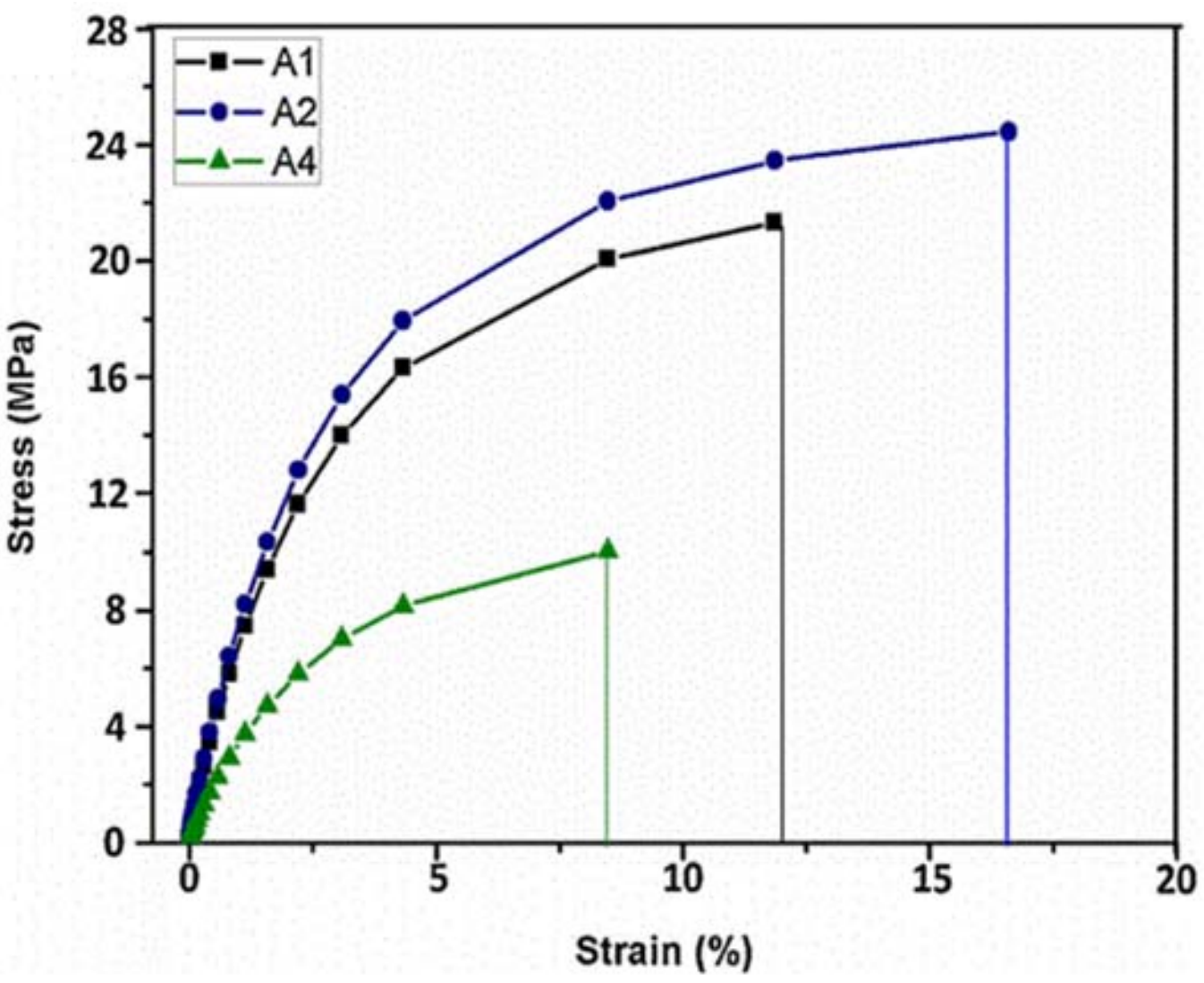

**Figure 21**. Stress strain behavior of electrolyte samples A1 (polymer-salt), A2 (0 wt. % IL) and A4 (10 wt. % IL) [Reproduced form Ref. 65, Copyright (2016), with permission from Elsevier].

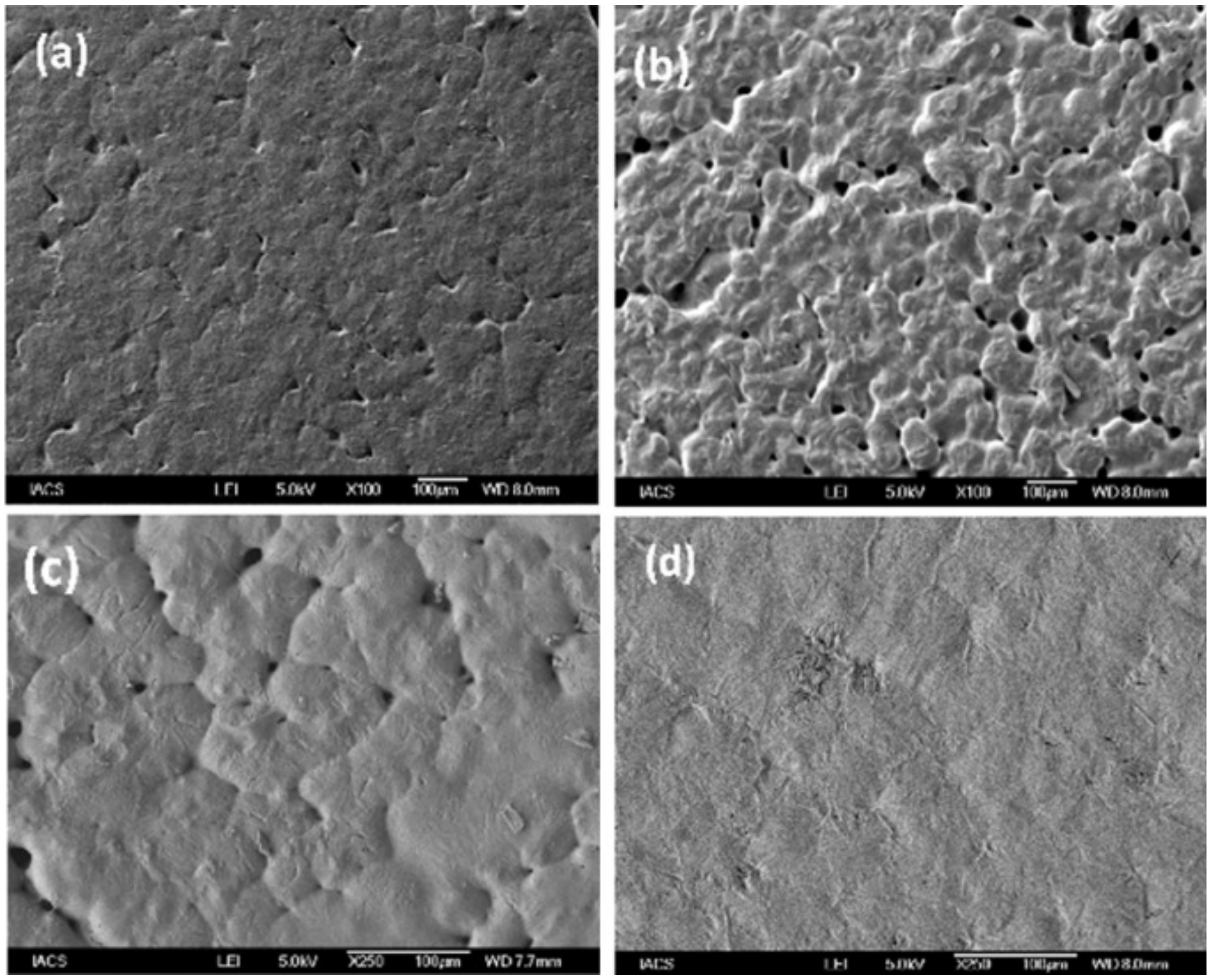

**Figure 22.** FE-SEM images of (a) PEO-$LiClO_4$ (b) PEO- $LiClO_4$ -30 wt.% EC (c) PEO- $LiClO_4$ -30 wt.% DMC (d) PEO- $LiClO_4$ -30 wt.% PEG polymer electrolytes [Reproduced form Ref. 66, Copyright (2016), with permission from Elsevier].

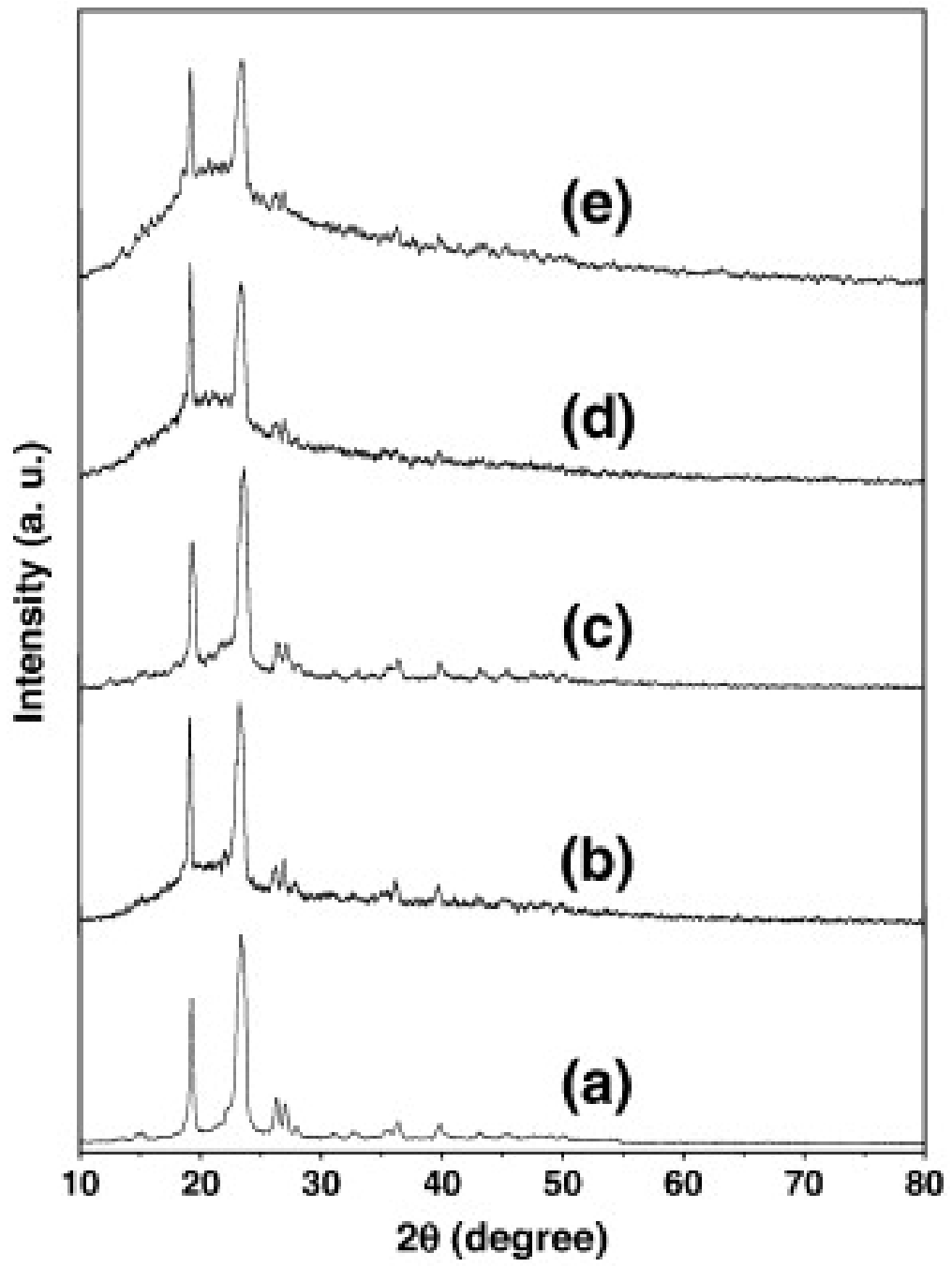

**Figure 23**. XRD pattern of (a) pure PEO, (b) $(PEO)_{25}$.LiTf complex and $(PEO)_{25}$.LiTf + x wt.% EMITf system for (c) x = 5, (d) x = 20, and (e) x = 40 **[**Reproduced from Ref. 67, Copyright (2011), with permission from Elsevier**].**

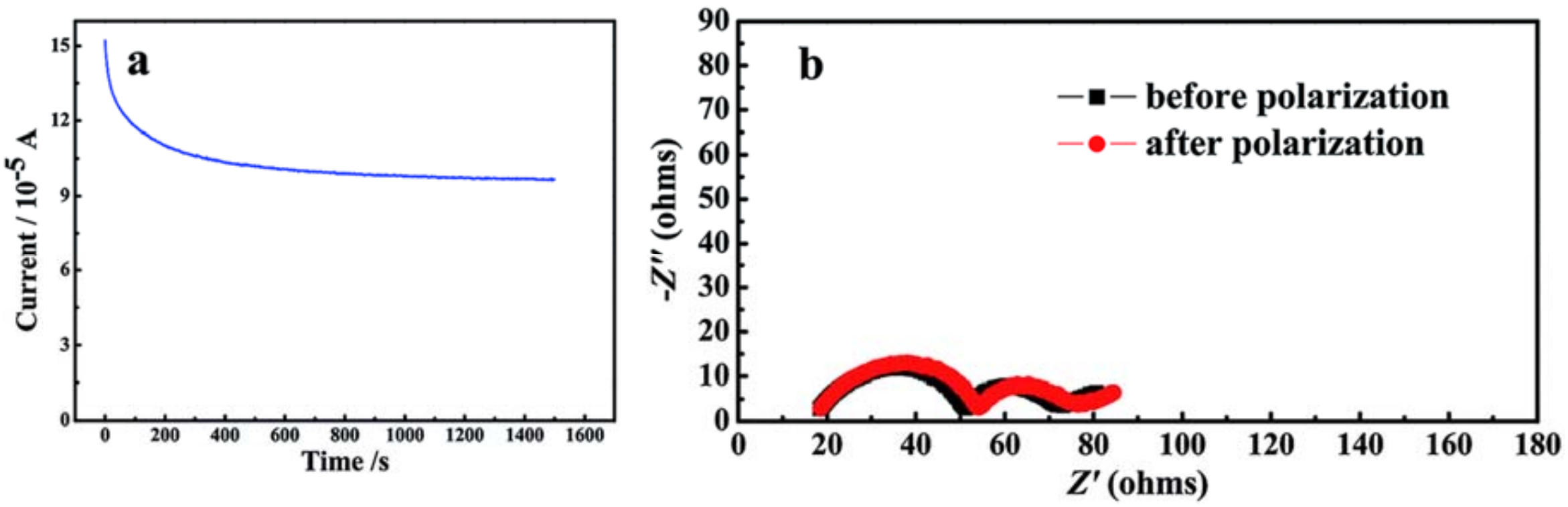

**Figure 24.** (a) Chronoamperometry profiles for the GPE at 25 °C in block cells using Li metal as both electrodes with step potential of 10 mV. (b) Nyquist profiles of the cell electrochemical impedance spectroscopy response before and after polarization [Reproduced from Ref. 68, Copyright (2017), with permission from Royal Society of Chemistry].

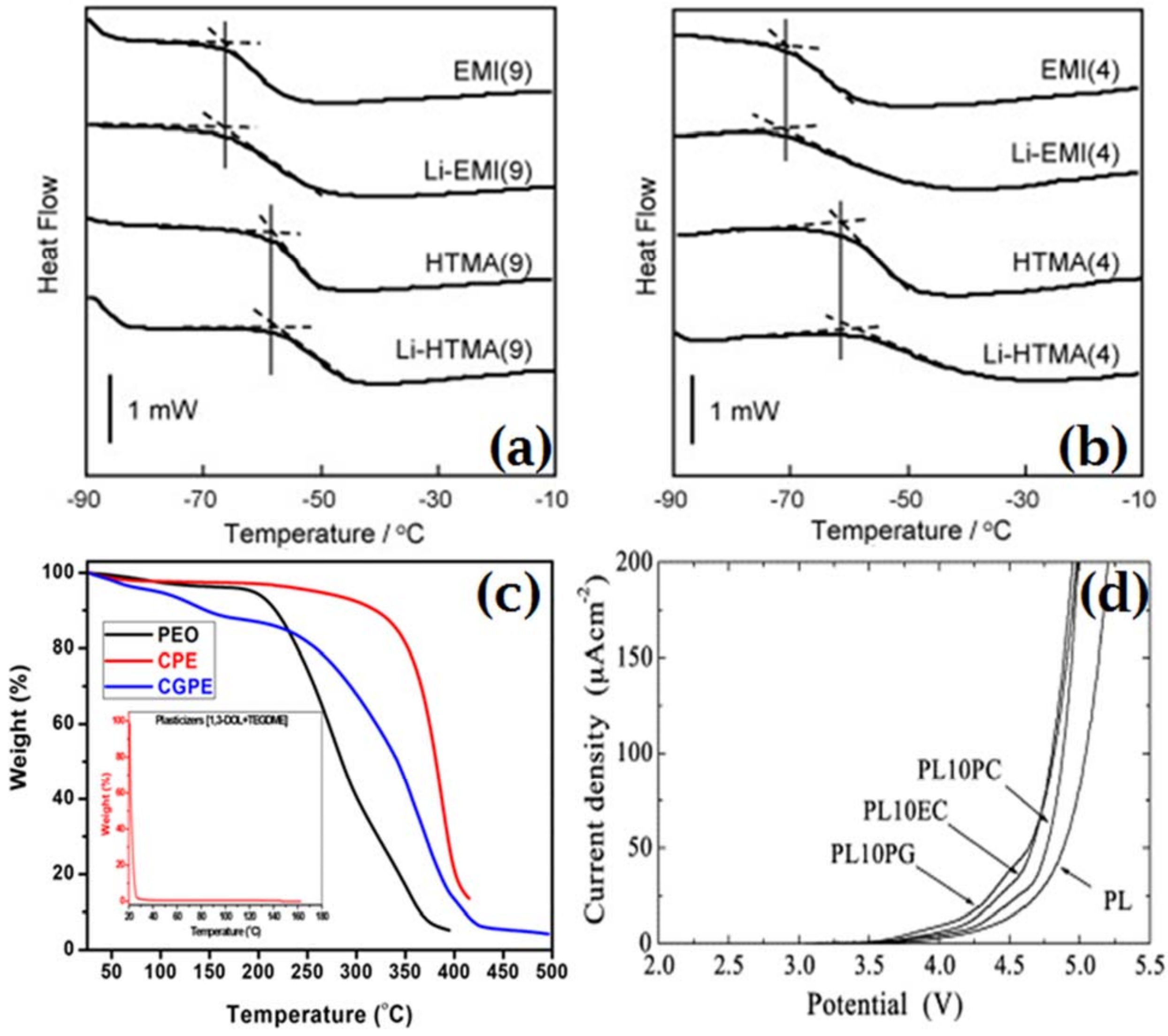

**Figure 25.** DSC profiles of ionic liquid/PEO-PMA gel electrolytes on heating. Heating rate: 10 ◦Cmin−1 **(a)** PEO-PMA matrix with *n* = 9, **(b)** PEO-PMA matrix with *n*=4 [Reproduced from 70, Copyright (2008), with permission from Elsevier], **(c)** Thermogravimetric analysis of PEO, CPE (0 % plasticizer), and CGPE (20 % plasticizer) [Reproduced from Ref. 71, Copyright (2017), with permission from John Wiley and Sons] and **(d)** Onset potentials for oxidative degradation of the Li/SPE/SS cells at 80 $^{o}$C **[**Reproduced from Ref. 73, Copyright (2002), with permission from Elsevier**].**

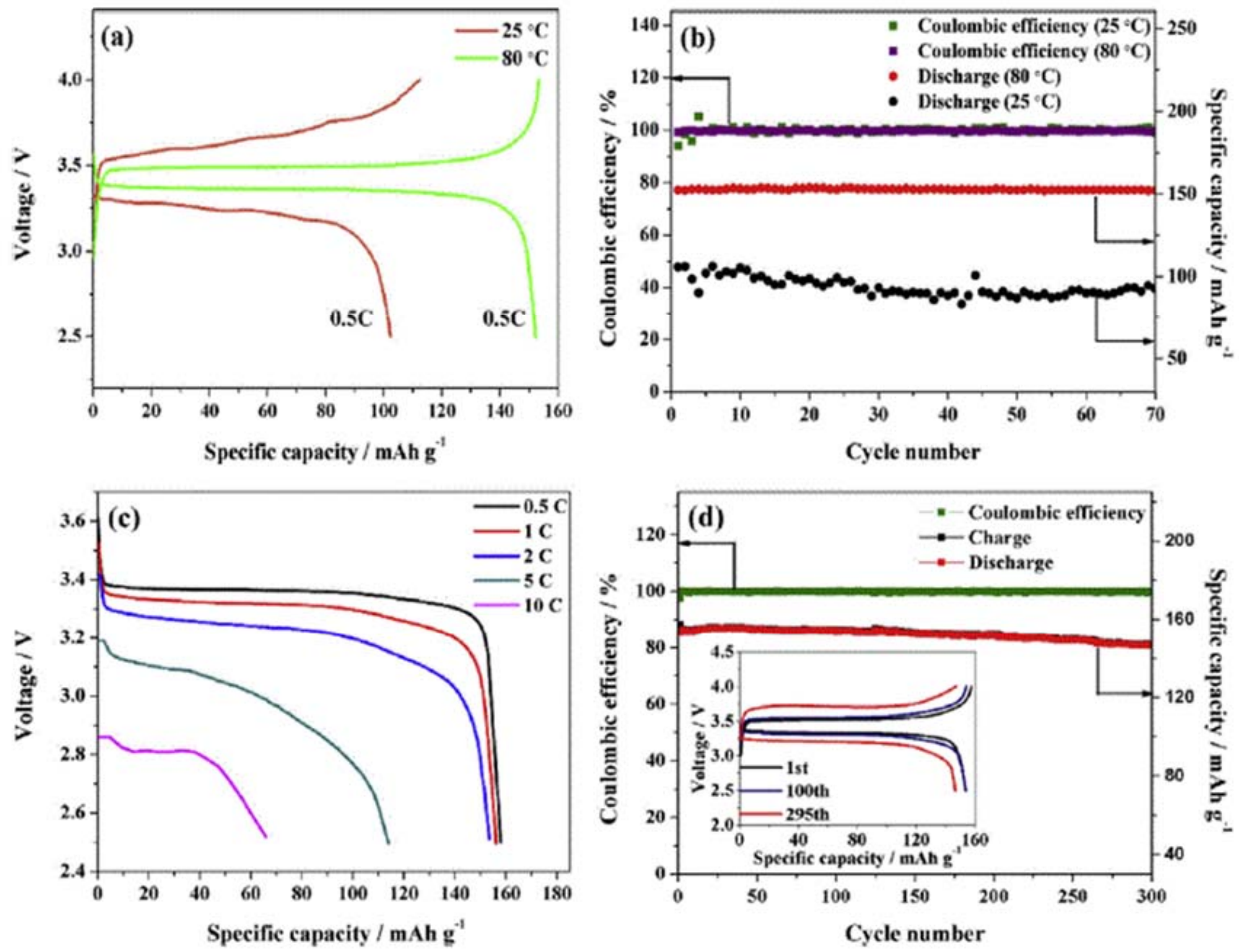

**Figure 26.** (a) Initial charge-discharge profiles and (b) Cycling performance of $LiFePO_4$/Li cells using $TiO_2$-grafted NHPE at 25 °C and 80 °C at 0.5 C. (c) Discharge curves for $LiFePO_4$/$TiO_2$-grafted NHPE/Li cell at various C-rates and 80 °C. (d) Cycling stability of $LiFePO_4$/$TiO_2$-grafted NHPE/Li battery at 1 C and 80 °C. Inset: charge/discharge curves at different cycles [Reproduced from Ref. 75, Copyright (2016), with permission from Elsevier].

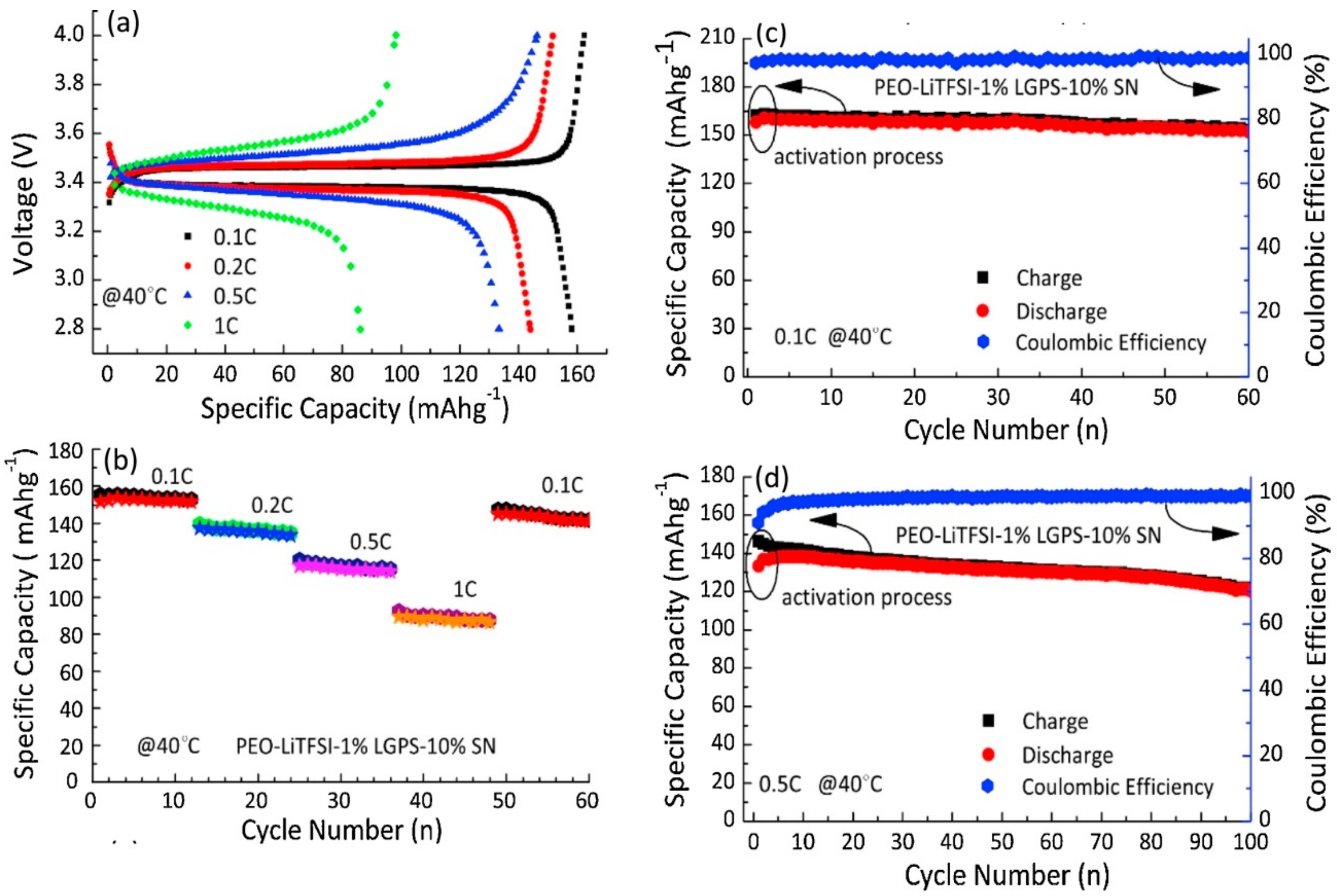

**Figure 27.** The cycling and rate performance under 40 °C for all-solid-battery Li/$PEO_{18}$-LiTFSI-1%LGPS-10%SN/$LiFePO_4$; (a) the initial charge and discharge curves under different rates (0.1 C, 0.2 C, 0.5 C, 1 C); (b) the rate cycling performance of the all-solid-battery cell; (c) cycling performance at 0.1C; (d) cycling performance at 0.5C **[**Reproduced from Ref. 78, Copyright (2016), with permission from Elsevier**].**

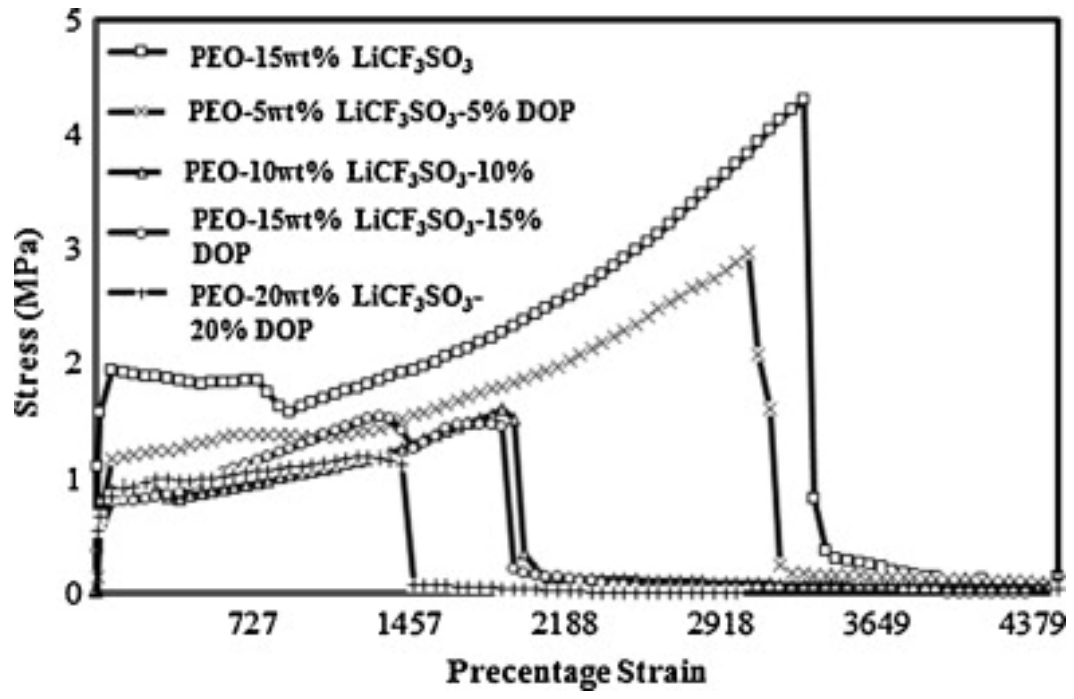

**Figure 28.** Tensile stress–strain behavior of PEO-$_{15}$ wt% ($LiCF_3SO_3$)-z (DOP) solid polymer electrolytes [Reproduced from Ref. 79, Copyright (2015), with permission from Elsevier].

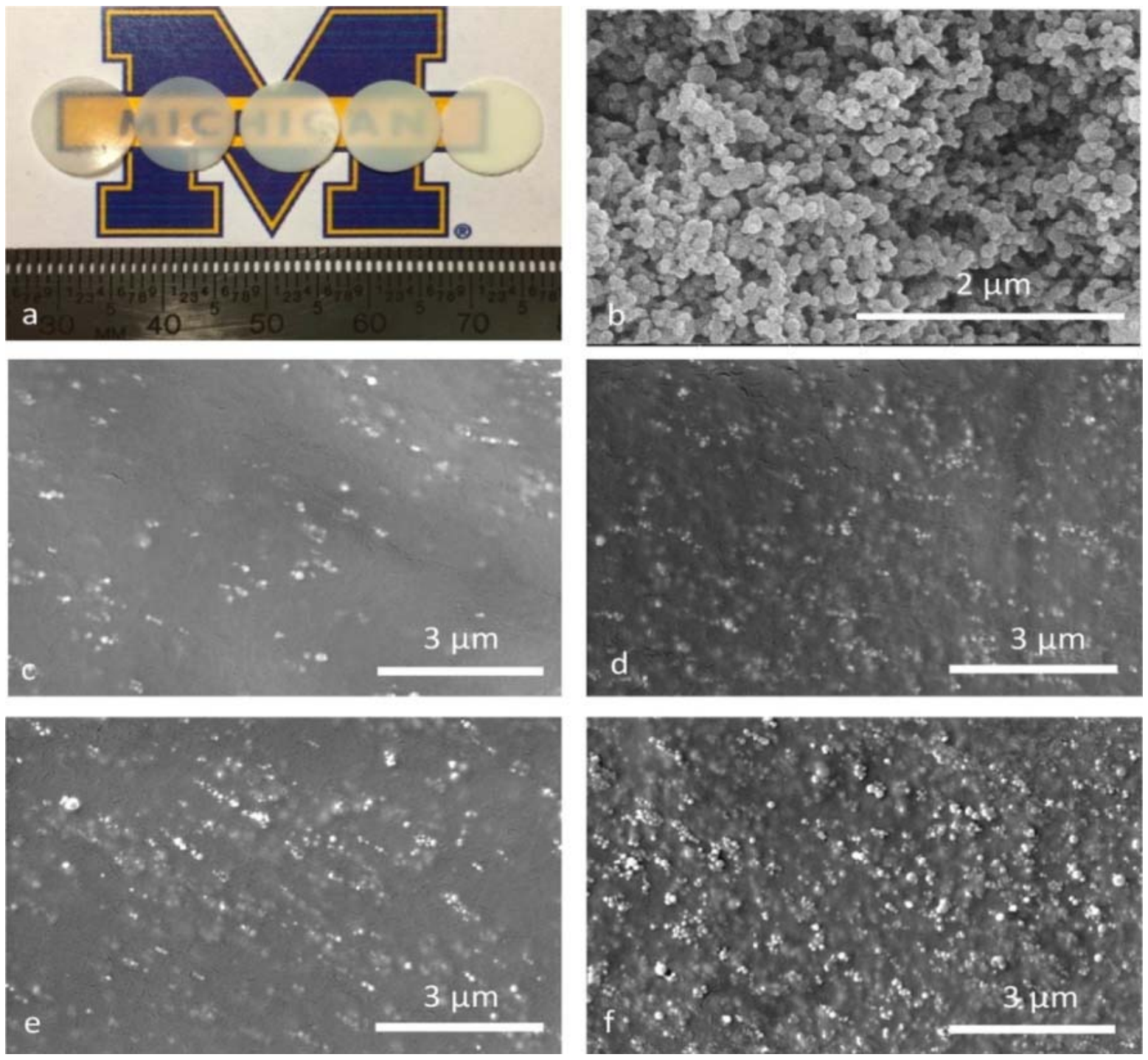

**Figure 29.** (a) Photograph of composite films; (b) SEM image of LATP particles, as collected from flame spray pyrolysis; SEM images of nanocomposite samples with (c) 5, (d) 10, (e) 15, and (f) 20 wt. % LATP nanoparticles [Reprinted with permission from Ref. 85, Copyright (2017) American Chemical Society].

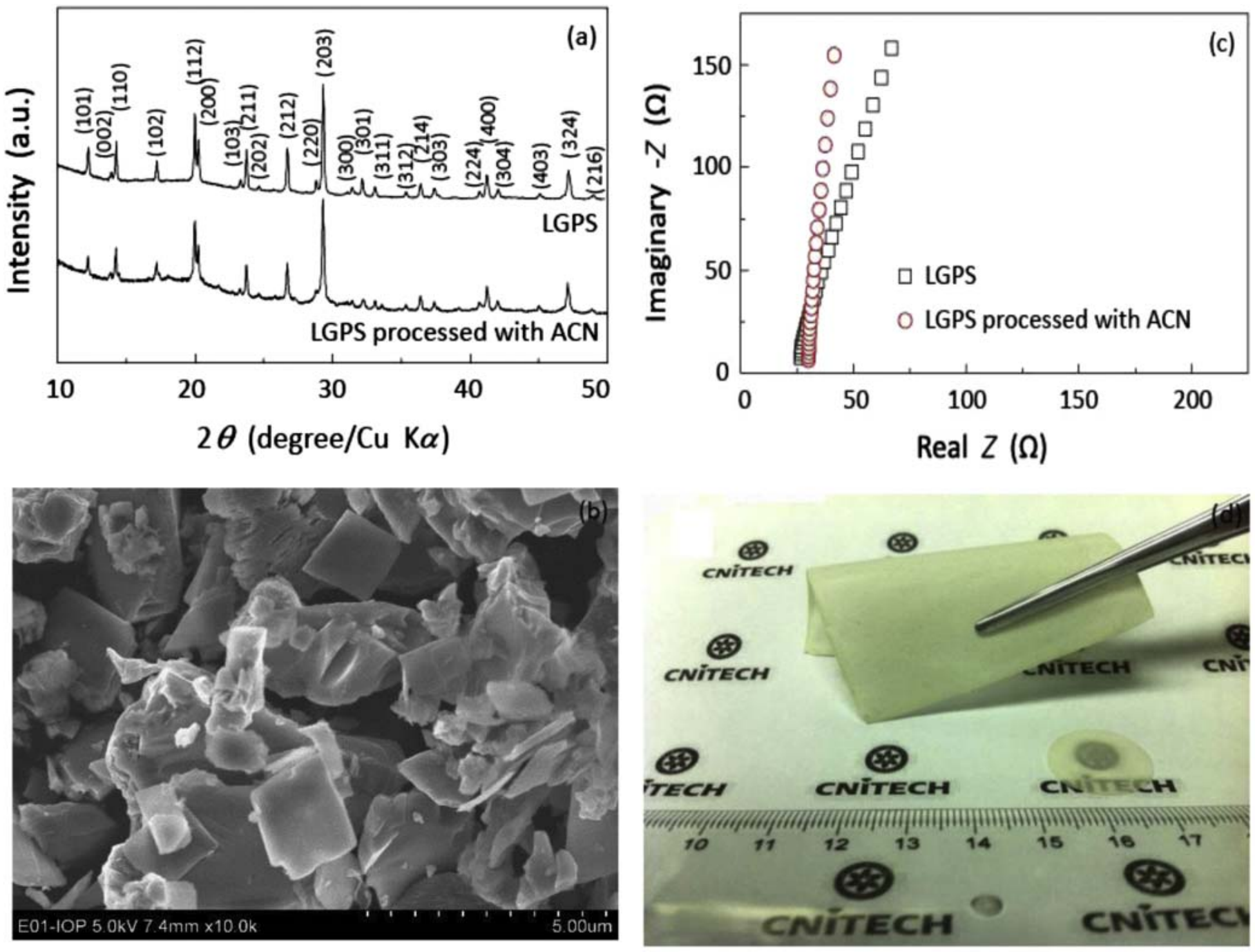

**Figure 30.** (a) XRD patterns of LGPS before and after processed with ACN; (b) FESEM image of the LGPS particles; (c) AC impedance spectra LGPS before and after processed with ACN and (d) a photo of SPE membrane [Reproduced from Ref. 86, Copyright (2016), with permission from Elsevier].

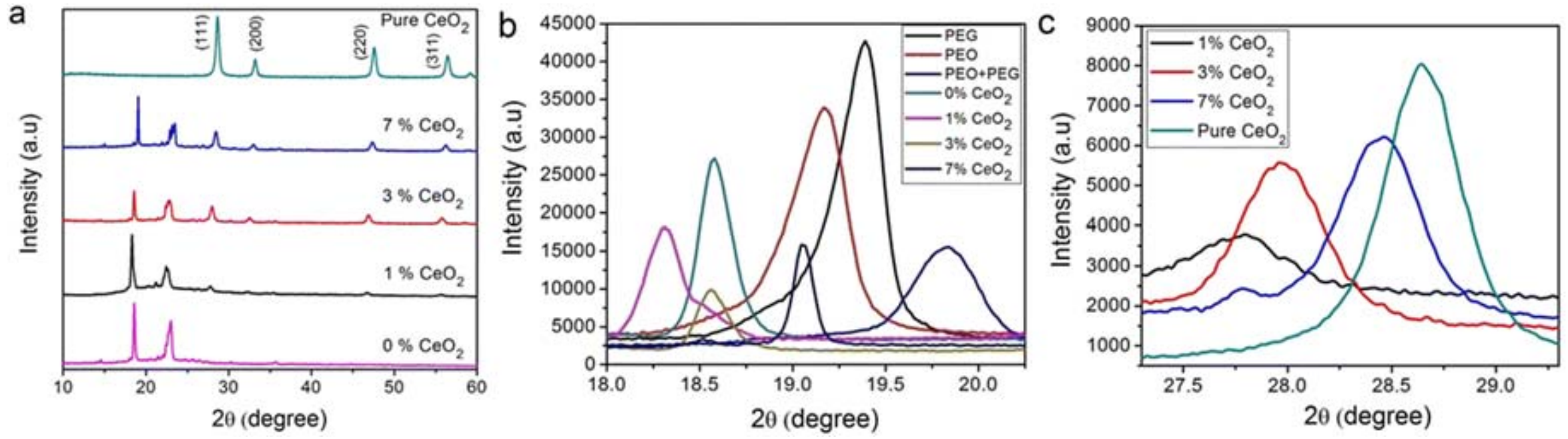

**Figure 31**. XRD patterns of a pure $CeO_2$ and PEO–PEG–$LiClO_4$–EC (68:16:11:5) with different weight percentage of $CeO_2$. b PEO/PEG characteristics peaks in pure PEO /PEG, blend polymer, and NCPEs. C Maximum intensity peak of $CeO_2$ in NCPE and its pure form [Reproduced from Ref. 88, Copyright (2015), with permission from Springer].

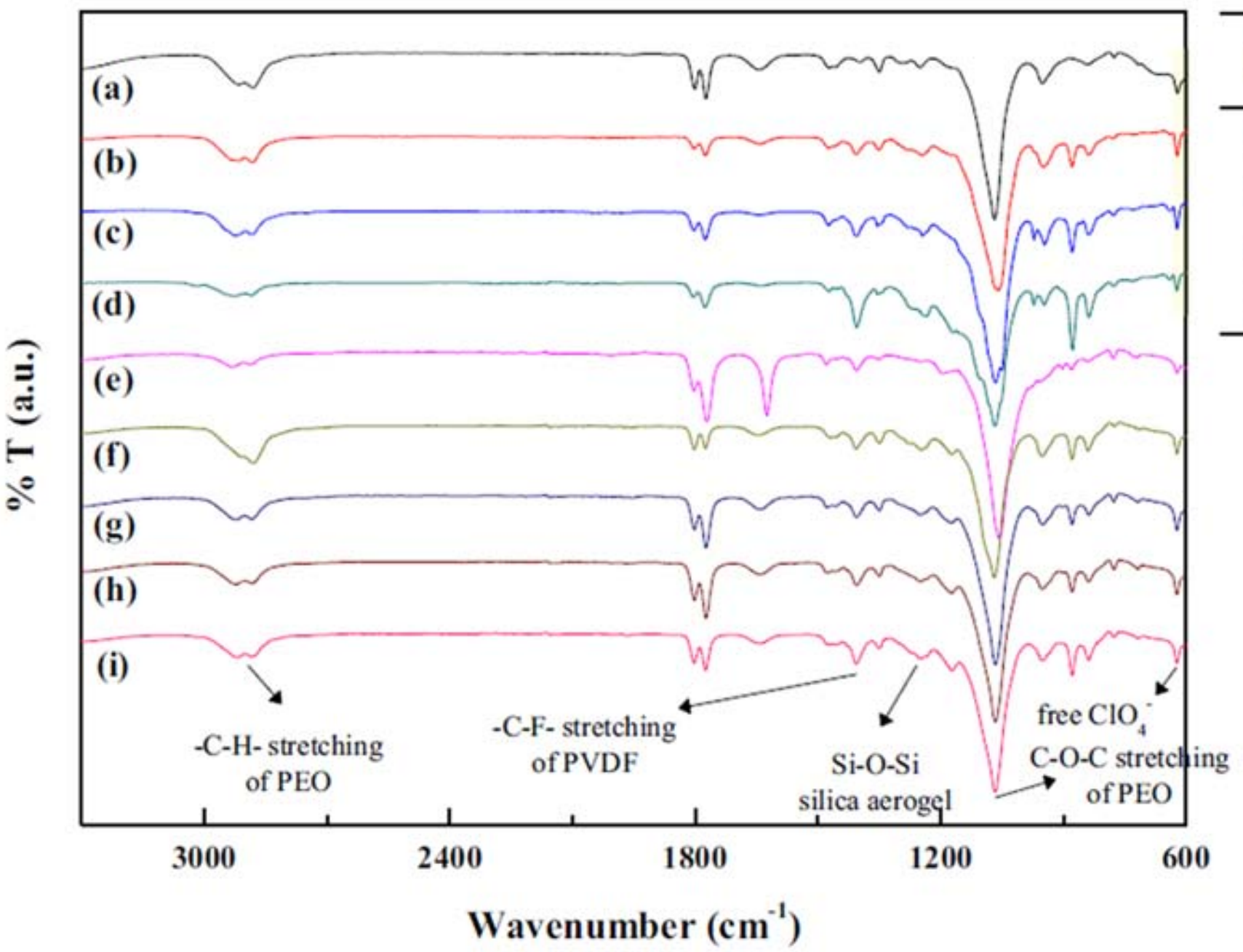

| Sample | Peak area (630-610cm$^{-1}$) |
|---|---|
| SPE-1 | 0.795 |
| SPE-2 | 0.818 |
| SPE-3 | 0.753 |
| SPE-4 | 0.592 |

**Figure 32**. FT-IR spectra of the solid polymer electrolytes (a) SPE-1, (b) SPE-2, (c) SPE-3, (d) SPE-4, (e) SPE-5, (f) SPE-6, (g) SPE-7, (h) SPE-8, (i) SPE-9 [Reproduced from Ref. 89, Copyright (2013), with permission from Elsevier].

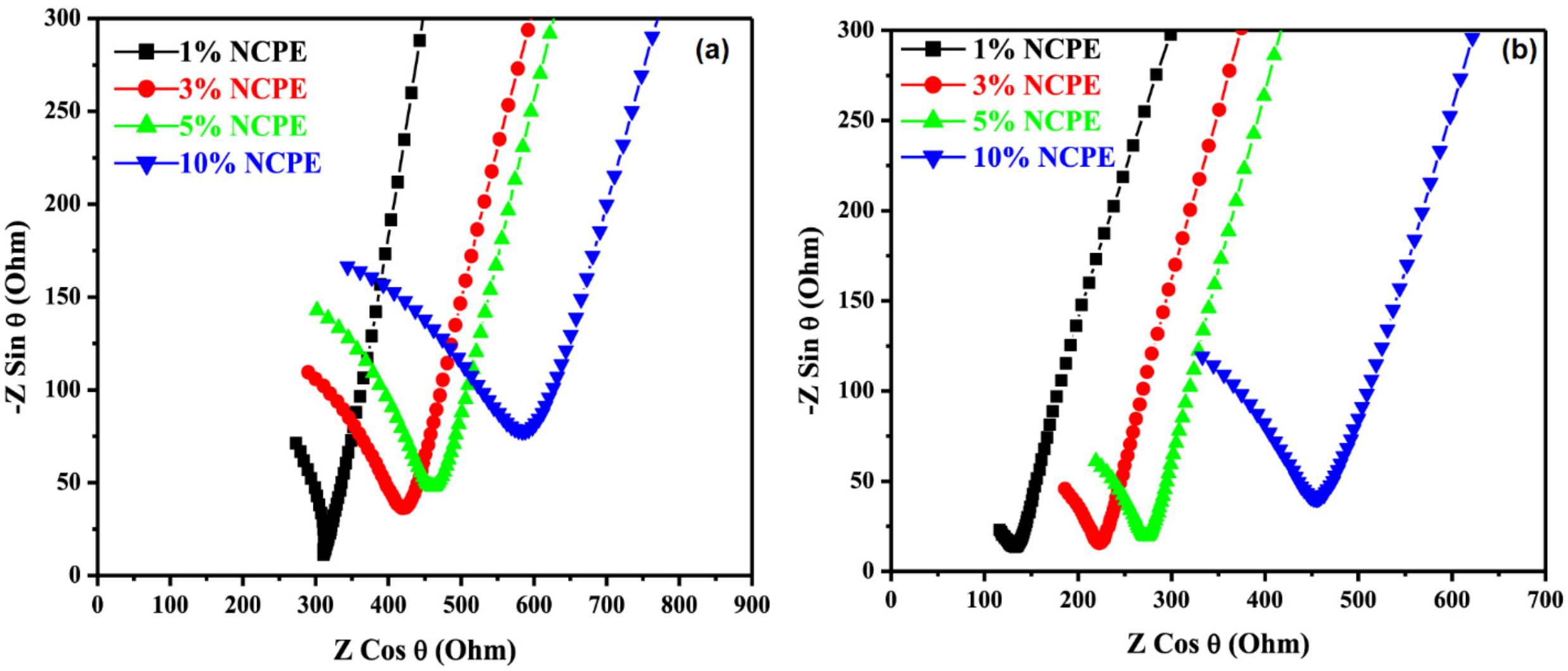

**Figure 33**. (a) Nyquist plot of the data obtained from AC impedance measurement of NCPE films prepared using chem-$SiO_2$ nanofillers. (b) Nyquist plot of the data obtained from AC impedance measurement of NCPE films prepared using epoxy-$SiO_2$ nanofillers. The plot gives the bulk resistance from which the Li+ ion conductivity is calculated. [Reproduced from Ref. 91, Copyright (2016), with permission from IOP Publishing]

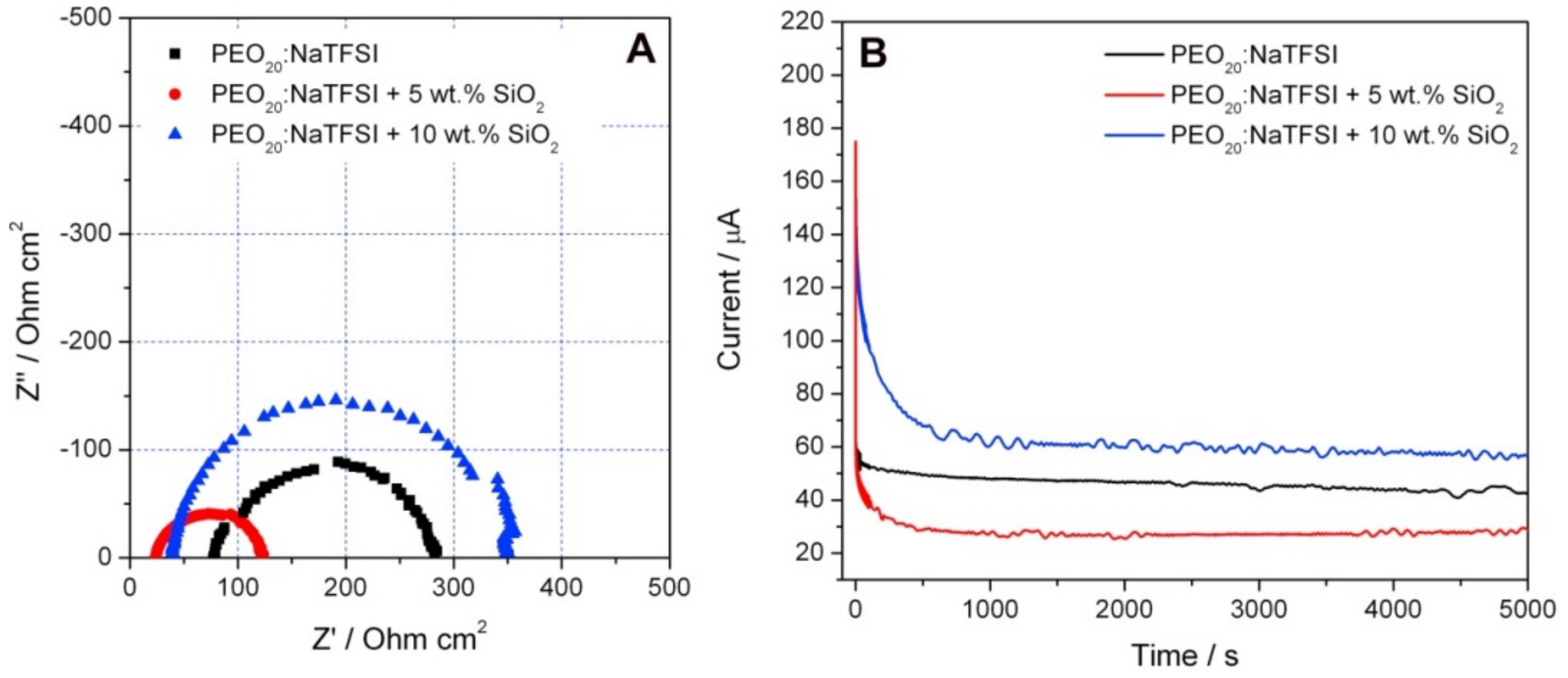

**Figure 34**. Nyquist plots of the $PEO_{20}$:NaTFSI polymer membranes with 0/5/10 wt.% of silica added at 75 C before the dc polarization (a) and their corresponding chrono-amperometric curves (b) [Reproduced from Ref. 93, Copyright (2014), with permission from Elsevier].

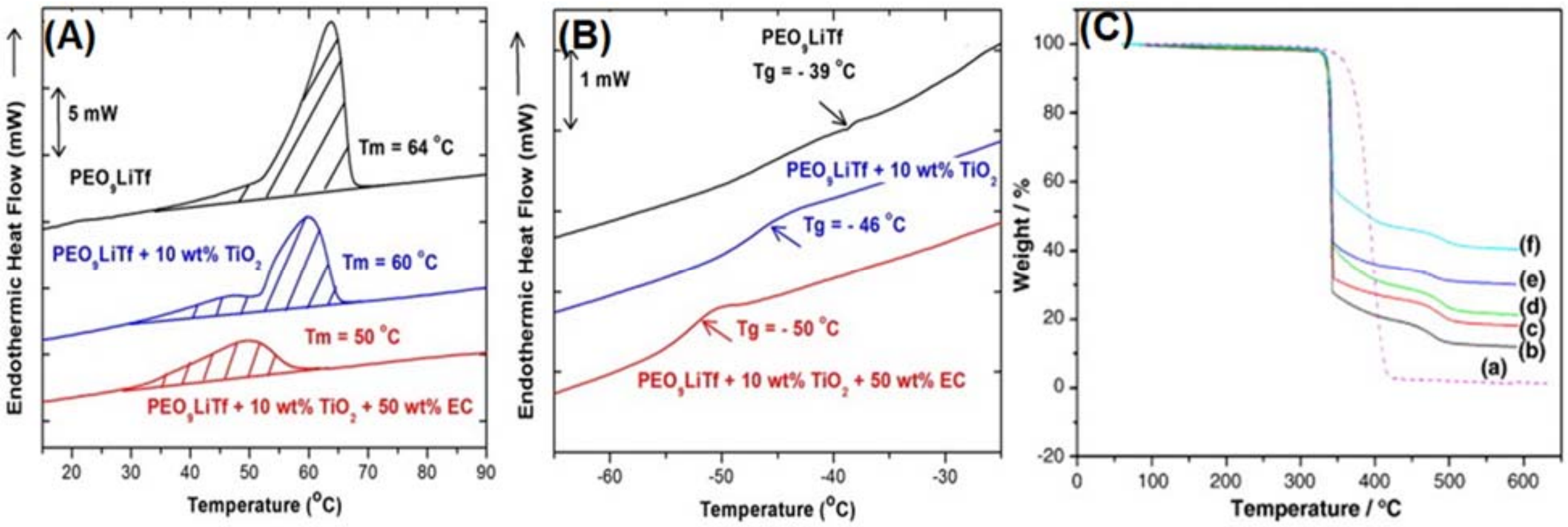

**Figure 35**. DSC endotherms took during the heating run for filler-free, 10 wt.% $TiO_2$ added and 10 wt.% $TiO_2$ + 50 wt.% EC added solid polymer electrolytes showing the variation of; **(A)** Crystallite melting temperature ($T_m$). **(B)** Glass transition temperature ($T_g$) [Reproduced from Ref. 95, Copyright (2014), with permission from Elsevier] and **(C)** TGA graphs of the CPEs containing (a) pure PEO, (b) no filler, (c) 5 wt.%, (d) 10 wt.%, (e) 15 wt.%, and (f) 20 wt.% of nano-sized $BaTiO_3$ [Reproduced from Ref. 96, Copyright (2013), with permission from Elsevier].

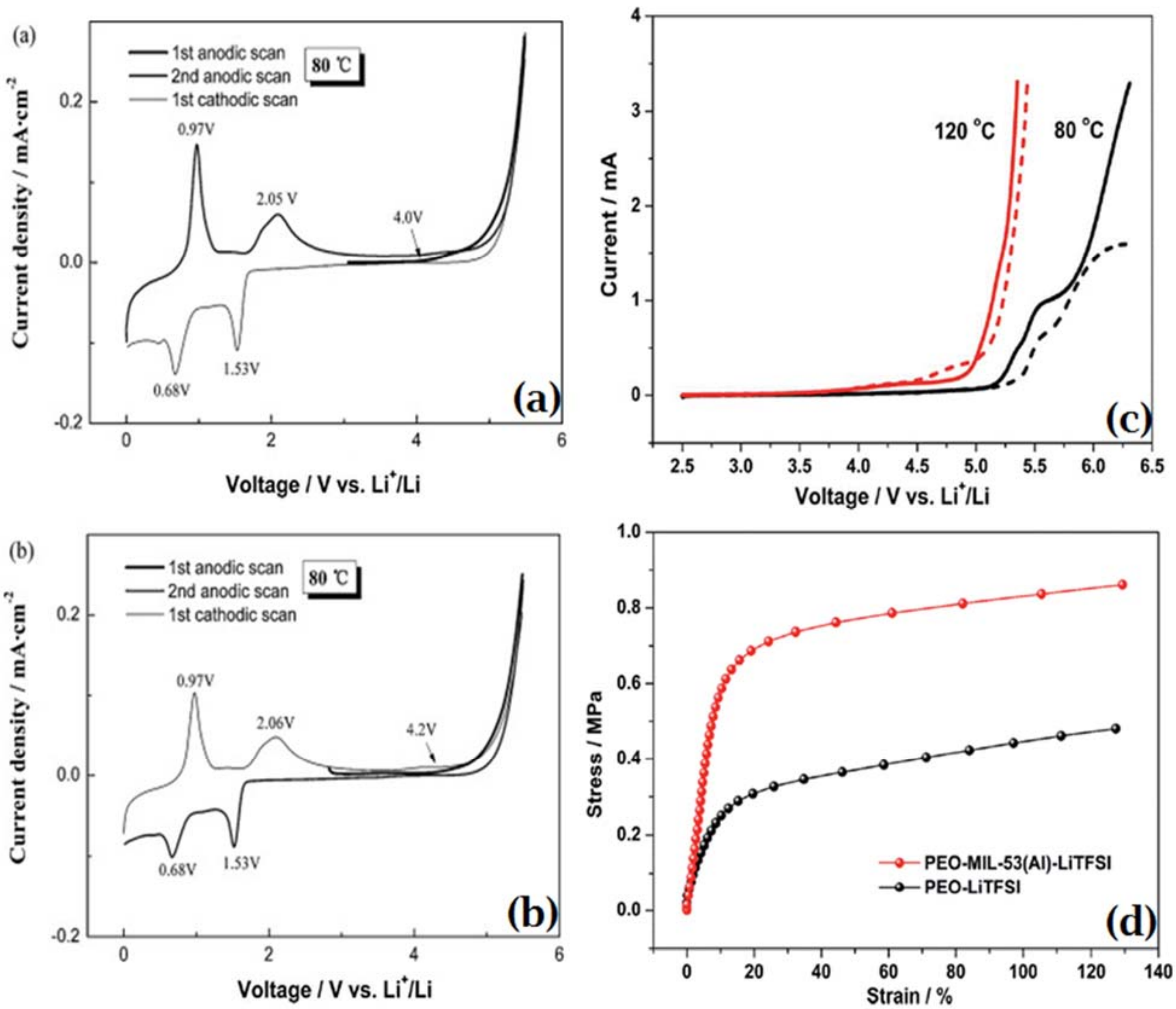

**Figure 36.** CV curves of $PEO_{20}$–LiBOB **(a)** and $PEO_{20}$–LiBOB–5 wt. % MgO **(b)** at 80 ◦C. The first anodic scan, first cathodic scan and second anodic scan are measured in the voltage range of 0–5.5V **[**Reproduced from Ref. 97, Copyright (2011), with permission from Elsevier**], (c)** Linear sweep voltammograms of SS/PEO-LiTFSI/Li (solid line) and SS/PEO-MIL-53(Al)-LiTFSI/Li (dotted line) batteries at 80 $^{o}$C (black) and 120 $^{o}$C (red). The electrolytes were swept in the potential range from 2.5 V to 6.5 V (vs. $Li/Li^{+}$) at a rate of 10 mV $s^{-}1$, and **(d)** Stress–strain curves of the PEO-MIL-53(Al)-LiTFSI and PEOLiTFSI electrolytes [Reproduced from Ref. 98, Copyright (2014), with permission from Royal Society of Chemistry].

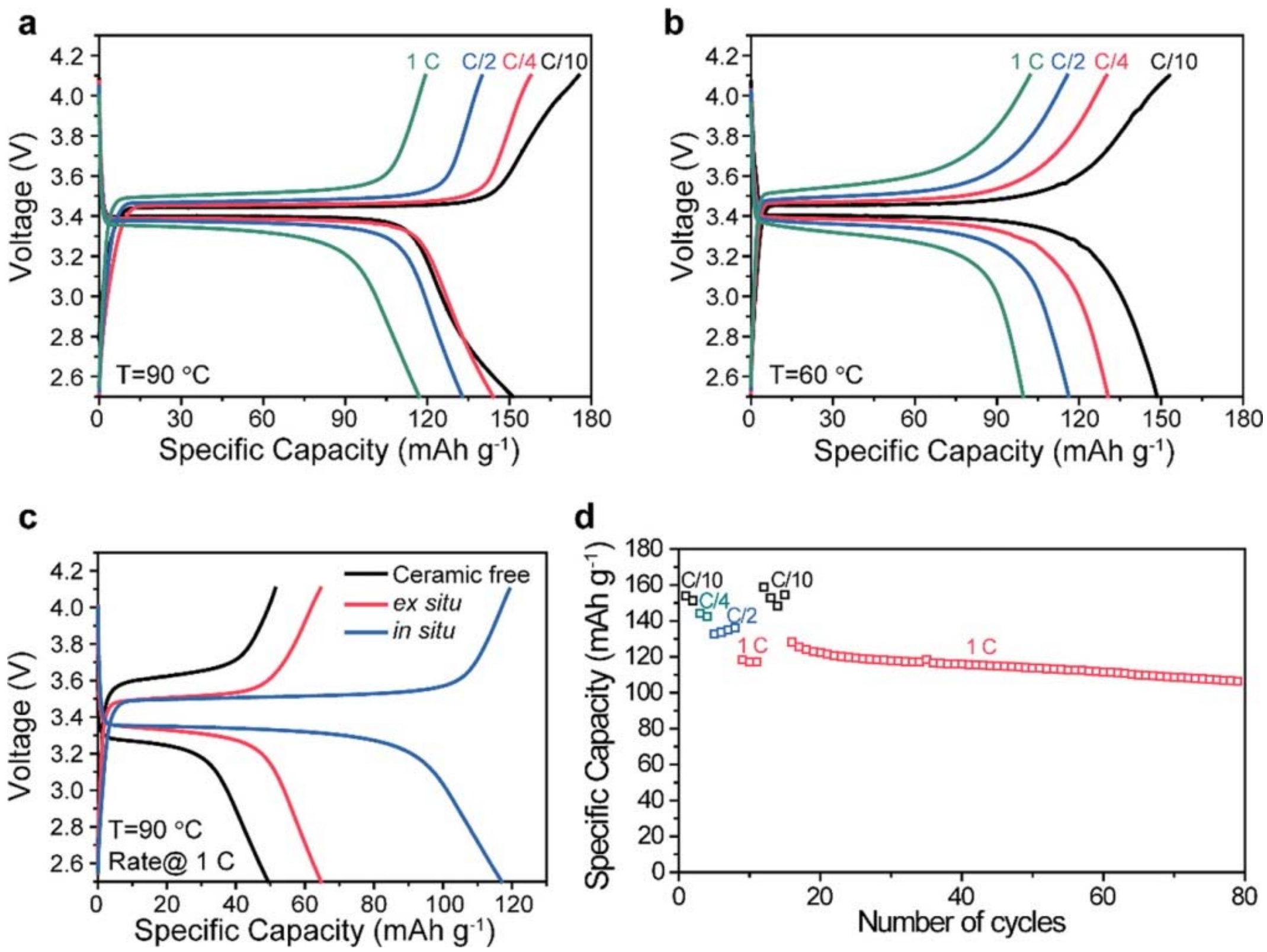

**Figure 37**. Electrochemical performance of all-solid-state lithium polymer batteries. (a, b) Rate capability of lithium polymer batteries with configuration of $LiFePO_4$ cathode/in situ CPE/lithium foil anode at 90 °C (a) and 60 °C (b). (c) Comparison on capacity retention of batteries with different electrolytes of ceramic-free SPE, ex situ CPE, and in situ CPE operating at 90 °C and the rate of 1 C. (d) Cycle performance of batteries with in situ CPE at 90 °C. [Reprinted with permission from (Ref. 99). Copyright (2015) American Chemical Society]

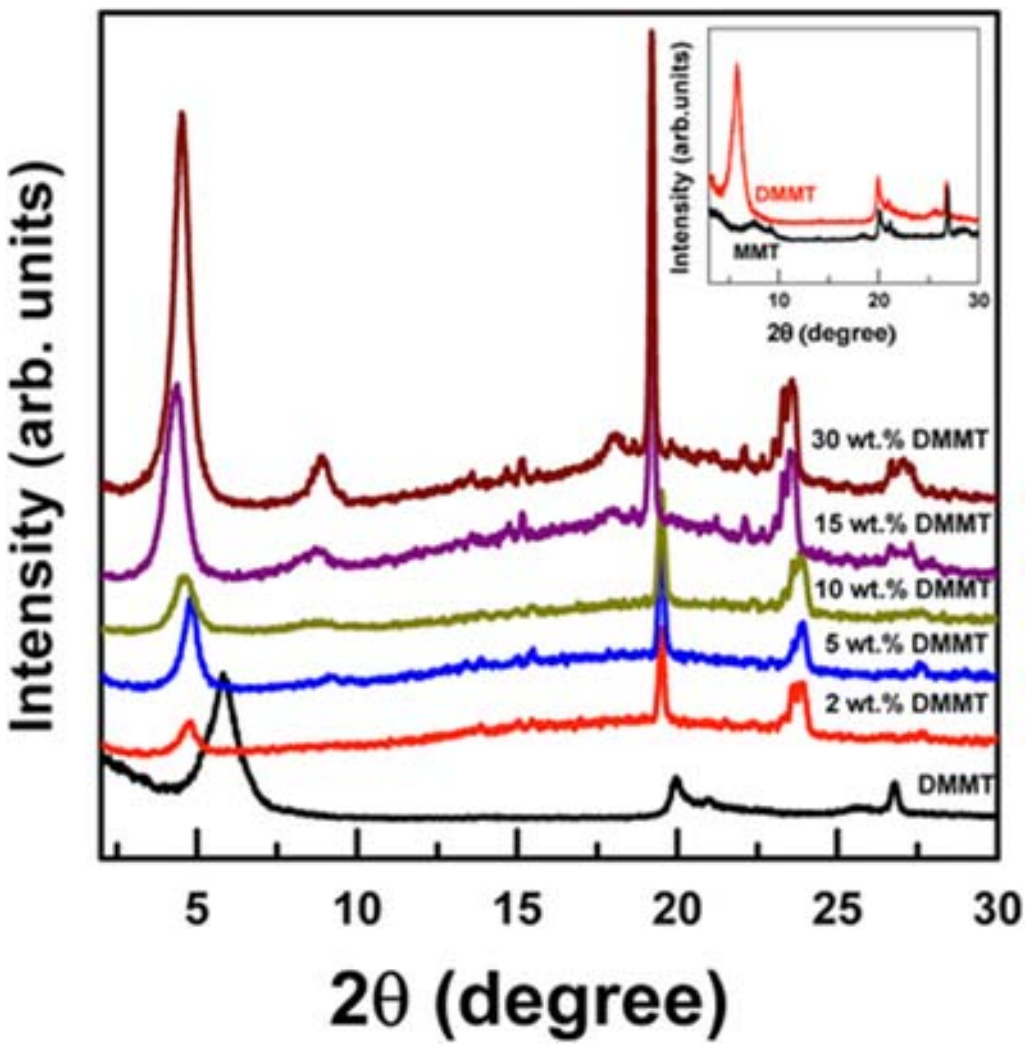

**Figure 38**. XRD patterns of DMMT and $PEO_{20}$-$LiAsF_6$+ x wt. % DMMT $x = 0$, 2, 5, 10, 15, and 30. *Inset* shows comparison of XRD patterns of MMT and DMMT [Reproduced from Ref. 108, Copyright (2015), with permission from Springer].

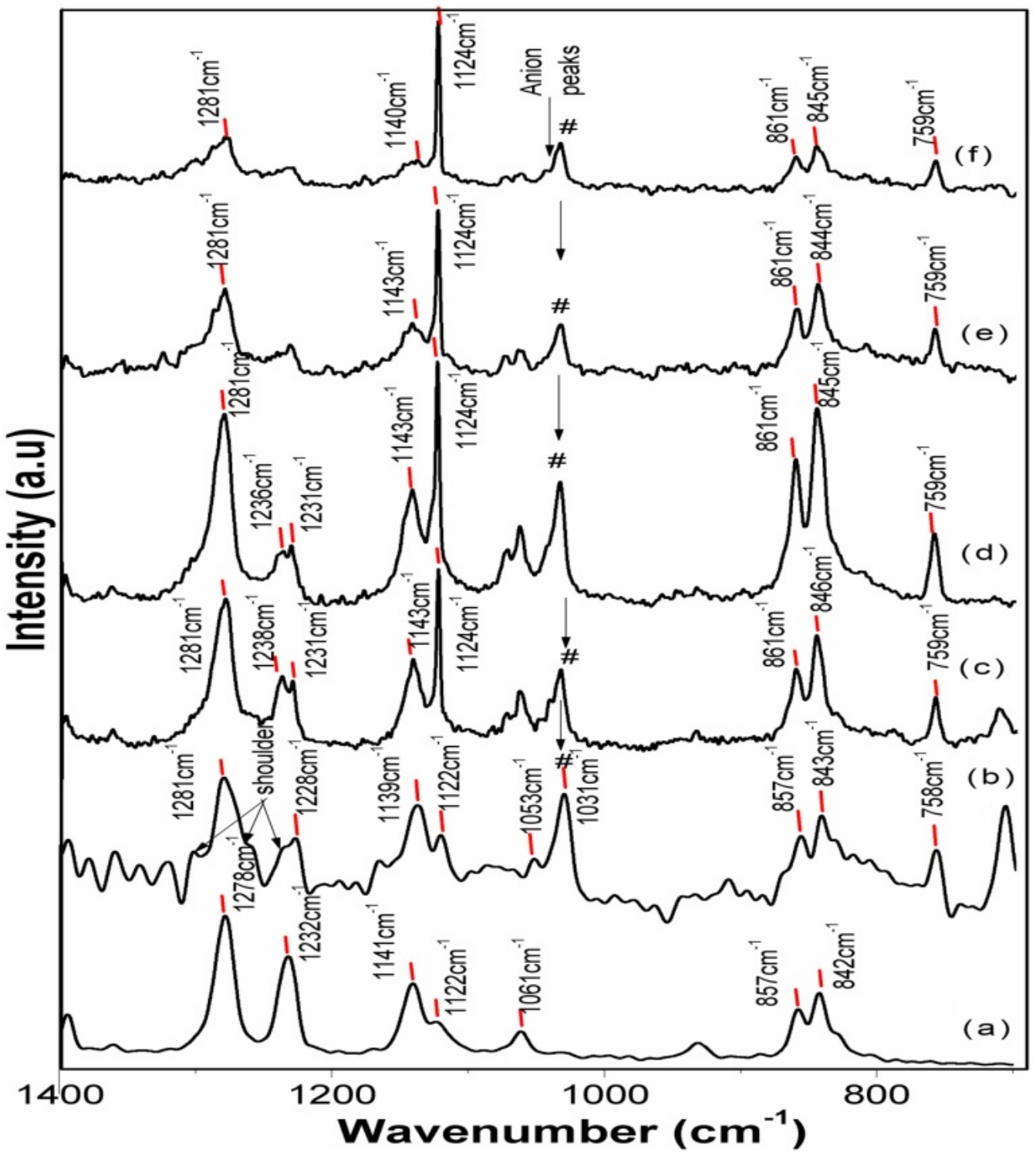

**Figure 39**. Raman spectrum of clay based PNC (a) PEO, (b) Polymer blend salt, (c) 1 wt. % clay, (d) 3 wt% clay, (e) 5 wt. % clay, and (f) 7 wt. % clay. Arrow indicates significant changes in the Raman profile [Reproduced from Ref. 109, Copyright (2017), with permission from Elsevier].

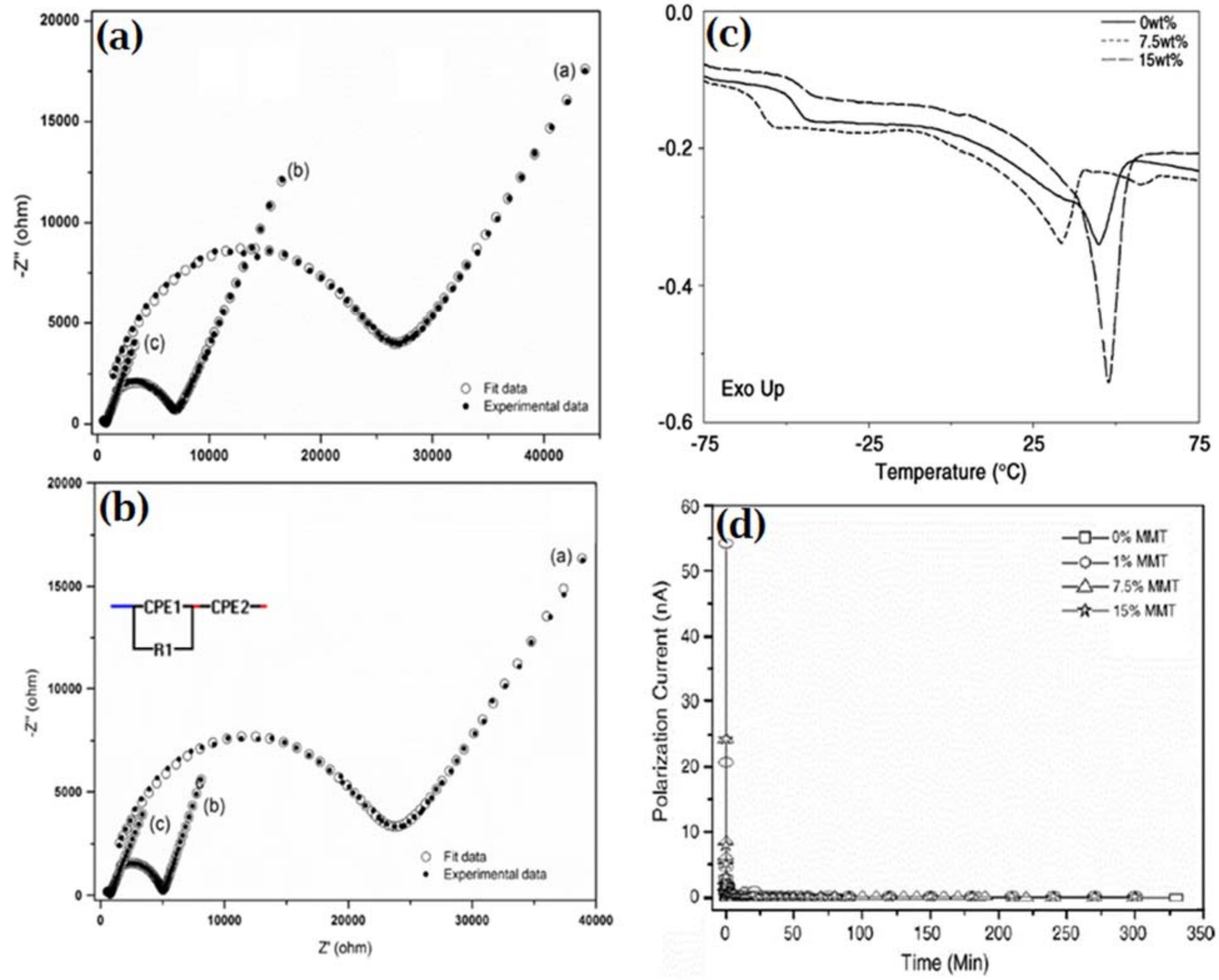

**Figure 40. (A)** Impedance spectra of nanocomposite electrolyte films (a) $(PEO)_{16}LiCBSM$, (b) $(PEO)_8LiCBSM$ and (c) $(PEO)_4LiCBSM$ at 298 K and **(B)** Impedance spectra of nanocomposite electrolyte films (a) $(PEO)_{16}LiCBSM$, (b) $(PEO)_8LiCBSM$ and (c) $(PEO)_4LiCBSM$ at 323 K [Reproduced from Ref. 110, Copyright (2016), with permission from Springer], **(C)** DSC thermograms of PEO/clay/$Li^+$ nanocomposite electrolytes **[**Reproduced from Ref. 112, Copyright (2007), with permission from John Wiley and Sons**]** and **(D)** Variation of polarization current as function of time under constant applied voltage (V = 50mV) for different concentration (x=0, 1, 7.5, 15) of DMMT clay [Reproduced from Ref. 111, Copyright (2009), with permission from Elsevier].

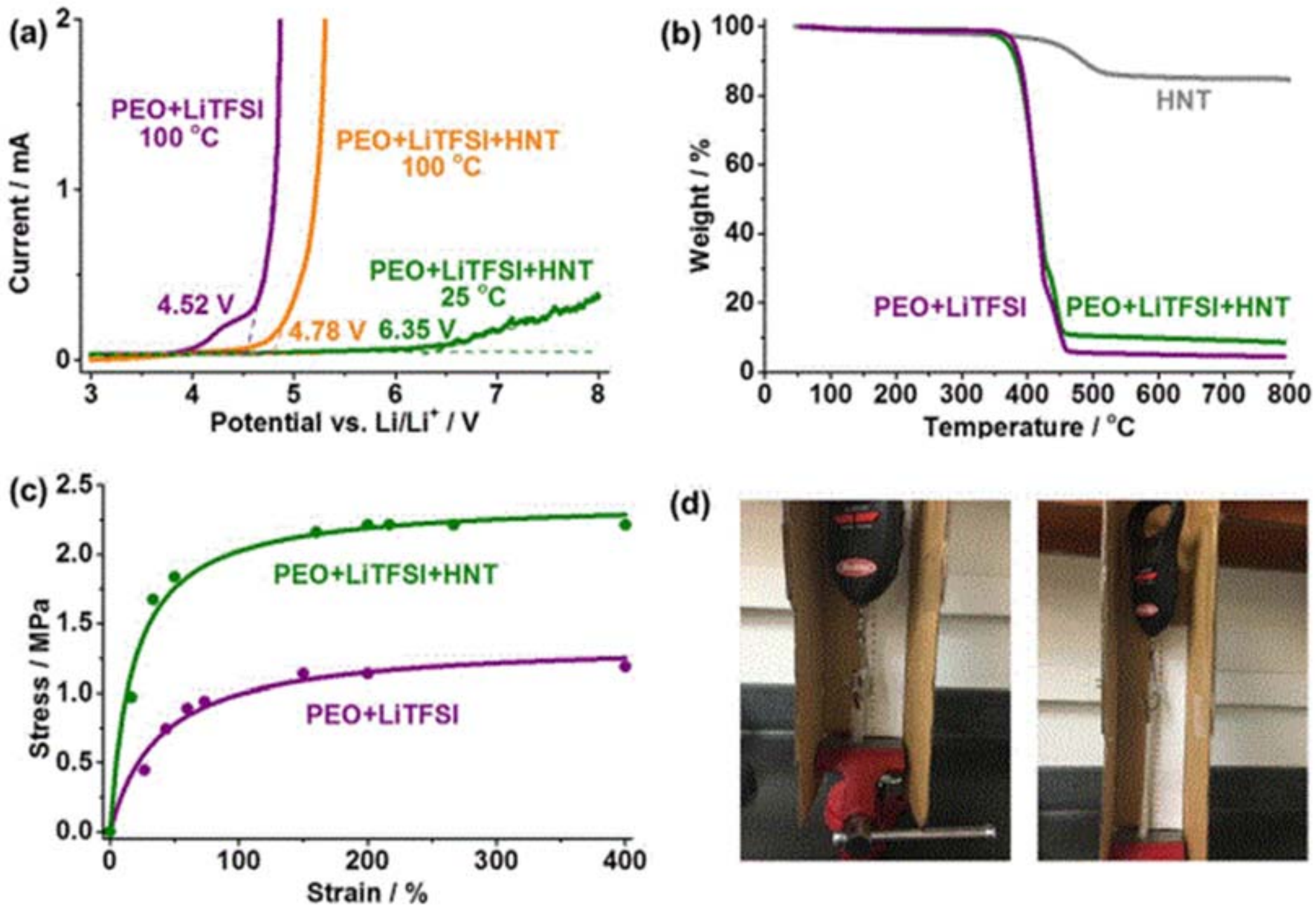

**Figure 41**. Electrochemical, thermal and mechanical stability of the HNT nanocomposite electrolyte. (a) Linear sweep voltammetry Li/PEO+LiTFSI+HNT/SS cells at 25 °C and 100 °C, and Li/PEO+LiTFSI/SS cells at 100 °C at a rate of 10 mV $s^{-1}$ . (b) Thermogravimetric analysis of PEO+LiTFSI+HNT, PEO+LiTFSI, and HNT. (c) Stress-strain curves of the PEO+LiTFSI +HNT and PEO+LiTFSI electrolytes. (d) Photos of the PEO+LiTFSI+HNT film at initial and final tension state [Reproduced from Ref. 113, Copyright (2017), with permission from Elsevier].

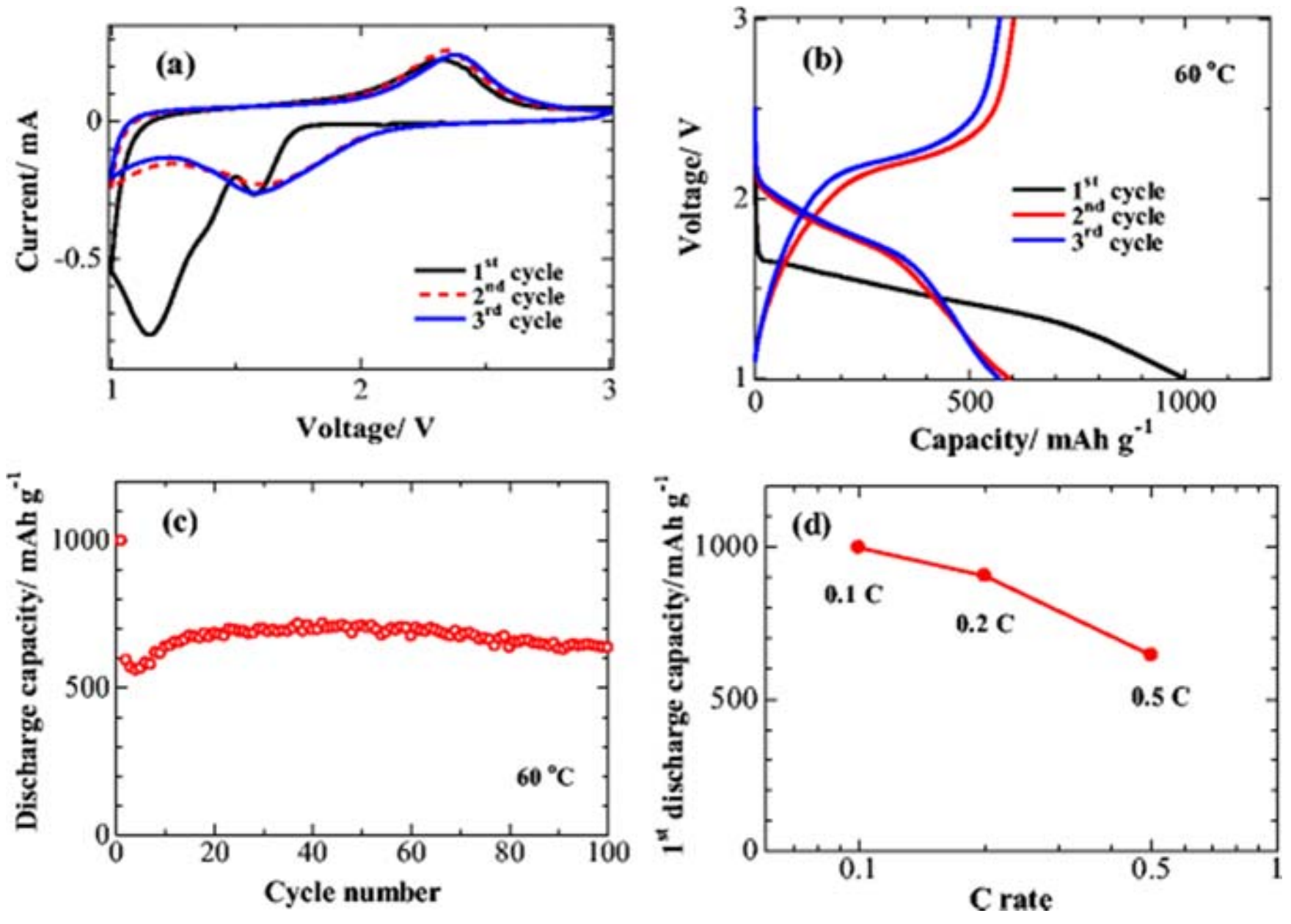

**Figure 42. a** Initial CV profiles of all solid-state Li/S cell at 60 °C; the measurement is conducted at a scan rate of 0.1 mV s−1 in the voltage range of 1.0 to 3.0 V vs. Li+/Li; **b** Charge/discharge profiles (at 0.1 C) of all solid-state Li/S cell at 60 °C; **c** Cycle performance (at 0.1 C) of all solid-state Li/S cell at 60 °C; **d** Rate capability of all solid-state Li/S cell at 60 °C **[**Reproduced from Ref. 114, Copyright (2015), with permission from Springer**].**

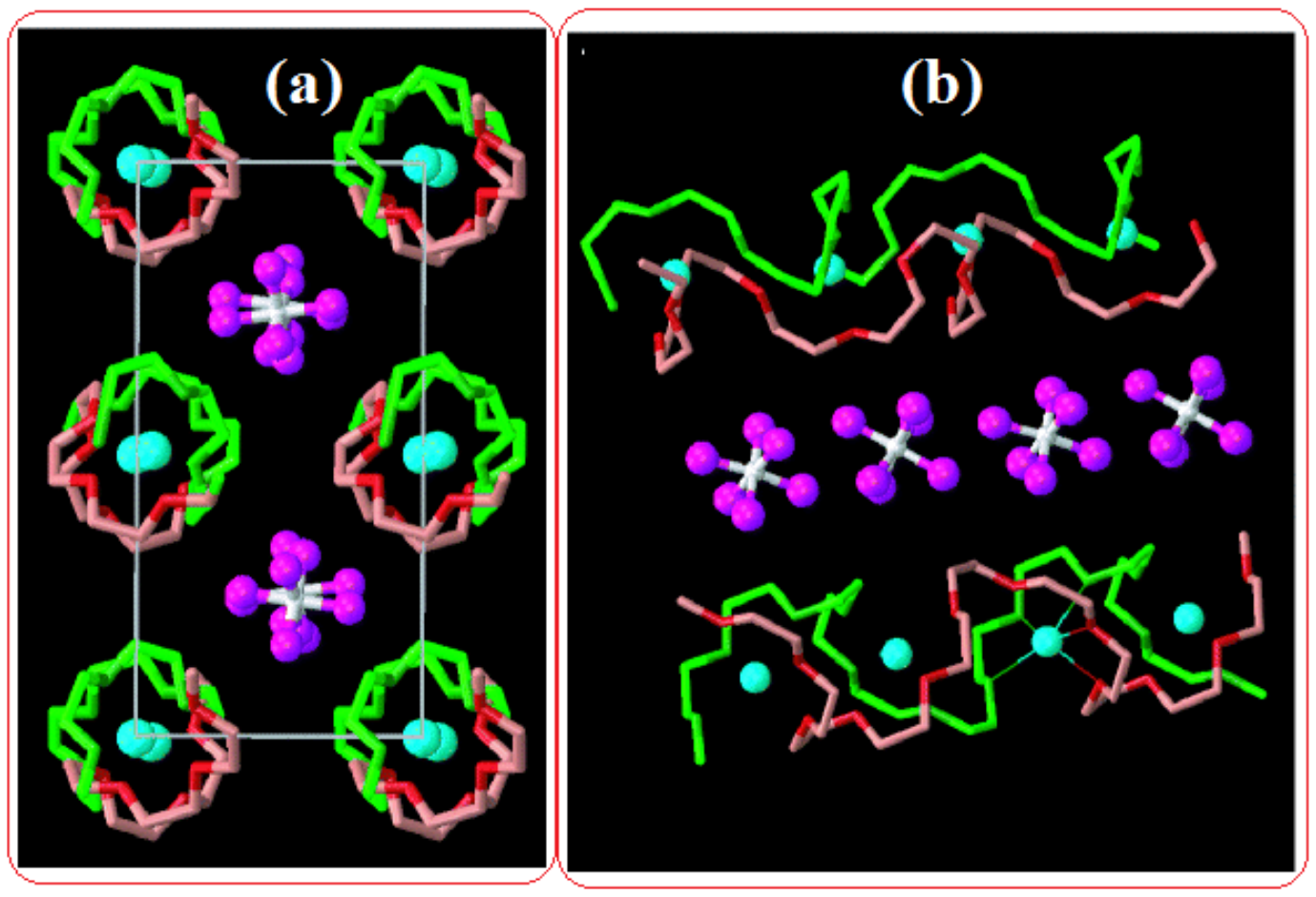

**Figure 43**. **a**, View of the $PEO_6$:$LiAsF_6$ structure along a, showing rows of $Li^+$ ions perpendicular to the page. Blue spheres, lithium; white spheres, arsenic; magenta, fluorine; light green, carbon in chain 1; dark green, oxygen in chain 1; pink, carbon in chain 2; red, oxygen in chain 2. **b,** View of the structure showing the relative positions of the chains and their conformations. Thin lines indicate coordination around the $Li^+$ cation. The lithium–ether oxygen distances (Å) for chain 1 are 2.07(5), 2.26(4), 2.28(4); for chain 2 they are 2.05(5), 2.14(6) [Reproduced from Ref. 119, Copyright (1999), with permission from Nature Publishing Group].

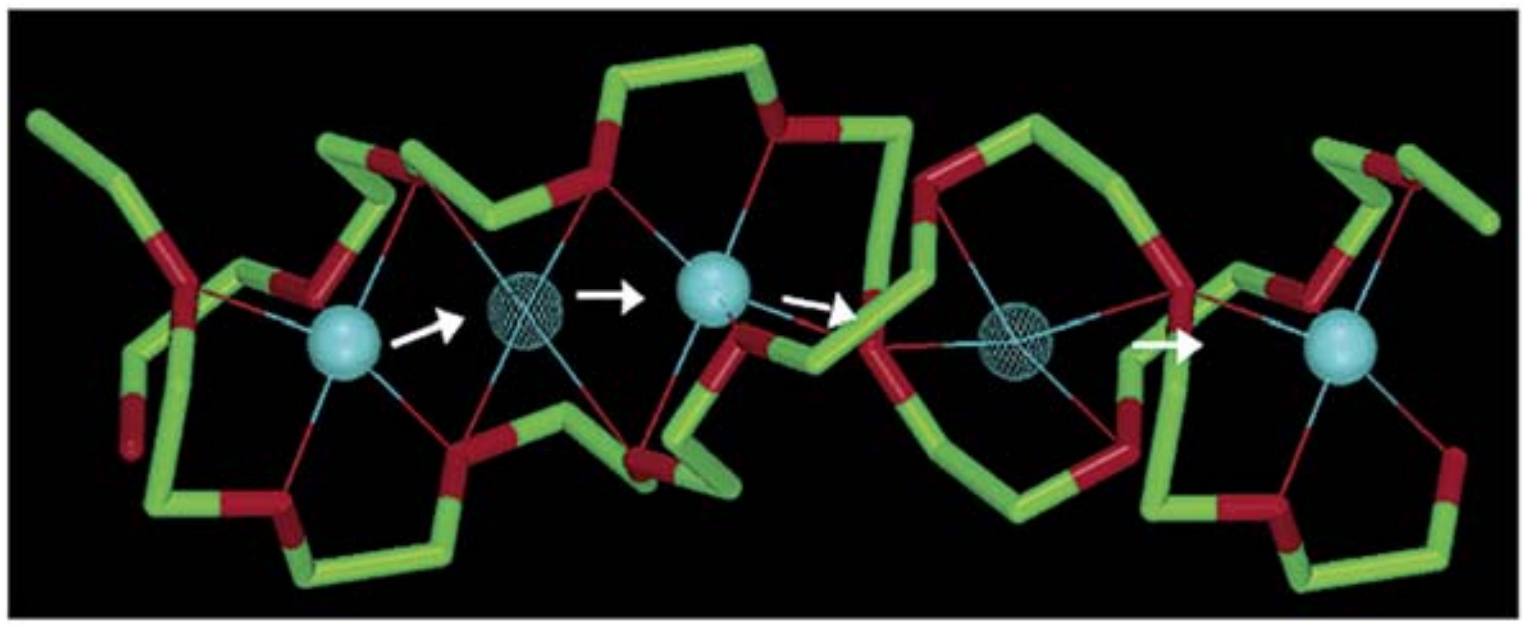

**Figure 44**. I Schematic diffusion pathway of the $Li^+$ cations in $PEO_6$:$LiPF_6$. Thin lines indicate coordination around the $Li^+$ cation; solid blue spheres, lithium in the crystallographic five-coordinate site (note that the fifth thin line is very short in this view); meshed blue spheres, lithium in the intermediate four-coordinate site; green, carbon; red, oxygen [Reprinted with permission from Ref. 120. Copyright (2003) American Chemical Society].

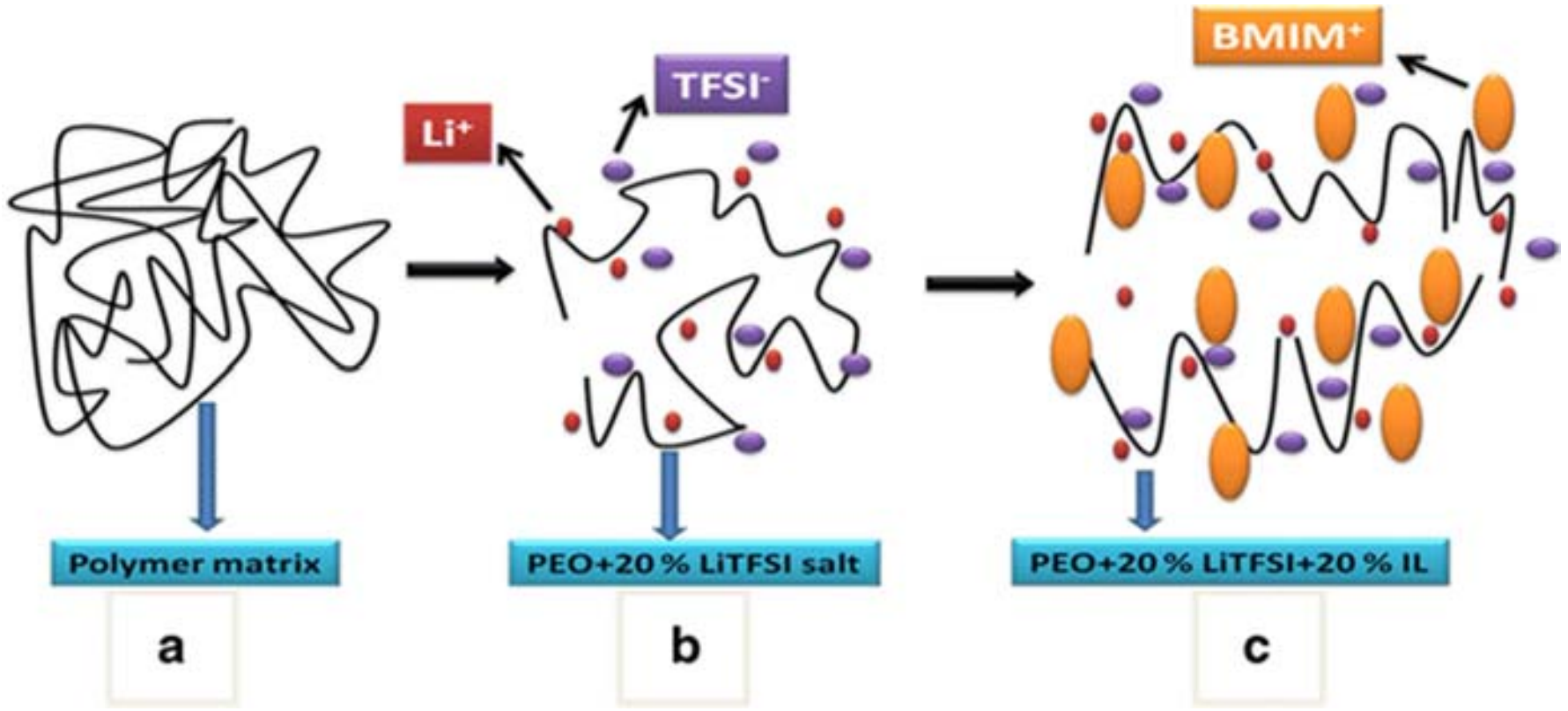

**Figure 45**. Schematic representation to understand the ionic transport behavior of pristine PEO and SPEs PEO + 20 wt. % LiTFSI and PEO + 20 wt. % LiTFSI + 20 wt.% BMIMTFSI at room temperature (30 °C) [Reproduced from Ref. 121, Copyright (2017), with permission from Springer].

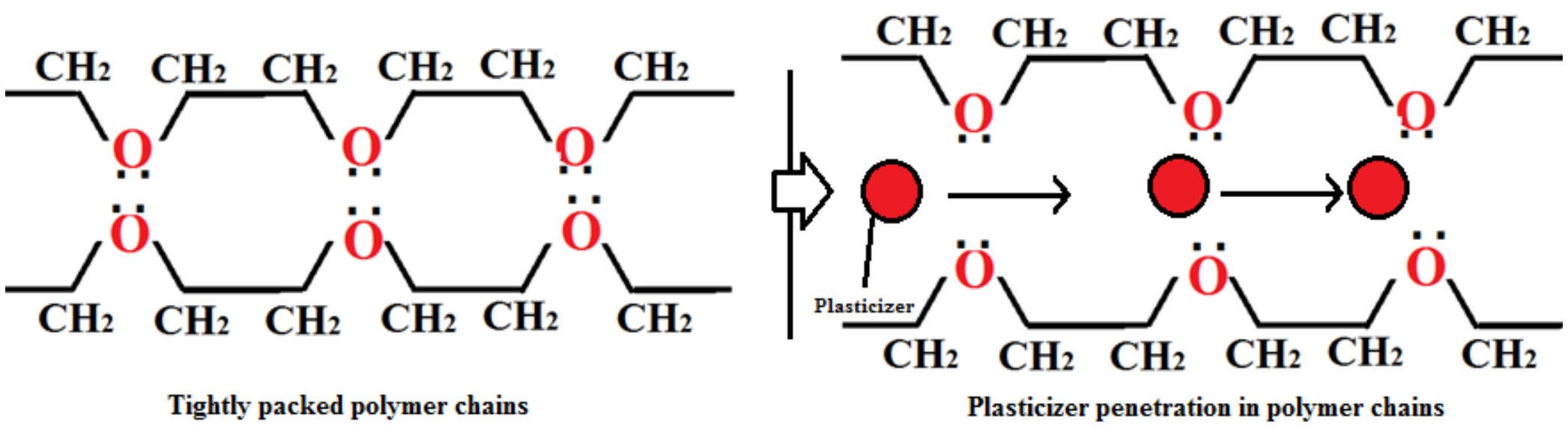

**Figure 46.** Effect of plasticizer in polymer matrix.

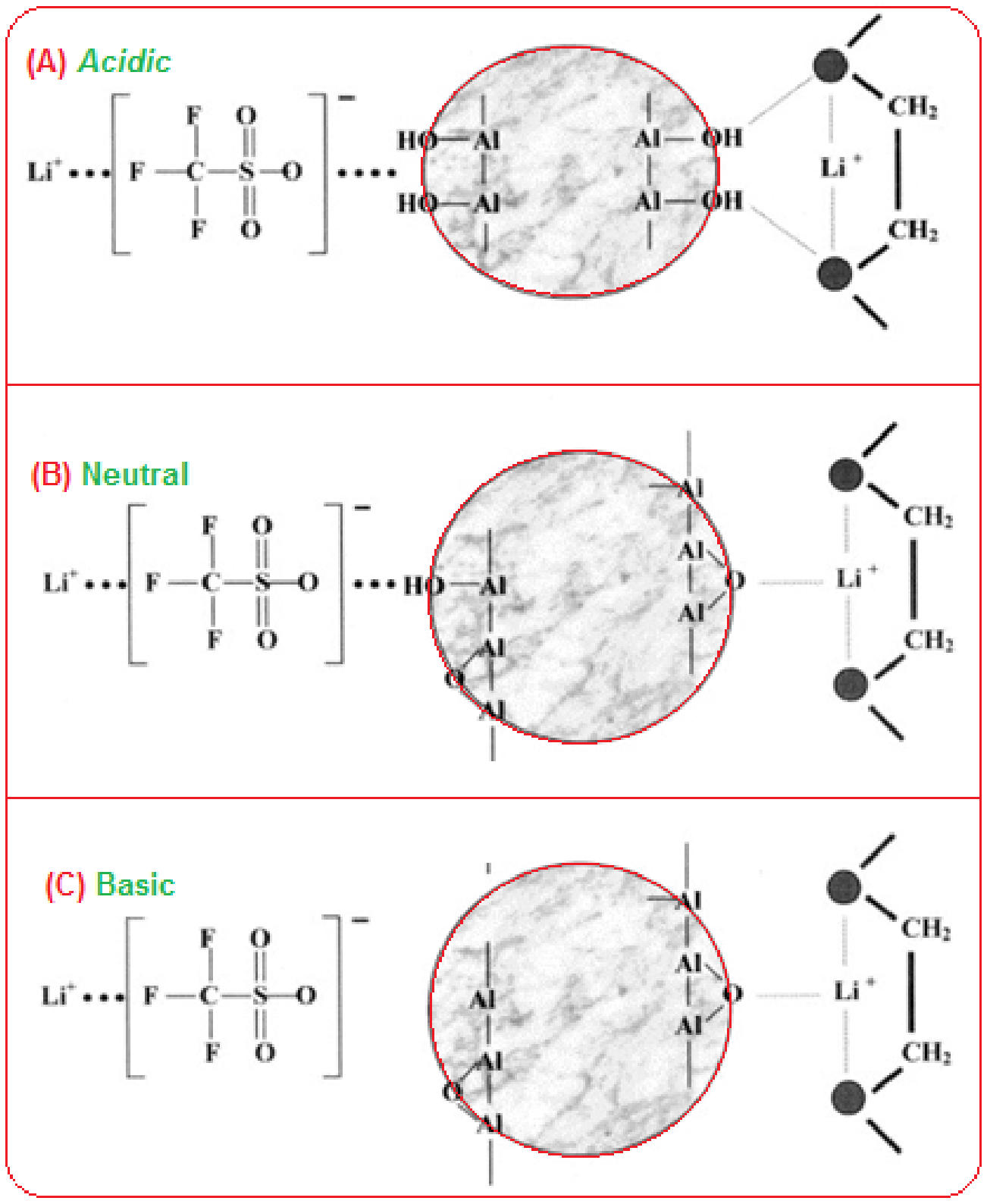

**Figure 47**. Pictorial model of the surface interactions between three forms of dispersed nanosized $Al_2O_3$ ceramic and the PEO–$LiSO_3CF_3$ electrolyte complex. (A) $Al_2O_3$ acidic; (B) $Al_2O_3$ neutral; (C) $Al_2O_3$ basic [Reproduced from Ref. 124, Copyright (2001),with permission from Elsevier].

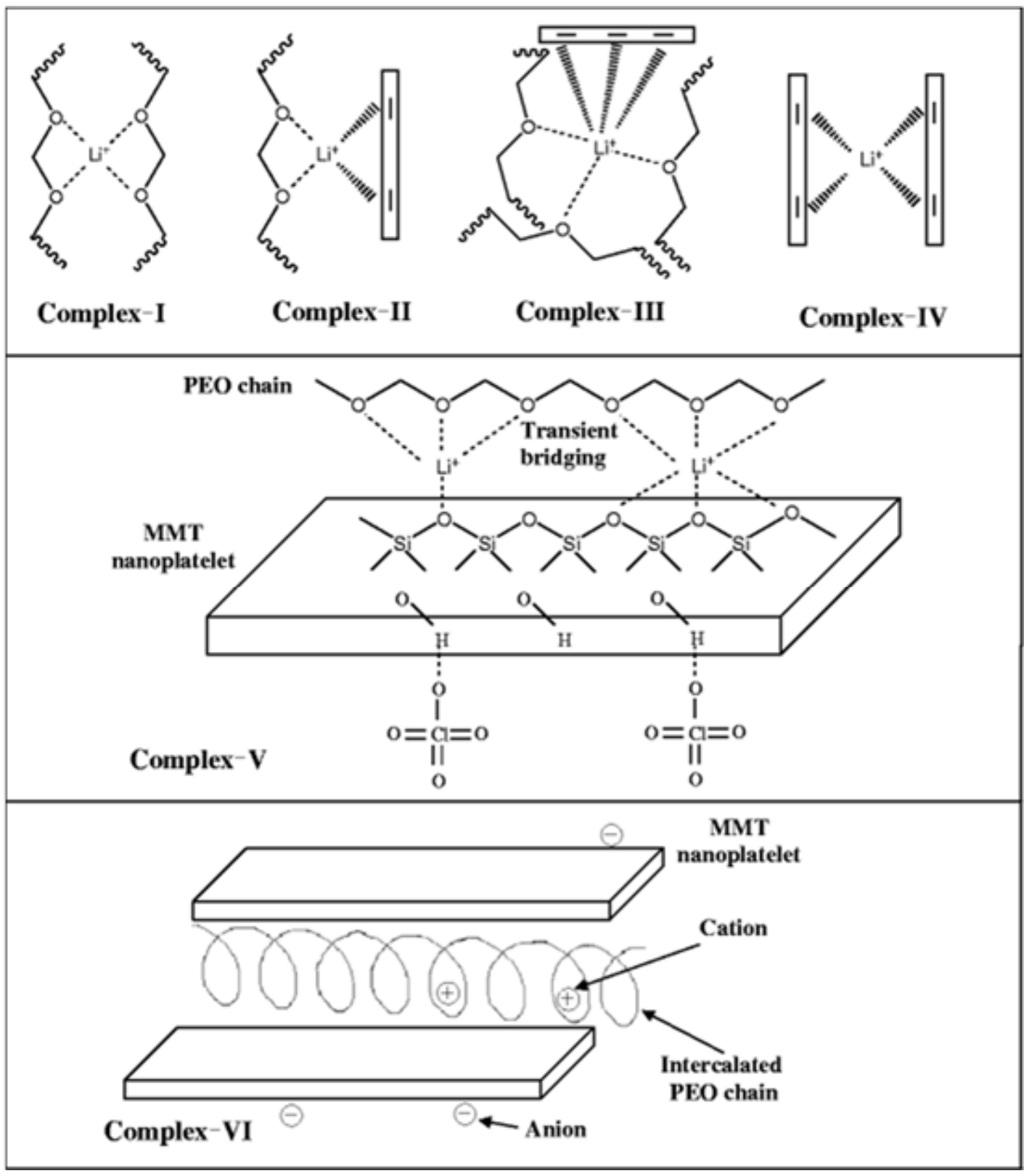

**Figure 48.** Schematic illustration of PEO–$Li^+$–MMT interactions: PEO complex with $Li^+$ (complex I), PEO and MMT complexes with $Li^+$ (complex II and III), MMT complex with $Li^+$ (complex IV), transient cross-linking of PEO and exfoliated MMT nano-platelet through $Li^+$ (complex V), $Li^+$ coordinated PEO intercalation in MMT gallery (complex VI) **[**Reproduced from Ref. 116, Copyright (2011), with permission from Springer**].**

**Table 1.** General requirements for separators used in lithium-ion batteries [Reproduced from Ref. 4, Copyright (2014), with permission from The Royal Society of Chemistry.].

| Parameter | Requirement |
|---|---|
| Chemical and electrochemical stabilities | stable for an extended period of time |
| Wettability | wet out quickly and completely |
| Mechanical property | > 1000 kg/cm (98.06 MPa) |
| Thickness | 20 – 25 μm |
| Pore size | < 1 μm |
| Porosity | 40-60% |
| Permeability (Gurley) | < 0.025 sec/μm |
| Dimensional stability | no curl up and lay flat |
| Thermal stability | < 5% shrinkage after 60 min at 90 °C |
| Shutdown | effectively shut down the battery at elevated temperatures |

Table 2. Merits, demerits and applications of the lithium ion Battery (LIB)

| Merits | Demerits | Applications |
|---|---|---|
| High energy density, and zero memory effect | Requires protection circuit to maintain current and voltage within limits | Solar and wind energy storage |
| Low internal resistance and self-discharge | Suffer from ageing effect | Electric Vehicles and electronics |
| Good coulombic efficiency and high OCV | Degrades at high temperature and when stored at high voltage | Defense Applications |
| Low maintaince cost and long life span | Not fully mature | Communication and space applications |
| Low effective capacity loss at high discharge rate | Transportation regulations required when shipping in larger quantities | Space Applications |
| Fast charging and slow discharging | Battery must be sealed properly | Energy Storage Systems |

Table 3. Fundamental characteristics of various constituents of the polymer electrolyte matrix.

| Polymer Host | Plasticizer |
|---|---|
| • Provide fast segmental motion of polymer chain<br>• Low glass transition temperature<br>• High molecular weight<br>• Low Viscosity<br>• High degradation temperature<br>• High Dielectric Constant<br>• Must have electron donor groups | • Low Melting Point<br>• High Boiling Point<br>• High dielectric constant<br>• Low viscosity<br>• Easily Available<br>• Economic<br>• Inert to both electrodes<br>• Good Safety and Nontoxic Nature |
| Solvent | Nanofiller |
| • Abundant in Nature<br>• Non Aqueous in Nature<br>• Low Melting Point<br>• Low Viscosity<br>• Large Flash Point<br>• High Dielectric Constant<br>• Good Solubility for Polymer and Salt | • High Polarity<br>• Low Melting and High Boiling Point<br>• Safe and Nontoxic<br>• Environmental friendly and cost effective<br>• Inert to All Cell Components.<br>• Act as Lewis Acid for Interaction With Polymer<br>• High Dielectric Constant for better dissociation of salt |
| Salt | Nanoclay |
| • Low Lattice Energy for More Availability of Free Ions<br>• High Ionic Conductivity<br>• High Mobility<br>• Broad Voltage Stability Window<br>• Low Ion Pair Formation at High Content<br>• Large Anion Size<br>• Small Cation size for fast migration between the electrodes<br>• High Thermal and Chemical Stability<br>• Large Ion Transference Number<br>• Inert Towards Cell Components | • Layered/unique structure with high aspect ratio (~1000).<br>• Complex rheological behavior<br>• Amorphous behavior with acid base properties<br>• Greater ability for intercalation and swelling<br>• Increase solubility of salts<br>• High swelling index (water and polar solvents)<br>• High cation exchange capacity (CEC) (~80 meq/100 g)<br>• High external/internal surface area (~31.82 $m^2$ $g^{-1}$)<br>• Appropriate interlayer charge (~0.55)<br>• Adjustable hydrophilic/hydrophobic balance |
| Ionic Liquid | |
| • Good Thermal stability<br>• Wide electrochemical stability<br>• Low Melting Point<br>• Low viscosity for fast transport<br>• Negligible Volatility<br>• Non-flammability<br>• Negligible Vapor Pressure<br>• High Ionic Conductivity<br>• High Polarity<br>• High Dielectric Constant | |

**Table 4.** Ionic conductivity, cation/ion transference number, thermal stability, voltage stability window and cyclic stability of polymer salt complex.

| Materials Used | Electrical Conductivity ($Scm^{-1}$) | Transference number (ion/cation) | Electrochemical Stability Window | Thermal Stability | Cyclic Stability | Ref. |
|---|---|---|---|---|---|---|
| PEO- LiDFOB | $3.18\times10^{-5}$ | - | - | 240 °C | - | 37 |
| PEO-PVP+ $LiNO_3$ | $1.13\times10^{-3}$ S $cm^{-1}$ | 0.332 | 5 V | 400 | - | 32 |
| PEO- LiBNFSI | $2.2\times10^{-4}$ S $cm^{-1}$ | 0.31 | 5.4 | - | - | 38 |
| PEO-PVP+ Mg $(NO_3)_2$ | $5.8\times10^{-4}$ | 0.99/0.33 | 4 | 380 | - | 39 |
| PEO- LiFSI | $1.3\times10^{-3}$ S $cm^{-1}$ at 80 °C | 0.14 | 5.3 | - | Specific charge capacity of 159 mAh $g^{-1}$ and discharge capacity of 146 mAh $g^{-1}$ at the first cycle, coulombic 99% after the first cycle. | 42 |
| PEO- LiTFSI | $1\times10^{-3}$ S $cm^{-1}$ at 80 °C | 0.18 | 5.7 | - | - | 42 |
| PEO- NaPCPI | 0.1 mS $cm^{-1}$ above 60 °C | - | 3 | 600 | - | 43 |
| PEO-NaTIM | - | - | 4.5 | 570 | - | 43 |
| PEO-NaTCP | 1 mS $cm^{-1}$ At 70 °C | - | 5 | 540 | - | 43 |
| PEO- LiTNFSI | $3.69\times10^{-4}$ S $cm^{-1}$ at 90 °C | - | 4 V | >350 °C | average discharge specific capacity of ~450 mAh $g^{-1}$ at 0.2 C for more than 200 cycles | 44 |
| PEO- LiFSI | $1.0\times10^{-4}$ S $cm^{-1}$ at 70 °C | - | - | - | Specific discharge capacity 100–400 $mAhg_{sulfur}^{-1}$ at 0.5C after 50 cycles. | 45 |

**Table 5.** Ionic conductivity, cation/ion transference number, thermal stability, voltage stability window and cyclic stability of ionic liquid and gel polymer electrolytes.

| Materials Used | Electrical Conductivity ($Scm^{-1}$) | Transference number (ion/cation) | Electrochemical Stability Window | Thermal Stability | Cyclic Stability | Ref. |
|---|---|---|---|---|---|---|
| PEO- LiTFSI-IL | $4.2\times10^{-5}$ S $cm^{-1}$ | 0.99/0.37 | 3.34 | 350 °C | - | 51 |
| PEO-LiDFOB- EMImTFSI | $1.85\times10^{-4}$ S $cm^{-1}$ | - | - | - | Initial specific capacity 155 mAh $g^{-1}$ up to 50 cycles. | 52 |
| PEO-LiTFSI- EMIMTFSI | $2.08\times10^{-4}$ S $cm^{-1}$ | 0.99/0.39 | 4.6 | - | The discharge capacity at first cycle is 56 mAh $g^{-1}$ and 120 mAh $g^{-1}$ in the $10^{th}$ cycle. The discharge efficiency (η) was more than 98 % after 100 cycles | 53 |
| $P(EO)_{20}$LiTFSI - BMPyTFSI | $6.9\times10^{-4}$ S $cm^{-1}$ at 40 ◦C | - | 4.8-5.3 | - | - | 54 |
| $P(EO)_{20}$LiTFSI+ 1.27PP1.3TFSI | $2.06\times10^{-4}$ S $cm^{-1}$, $8.68\times10^{-4}$ S $cm^{-1}$ at 60 °C | 0.339 | 4.5-4.7 | 200 °C | Capacity 120 mAh/g with coulomb efficiency greater than 99%. | 56 |
| PEO-NaMS- BMIM-MS | $1.05\times10^{-4}$ S $cm^{-1}$ | 0.99/0.46 | 4-5 | 300 °C | - | 57 |
| PEO-LiTFSI- $PYR_{13}$FSI | $3.4\times10^{-4}$ (-20 °C), $2.43\times10^{-3}$ (20 °C) S $cm^{-1}$, $9.1\times10^{-3}$ S $cm^{-1}$ (60 °C) | - | 4.5 | 200 | Capacity 107 mAh$g^{-1}$ | 59 |
| PEO-LiTFSI-$PYR_{13}$FSI-EC | $1.5\times10^{-4}$ and $1.6\times10^{-3}$ S $cm^{-1}$ at -20 and 20 °C | - | - | - | - | 60 |
| PEO-LiTFSI-BMITFSI | $3.2\times10^{-4}$ $Scm^{-1}$ (25 °C) and $3.2\times10^{-3}$ $Scm^{-1}$ (80 °C) | - | - | - | initial discharge capacity of 90 mAh/g and increases to 140 mAh $g^{-1}$ after 5 cycles. | 61 |
| PEO-LiTFSI- PYR1ATFSI | >$10^{-4}$ S/cm | - | - | - | capacity of 125 mAh $g^{-1}$ and 100 mAh $g^{-1}$ at 30 °C and 25 °C | 62 |
| $PEO_{25}$·LiTf+40 wt.%. EMITf | $10^{-4}$ S $cm^{-1}$ | - | 4 | - | 3.1 F $g^{-1}$, Columbic efficiency (η) ~90 %, | 64 |
| PEO-$Mg(Tf)_2$-EMITf | - | - | 4.8 | - | 2.4 F $g^{-1}$ , Columbic efficiency (η) ~90 %, | 64 |
| $(PEO)_8$LiTFSI- EMImTFSI | $10^{-2}$ S $cm^{-1}$ | - | 3.97 | - | - | 65 |
| PEO-LiTf- EMITf | 3 × $10^{-4}$ S $cm^{-1}$ | 0.99 | 4.9 | - | - | 67 |
| PEO- $LiPF_6$ as liquid electrolyte solution | 3.3×$10^{-3}$ S $cm^{-1}$ | 0.76 | 4.9 | - | reversible capacity 133 $mAhg^{-1}$ and discharge capacity 103 $mAhg^{-1}$ after 500 cycles. Capacity retention 81%. | 68 |
| $PEO_{20}$LiTFSI + EMI-TFSI | 2.67×$10^{-4}$ S $cm^{-1}$ at 40 °C | - | 5.2 | - | specific capacity of 132 $mAhg^{-1}$ at a 0.05 C rate | 69 |
| PEGMEM/SMA | 1.10×$10^{-4}$ S $cm^{-1}$ at 30 °C | - | - | - | initial discharge capacity of 153.5 $mAhg^{-1}$ with capacity retention ratio of 96 % after 300 cycles | 75 |
| PEO-LiTFSI-1% LGPS-SN | 1.58×$10^{-3}$ $Scm^{-1}$ at 80 °C and 9.10×$10^{-5}$ S $cm^{-1}$ at 25 °C | - | 5.5 | 255 | discharge specific 152.1 $mAhg^{-1}$ after 60 cycles with 99.5 % coulombic efficiency | 78 |

**Table 6.** Ionic conductivity, cation/ion transference number, thermal stability, voltage stability window and cyclic stability of dispersed and intercalated type solid polymer electrolytes.

| **Materials Used** | **Electrical Conductivity ($S\ cm^{-1}$)** | **Transference number (ion/cation)** | **Electrochemical Stability Window (V)** | **Thermal Stability (°C)** | **Cyclic Stability** | **Ref.** |
|---|---|---|---|---|---|---|
| PEO-$LiClO_4$+ LATP | $1.71 \times 10^{-4}$ S $cm^{-1}$ at 20 °C | - | - | - | | 85 |
| PEO- LiTF-LGPS | $1.21\times10^{-3}$ S $cm^{-1}$ at 80 °C | - | 5.2 | - | Capacities of 158, 148, 138 and 99 mAh $g^{-1}$ with current rate 0.1 C, 0.2 C, 0.5 C and 1 C at 60 °C | 86 |
| PEO-$LiClO_4$-LAGP | $1.0 \times 10^{-5}$ S $cm^{-1}$ | - | 4.75 | - | initial discharge capacity was 137.6 mAh $g^{-1}$ and 142.2 mAh $g^{-1}$ for the 20th cycle, initial capacities of 133.0~138.5 mAh $g^{-1}$ to 113.4~121.5 mAh $g^{-1}$ at the 100th cycle | 90 |
| PEO-PVdF- $BaTiO_3$ | $1.2\times10^{-4}$ S $cm^{-1}$ | - | 4.7 | 330 | - | 96 |
| $PEO_{20}$–LiBOB-MgO | - | - | 4.2 | - | charge capacity and discharge capacity 168.8 and 156.8 mAh $g^{-1}$, respectively, and discharge capacity degrades to 142.5 mAh $g^{-1}$ after 20 cycles | 97 |
| PEO-LiTFSI- MIL-53(Al) | $3.39\times10^{-3}$ S $cm^{-1}$ at 120 °C | 0.343 | 5.13 | - | Initial discharge capacity 127.1 mA h $g^{-1}$ (at 5 C) at 80 °C and 136.4 mA h $g^{-1}$ at 120 °C. After 300 cycles 116.0 mA h $g^{-1}$ at 80 °C and 129.2 mA h $g^{-1}$ at 120 °C. | 98 |
| (PEO)-$LiClO_4$+$MUSiO_2$ | $1.2 \times 10^{-3}$ at 60 °C | - | 4.7 V for ex-situ CPE and is 5.5 V for in-situ CPE | - | capacity retention 120 mAh $g^{-1}$ was observed for in situ CPE and 65 mAh $g^{-1}$ ex situ CPE | 99 |
| PEO-$LiClO_4$- $BaTiO_3$ | $1\times10^{-5}$ S/cm at 25 °C and $1.2\times10^{-3}$ S/cm at 70 °C | 0.37 | 4 | - | - | 103 |
| PEO-$LiClO_4$-MMT | $9.43\times10^{-4}$ $Scm^{-1}$ | 0.99/0.50 | 3 | - | - | 111 |
| PEO-LiTFSI-HNT | $1.11\times10^{-4}$ at 25 °C,, $1.34\times10^{-3}$ at 25 °C, and $2.14\times10^{-3}$ S $cm^{-1}$ at 100 °C | 0.40 | 6.35, 4.78 | 430 °C | Discharge capacities 745 ± 21 $mAhg^{-1}$ in the 100 discharge/charge cycles with 87 % capacity retention. | 113 |
| PEO- LiTFSI-MMT | $2.75\times10^{-5}$ S $cm^{-1}$ at 25 °C and $3.22\times10^{-4}$ S $cm^{-1}$ at 60 °C | 0.45 | 4 | - | The specific capacity of 998 mAh $g^{-1}$ in the first discharge and a reversible capacity of 591 mAh $g^{-1}$ is achieved in the second cycle. After 100 cycles, the reversible specific discharge capacity of 634 mAh $g^{-1}$ | 114 |